\documentclass[journal]{IEEEtran}

\usepackage{pgfplots} 
\usepackage{pgfplotstable}
\usepackage{graphicx}
\usepackage{subcaption}
\usepackage{caption}

\usepackage{dutchcal}
\usepackage{tikz}
\usepackage{adjustbox}
\usepackage{multicol}
\usepackage{multirow}

\usepackage{tikz,xcolor,pifont}
\usetikzlibrary{patterns,angles,calc,arrows.meta}
\usepackage[
    n, lambda,
    operators,
    advantage,
    sets,
    adversary,
    landau,
    probability,
    notions,    
    logic,
    ff,
    mm,
    primitives,
    events,
    complexity,
    asymptotics,
    keys]{cryptocode}

\usepackage{tikz,xcolor,pifont}
\usetikzlibrary{patterns,angles,calc,arrows.meta}

\usepackage{enumitem}
\usepackage{tabularx}
\usepackage{cite}
\usepackage{float}
\usepackage{acronym} 
\usepackage{multicol}

\ifCLASSINFOpdf
\else
\fi
\usepackage{amsmath}
\usepackage{graphicx}
\usepackage{mathtools}  
\usepackage{amssymb}
\usepackage{tabulary}
\usepackage{booktabs}
\usepackage{booktabs,chemformula}
\usepackage{verbatim}
\usepackage{pifont}

\usepackage{algorithmic}
\usepackage[ruled,vlined]{algorithm2e}

\usepackage{tikz}\usetikzlibrary{patterns,angles,calc,arrows.meta}
\usepackage{xcolor}
\usepackage{pifont}
\usepackage[utf8]{inputenc}
\usepackage[left=2cm, right=2cm, top=2cm]{geometry}
\definecolor{bluef}{HTML}{132380}
\definecolor{titleColor}{rgb}{1.0,0.86,0.6}
\definecolor{nuageColor}{rgb}{0.57,0.73,0.88}

\newcommand\mycommfont[1]{\footnotesize\ttfamily\textcolor{blue}{#1}}

\SetCommentSty{\mycommfont}
\SetKwInput{KwInput}{Input}
\SetKwInput{KwOutput}{Output}
\SetKwInput{KwData}{S1(CA)}

\newcommand{\cmark}{\ding{51}}%
\newcommand{\xmark}{\ding{55}}%

\newcolumntype{L}{>{\centering\arraybackslash}m{3cm}}
\usepackage[acronym]{glossaries}

\makeglossaries
\newacronym{sama}{SAMA}{Secure and Privacy Preserving Data Aggregation}
\newacronym{he}{HE}{Homomorphic Encryption}
\newacronym{fhe}{FHE}{Fully Homomorphic Encryption}
\newacronym{phe}{PHE}{Partial Homomorphic Encryption}
\newacronym{pe}{PE}{Paillier Encryption}

\newacronym{vp}{VP-HE}{Variant Paillier Homomorphic Encryption} 
\newacronym{abe}{ABE}{Attribute Based Encryption}
\newacronym{cpabe}{CP-ABE}{Ciphertext-Policy Attribute-Based Encryption}
\newacronym{ka}{KA}{Key Authority}
 \newacronym{do}{DO}{Data Owner}
  \newacronym{dr}{DR}{Data Requester}
    \newacronym{do-do}{DO-DO}{Data owner access own agg. result}
 \newacronym{drs-do}{DRs-DO}{Data requesters access single data owner agg. result}
  \newacronym{drs-dos}{DRs-DOs}{Data requesters access multiple data owners agg. result}
\newacronym{sp}{SP}{Service Provider}
\newacronym{cp}{CP}{Computation party}

\definecolor{UniBlau}{cmyk}{1,0.7,0,0}

\usepackage{soul}
\newcommand{\added}[1]{{\color{blue} #1}}

\newcommand{\substituted}[2]{\st{#1} {\color{blue} #2}}

\begin{document}
\definecolor{bluef}{HTML}{132380}
\definecolor{titleColor}{rgb}{1.0,0.86,0.6}
\definecolor{nuageColor}{rgb}{0.57,0.73,0.88}

\def\phonexx{\begin{tikzpicture} \fill[black!15, rounded corners=20] (0,0) rectangle (5,10); \fill[fill=bluef!75] (0,1.25) rectangle (5,8.75); \node at (1.7,0.65)[]{\begin{tikzpicture}[scale=1.15] \fill[black!35] (0,0) circle(0.115cm and 0.15cm); \end{tikzpicture}}; \node at (2.1,0.65)[]{\begin{tikzpicture}[scale=1.15] \fill[black!35] (0,0) circle(0.115cm and 0.15cm); \end{tikzpicture}}; \node at (2.5,0.65)[]{\begin{tikzpicture}[scale=1.15] \fill[black!35] (0,0) circle(0.115cm and 0.15cm); \end{tikzpicture}}; \node at (2.9,0.65)[]{\begin{tikzpicture}[scale=1.15] \fill[black!35] (0,0) circle(0.115cm and 0.15cm); \end{tikzpicture}}; \node at (3.3,0.65)[]{\begin{tikzpicture}[scale=1.15] \fill[black!35] (0,0) circle(0.115cm and 0.15cm); \end{tikzpicture}}; \node at (1.7,9.4)[]{\begin{tikzpicture}[scale=1.15] \fill[black!35] (0,0) circle(0.115cm and 0.15cm); \end{tikzpicture}}; \node at (2.1,9.4)[]{\begin{tikzpicture}[scale=1.15] \fill[black!35] (0,0) circle(0.115cm and 0.15cm); \end{tikzpicture}}; \node at (2.5,9.4)[]{\begin{tikzpicture}[scale=1.15] \fill[black!35] (0,0) circle(0.115cm and 0.15cm); \end{tikzpicture}}; \node at (2.9,9.4)[]{\begin{tikzpicture}[scale=1.15] \fill[black!35] (0,0) circle(0.115cm and 0.15cm); \end{tikzpicture}}; \node at (3.3,9.4)[]{\begin{tikzpicture}[scale=1.15] \fill[black!35] (0,0) circle(0.115cm and 0.15cm); \end{tikzpicture}}; \node at (0.5,9.4)[]{\begin{tikzpicture}[scale=1.15] \fill[black!35] (0,0) circle(0.115cm and 0.15cm); \end{tikzpicture}}; \node at (4.5,9.4)[]{\begin{tikzpicture}[scale=1.15] \fill[black!35] (0,0) circle(0.115cm and 0.15cm); \end{tikzpicture}}; \node at (0.7,7.9)[rotate=50, scale=0.6]{\begin{tikzpicture} \fill[blue!60, rounded corners=7] (0,0) rectangle (2,0.5);	\end{tikzpicture}}; \node at (1.1,7.5)[rotate=50, scale=0.6]{\begin{tikzpicture} \fill[blue!60, rounded corners=7] (0,0) rectangle (2,0.5); \fill[blue!60, rounded corners=7] (2.2,0) rectangle (2.7,0.5); \fill[blue!60, rounded corners=7] (2.9,0) rectangle (4.5,0.5);	 \end{tikzpicture}}; \node at (1.55,7.05)[rotate=50, scale=0.65]{\begin{tikzpicture} \fill[blue!60, rounded corners=7] (0,0) rectangle (3.5,0.5); \fill[blue!60, rounded corners=7] (3.7,0) rectangle (4.2,0.5); \fill[blue!60, rounded corners=7] (4.5,0) rectangle (5.5,0.5); \end{tikzpicture}}; \end{tikzpicture}}

\def\ManAvatarxx{\begin{tikzpicture}[fill=black!80] \fill[black!85] (0,4.5)[rounded corners=35] -- (-2,4.5)[rounded corners=11] -- (-2,0.1) -- (-1.2,0.1)[rounded corners=0] -- (-1.2,3) -- (-1,3)[rounded corners=11] -- (-1,-3.5) -- (-0.1,-3.5)[rounded corners=0] -- (-0.1,0) -- (0,0)[rounded corners=0] -- (0.1,0)[rounded corners=11] -- (0.1,-3.5) -- (1,-3.5)[rounded corners=0] -- (1,3) -- (1.2,3)[rounded corners=11] -- (1.2,0.1) -- (2,0.1)[rounded corners=35] -- (2,4.5) -- (0,4.5); \fill[fill=black!85] (0,5.75) circle(1cm); \end{tikzpicture}}

\def\RoundAssociatorxx#1#2{\begin{tikzpicture} \draw[line width=1mm, rounded corners=25, dashed, red] (0,0) rectangle (3,3); \node at (1.5,1.5)[]{#1}; \draw[line width=1mm, rounded corners=25, dashed, red] (0,4) rectangle (3,7); \node at (1.5,5.5)[]{#2}; \draw[line width=1mm, red, dashed] (0,1.5) -- (-3,3.5) -- (0,5.5); \node at (-0.25,1.75)[rotate=60]{\begin{tikzpicture} \fill[black!50] (0,0) circle(0.35cm and 0.4cm); \node at (0,0)[]{\begin{tikzpicture}[scale=0.7] \draw[line width=1mm, red] (0,0) circle(0.35cm and 0.4cm); \end{tikzpicture}}; \node at (0,0)[]{\begin{tikzpicture}[scale=0.4] \fill[white] (0,0) circle(0.35cm and 0.4cm); \end{tikzpicture}}; \end{tikzpicture}}; \node at (-0.25,5.25)[rotate=-60]{\begin{tikzpicture} \fill[black!50] (0,0) circle(0.35cm and 0.4cm); \node at (0,0)[]{\begin{tikzpicture}[scale=0.7] \draw[line width=1mm, red] (0,0) circle(0.35cm and 0.4cm); \end{tikzpicture}};\node at (0,0)[]{\begin{tikzpicture}[scale=0.4] \fill[white] (0,0) circle(0.35cm and 0.4cm); \end{tikzpicture}}; \end{tikzpicture}}; \draw[line width=0.mm, -Stealth] (3.2,1.5) -- (6,3.2); \draw[line width=0.6mm, -Stealth] (3.2,5.5) -- (6,3.8);\end{tikzpicture}}

\def\MagnificClockxx#1{\begin{tikzpicture} \node at (0,0)[scale=#1]{\begin{tikzpicture} \node at (0,0)[]{\begin{tikzpicture} \fill[black!45, rounded corners=5] (0,0) rectangle (1.5,4); \end{tikzpicture}}; \node at (0.2,0.5)[]{\begin{tikzpicture} \fill[black!35, rounded corners=2] (0,0) rectangle (2,0.5); \end{tikzpicture}}; \node at (0,0)[]{\begin{tikzpicture} \fill[black!65, rounded corners=12] (0,0) rectangle (2.2,2.5); \end{tikzpicture}}; \node at (0,0)[scale=0.9]{\begin{tikzpicture} \fill[black!80, rounded corners=11] (0,0) rectangle (2.2,2.5); \end{tikzpicture}}; \node at (0,0)[scale=0.7, xscale=0.85]{\begin{tikzpicture} \fill[white] (0,0) circle(0.54cm); \draw[black!80] (0.9,0) arc(0:{7*360/8}:0.9cm)node[]{\begin{tikzpicture} \fill[green!70] (0,0) circle(0.25cm); \end{tikzpicture}}; \draw[black!80] (0.9,0) arc(0:{6*360/8}:0.9cm)node[]{\begin{tikzpicture} \fill[red!70] (0,0) circle(0.25cm); \end{tikzpicture}}; \draw[black!80] (0.9,0) arc(0:{5*360/8}:0.9cm)node[]{\begin{tikzpicture} \fill[blue!70] (0,0) circle(0.25cm); \end{tikzpicture}}; \draw[black!80] (0.9,0) arc(0:{4*360/8}:0.9cm)node[]{\begin{tikzpicture} \fill[green!70] (0,0) circle(0.25cm); \end{tikzpicture}}; \draw[black!80] (0.9,0) arc(0:{3*360/8}:0.9cm)node[]{\begin{tikzpicture} \fill[blue!70] (0,0) circle(0.25cm); \end{tikzpicture}}; \draw[black!80] (0.9,0) arc(0:{2*360/8}:0.9cm)node[]{\begin{tikzpicture} \fill[red!50] (0,0) circle(0.25cm); \end{tikzpicture}}; \draw[black!80] (0.9,0)node[]{\begin{tikzpicture} \fill[blue!70] (0,0) circle(0.25cm); \end{tikzpicture}} arc(0:{360/8}:0.9cm)node[]{\begin{tikzpicture} \fill[green!70] (0,0) circle(0.25cm); \end{tikzpicture}}; \end{tikzpicture}}; \end{tikzpicture}}; \end{tikzpicture}}

\def\VitesseBulletxx#1{\begin{tikzpicture} \node at (0,0)[scale=#1]{\begin{tikzpicture} \draw[black!75, line width=0.3mm] (0,0) circle(1.2cm and 0.35cm); \fill[black!45] (0,0) circle(1.2cm and 0.35cm); \node at (0,0.22)[]{\begin{tikzpicture} \fill[blue!45] (0,0) rectangle (1.8,0.9); \end{tikzpicture}}; \fill[black!60] (-1.2,0) arc(180:0:1.2cm and 0.35cm) -- (1.2,0.85) arc(0:180:1.2cm and 0.35cm) -- (-1.2,0); \draw[thick, line width=0.5mm] (-1,0.3) -- (-1,0.9); \draw[thick, line width=0.5mm] (1,0.3) -- (1,0.9); \node at (-0.85,0.6)[color=white]{\scriptsize $\bullet$}; \node at (-0.5,0.7)[color=red, scale=1.6]{\ding{170}}; \node at (0.05,0.73)[color=red, scale=1.5, xscale=0.7]{$\sf 120$}; \node at (0.68,0.7)[scale=0.3]{\begin{tikzpicture} \draw[line width=1.5mm, rounded corners=2, red] (-0.06,0) -- (-0.35,-0.3) -- (-0.4,-0.75); \draw[line width=1.5mm, rounded corners=2, red] (0.06,0) -- (0.25,-0.4) -- (0.7,-0.4); \draw[line width=1.5mm, rounded corners=2, red] (-0.06,0.68) -- (-0.35,0.5) -- (-0.7,0.8); \draw[line width=1.5mm, rounded corners=2, red] (0.06,0.7) -- (0.45,0.65) -- (0.65,0.3); \fill[red] (0,1.03) circle(5pt); \fill[red, rounded corners=2] (-0.15,-0.06) rectangle (0.15,0.8); \end{tikzpicture}}; \draw[line width=0.3mm] (1.2,0) arc(0:180:1.2cm and 0.35cm); \end{tikzpicture}}; \end{tikzpicture}}

\def\Tishortxx#1{\begin{tikzpicture} \node at (0,0)[scale=#1]{\begin{tikzpicture} \fill[black!10, rounded corners=3] (-1.45,1.4) -- (-1.5,0) to [bend right=20](1.5,0) -- (1.4,2.8) to [bend left=25](1,3.4) -- (0.95,4.2) -- (0.5,4.2) .. controls(0.2,3.7) and (-0.2,3.7) .. (-0.5,4.2) -- (-0.95,4.2) -- (-1,3.4) to [bend left](-1.4,2.8) -- (-1.45,1.4); \draw[line width=0.5mm, rounded corners=3] (-1.45,1.4) -- (-1.5,0) to [bend right=20](1.5,0) -- (1.4,2.8) to [bend left=25](1,3.4) -- (0.95,4.2) -- (0.5,4.2) .. controls(0.2,3.7) and (-0.2,3.7) .. (-0.5,4.2) -- (-0.95,4.2) -- (-1,3.4) to [bend left](-1.4,2.8) -- (-1.45,1.4); \fill[black!75] (0.5,4.2) .. controls(0.2,3.7) and (-0.2,3.7) .. (-0.5,4.2) -- (-0.75,4.2) .. controls(-0.6,3) and (0.6,3) .. (0.75,4.2); \draw[line width=0.5mm] (0.15,-0.3) -- (0.15,2);\node at (0.39,1.85)[scale=0.7]{\begin{tikzpicture} \fill[black!75, rounded corners=2] (0,0.2) -- (0,0) to [bend right=22](0.7,0) -- (0.7,0.4) to [bend left=22](0,0.4) -- (0,0.2); \end{tikzpicture}}; \draw[line width=0.5mm] (-0.15,-0.3) -- (-0.15,2); \node at (-0.39,1.85)[scale=0.7]{\begin{tikzpicture} \fill[black!75, rounded corners=2] (0,0.2) -- (0,0) to [bend right=22](0.7,0) -- (0.7,0.4) to [bend left=22](0,0.4) -- (0,0.2); \end{tikzpicture}}; \draw[line width=0.5mm, rounded corners=2] (0.35,-0.28) -- (0.35,1.21) -- (0.725,1.21) -- (0.8,2.75); \node at (0.9,2.75)[scale=0.7]{\begin{tikzpicture} \fill[black!75, rounded corners=2] (0,0.2) -- (0,0) to [bend right=22](0.7,0) -- (0.7,0.4) to [bend left=22](0,0.4) -- (0,0.2); \end{tikzpicture}}; \draw[line width=0.5mm, rounded corners=2] (-0.35,-0.28) -- (-0.35,1.21) -- (-0.725,1.21) -- (-0.8,2.75); \node at (-0.9,2.75)[scale=0.7]{\begin{tikzpicture} \fill[black!75, rounded corners=2] (0,0.2) -- (0,0) to [bend right=22](0.7,0) -- (0.7,0.4) to [bend left=22](0,0.4) -- (0,0.2); \end{tikzpicture}}; \draw[line width=0.5mm] (-0.5,-0.26) -- (-0.5,0.3) -- (-0.85,0.3) -- (-0.85,0.7); \node at (-1.08,0.7)[scale=0.7]{\begin{tikzpicture} \fill[black!75, rounded corners=2] (0,0.2) -- (0,0) to [bend right=22](0.7,0) -- (0.7,0.4) to [bend left=22](0,0.4) -- (0,0.2); \end{tikzpicture}}; \node at (1.2,0.6)[scale=0.5]{\begin{tikzpicture} \fill[black!10] (0,0) circle(1.15cm); \draw[thick] (0,0) circle(1cm); \draw[thick] (0,0) circle(0.9cm); \draw[thick] (0,0) circle(0.8cm); \fill[black!10] (-0.1,-1.1) rectangle (0.1,1.1); \fill[black!10] (-1.1,-0.1) rectangle (1.1,0.1); \draw[thick, rounded corners=2] (-0.1,-0.7) rectangle (0.1,0.7); \draw[thick, rounded corners=2] (-0.5,0.3) rectangle (0.5,0.5); \end{tikzpicture}}; \end{tikzpicture}}; \end{tikzpicture}}

\def\Nuagexx#1#2{\begin{tikzpicture}\node at (0,0)[scale=#2]{\begin{tikzpicture} \draw[black!40,line width=0.8mm, fill=nuageColor] (-1,1)[rounded corners=15] -- (-1.5,1) -- (-1.5,0) -- (1.5,0) -- (1.5,1)[rounded corners=0] -- (1,1) .. controls(1,1.8) and (-0.25,2) .. (-0.5,1.4) to [bend right=70](-1,1); \draw[line width=0.8mm, black!40] (-0.5,1.4) to [bend left](-0.22,1.1); \draw[line width=0.8mm, black!40] (1,1) -- (0.7,1); \node at (0,0.15)[scale=0.65]{\begin{tikzpicture} \node at (0.75,1.1)[]{\begin{tikzpicture} \draw[line width=1.5mm, rounded corners=13, white] (0,0) -- (0,1.2) -- (0.9,1.2) -- (0.9,0); \end{tikzpicture}}; \fill[blue!20] (0,0) rectangle (1.5,1.15); \node at (0.75,0.575)[]{\begin{tikzpicture} \fill[black!70, rounded corners=2] (0,0) rectangle (0.15,0.5); \fill[black!70] (0.075,0.4) circle(0.2cm); \end{tikzpicture}}; \end{tikzpicture}}; \node at (0,0.9)[scale=1.2]{\Large $\mbox{CSP}_{\mbox{\scriptsize #1}}$}; \end{tikzpicture}};\end{tikzpicture}}

\def\Titledefxx#1#2{\begin{tikzpicture}[xscale=#2, yscale=1.3] \fill[titleColor, rounded corners=5] (0,0) rectangle (3.5,0.5); \node at (1.75,0.25)[color=black]{\sf #1}; \end{tikzpicture}}

\def\Keyxx#1{\begin{tikzpicture} \node at (0,0)[scale=#1]{\begin{tikzpicture}[xscale=1.15, yscale=1.1] \filldraw[line width=0.8mm, fill=titleColor] (1,-0.25) -- (2.5,-0.25) arc(-155:0:0.9cm and 0.8cm) arc(0:165:0.9cm and 0.8cm) -- (2.2,0.35) -- (2,0.2) -- (1.8,0.4) -- (1.6,0.2) -- (1.4,0.45) -- (1.2,0.2) -- (1,0.45) -- (0.75,0.1) -- (1,-0.25); \filldraw[line width=0.8mm, fill=white] (3.7,0.1) circle(6pt and 5pt); \end{tikzpicture}}; \end{tikzpicture}}

\def\FirstHomexx#1{\begin{tikzpicture} \node at (0,0)[scale=#1]{\begin{tikzpicture} \draw[line width=1mm, fill=bluef!55] (4,0) rectangle (8,4.5); \draw[line width=0.8mm] (4,3.9) -- (6.1,3.9) (6.3,3.9) -- (6.6,3.9) (6.9,3.9) -- (8,3.9); \node at (4.8,0.8)[scale=0.65]{\begin{tikzpicture} \draw[line width=0.5mm, fill=white] (0.5,1.5) rectangle (1.5,2.25); \end{tikzpicture}}; \node at (4.8,1.6)[scale=0.65]{\begin{tikzpicture} \draw[line width=0.5mm, fill=white] (0.5,1.5) rectangle (1.5,2.25); \end{tikzpicture}}; \node at (4.8,2.4)[scale=0.65]{\begin{tikzpicture} \draw[line width=0.5mm, fill=white] (0.5,1.5) rectangle (1.5,2.25); \end{tikzpicture}}; \node at (4.8,3.2)[scale=0.65]{\begin{tikzpicture} \draw[line width=0.5mm, fill=white] (0.5,1.5) rectangle (1.5,2.25); \end{tikzpicture}}; \node at (6,0.8)[scale=0.65]{\begin{tikzpicture} \draw[line width=0.5mm, fill=white] (0.5,1.5) rectangle (1.5,2.25); \end{tikzpicture}}; \node at (6,1.6)[scale=0.65]{\begin{tikzpicture} \draw[line width=0.5mm, fill=white] (0.5,1.5) rectangle (1.5,2.25); \end{tikzpicture}}; \node at (6,2.4)[scale=0.65]{\begin{tikzpicture} \draw[line width=0.5mm, fill=white] (0.5,1.5) rectangle (1.5,2.25); \end{tikzpicture}}; \node at (6,3.2)[scale=0.65]{\begin{tikzpicture} \draw[line width=0.5mm, fill=white] (0.5,1.5) rectangle (1.5,2.25); \end{tikzpicture}}; \node at (7.2,0.8)[scale=0.65]{\begin{tikzpicture} \draw[line width=0.5mm, fill=white] (0.5,1.5) rectangle (1.5,2.25); \end{tikzpicture}}; \node at (7.2,1.6)[scale=0.65]{\begin{tikzpicture} \draw[line width=0.5mm, fill=white] (0.5,1.5) rectangle (1.5,2.25); \end{tikzpicture}}; \node at (7.2,2.4)[scale=0.65]{\begin{tikzpicture} \draw[line width=0.5mm, fill=white] (0.5,1.5) rectangle (1.5,2.25); \end{tikzpicture}}; \node at (7.2,3.2)[scale=0.65]{\begin{tikzpicture} \draw[line width=0.5mm, fill=white] (0.5,1.5) rectangle (1.5,2.25); \end{tikzpicture}}; \node at (5.9,4.5)[]{\begin{tikzpicture} \filldraw[line width=1mm, fill=black!25, rounded corners=8] (-0.3,0) rectangle (4.6,0.5); \end{tikzpicture}}; \fill[black!25] (0,0) rectangle (4,0.85); \fill[bluef!55] (0,0.85) rectangle (4,6); \draw[line width=0.6mm] (0,0.85) -- (0.4,0.85) (0.8,0.85) -- (3.2,0.85) (3.6,0.85) -- (4,0.85); \draw[line width=0.6mm] (0,5.5) -- (2,5.5) (2.4,5.5) -- (3,5.5) (3.4,5.5) -- (4,5.5); \node at (1,1.75)[scale=0.85]{\begin{tikzpicture} \draw[line width=0.5mm, fill=white] (0.5,1.5) rectangle (1.5,2.25); \end{tikzpicture}}; \node at (1,2.75)[scale=0.85]{\begin{tikzpicture} \draw[line width=0.5mm, fill=white] (0.5,1.5) rectangle (1.5,2.25); \end{tikzpicture}}; \node at (1,3.75)[scale=0.85]{\begin{tikzpicture} \draw[line width=0.5mm, fill=white] (0.5,1.5) rectangle (1.5,2.25); \end{tikzpicture}}; \node at (1,4.75)[scale=0.85]{\begin{tikzpicture} \draw[line width=0.5mm, fill=white] (0.5,1.5) rectangle (1.5,2.25); \end{tikzpicture}}; \node at (3,1.75)[scale=0.85]{\begin{tikzpicture} \draw[line width=0.5mm, fill=white] (0.5,1.5) rectangle (1.5,2.25); \end{tikzpicture}}; \node at (3,2.75)[scale=0.85]{\begin{tikzpicture} \draw[line width=0.5mm, fill=white] (0.5,1.5) rectangle (1.5,2.25); \end{tikzpicture}}; \node at (3,3.75)[scale=0.85]{\begin{tikzpicture} \draw[line width=0.5mm, fill=white] (0.5,1.5) rectangle (1.5,2.25); \end{tikzpicture}}; \node at (3,4.75)[scale=0.85]{\begin{tikzpicture} \draw[line width=0.5mm, fill=white] (0.5,1.5) rectangle (1.5,2.25); \end{tikzpicture}}; \draw[line width=1mm] (0,0) rectangle (4,6); \node at (2,6.15)[]{\begin{tikzpicture} \filldraw[line width=1mm, fill=black!25, rounded corners=8] (-0.3,0) rectangle (4.6,0.5); \end{tikzpicture}}; \draw[line width=1.1mm, fill=white, rounded corners=20] (1.35,0) -- (1.35,1.5) -- (2.65,1.5) -- (2.65,0); \draw[line width=1mm] (-0.8,0) -- (8.8,0); \end{tikzpicture}}; \end{tikzpicture}}

\def\ContactIconeFirstxx#1#2{
	\begin{tikzpicture}
		\node at (0,0)[scale=#1]{
			\begin{tikzpicture}
				\draw[line width=0.6mm, rounded corners=35, fill=white] (-4.5,-4.4) rectangle (6,3); \node at (0.75,-3.5)[scale=#2]{\Huge \sf Insurance Staff}; \node at (-2,-1)[scale=0.55]{\begin{tikzpicture} \fill[black] (0,0)[rounded corners=15] -- (2.3,0)[rounded corners=2] to [bend right](0.7,3)[rounded corners=2] -- (0,1.75)[rounded corners=2] -- (-0.7,3)[rounded corners=15] to [bend right](-2.3,0)[rounded corners=2] -- (0,0); \fill[black] (0,2.95) -- (-0.17,2.45) -- (0,2.15) -- (0.17,2.45) -- (0,2.95); \fill[white] (0,0.15) -- (1.4,0.15) -- (1.6,1.5) -- (-1.6,1.5) -- (-1.4,0.15) -- (0,0.15); \fill[black] (0,4) circle(0.9cm); \draw[line width=0.4cm] (-2.8,-0.4) -- (2.8,-0.4); \end{tikzpicture}}; \node at (0.8,0.7)[scale=0.55]{\begin{tikzpicture} \fill[black] (0,0)[rounded corners=3] -- (1.2,0)[rounded corners=3] -- (1.2,0.7)[rounded corners=3] to [bend right](1.65,0.9)[rounded corners=8] -- (1.65,3.25)[rounded corners=1] to [bend right=20](0.5,4) -- (0,3.5) -- (-0.5,4)[rounded corners=8] to [bend right=20](-1.65,3.25)[rounded corners=3] -- (-1.65,0.9)[rounded corners=3] to [bend right](-1.2,0.7) -- (-1.2,0) -- (0,0); \fill[black] (0,5) circle(0.85cm); \node at (1.2,1)[scale=0.7]{\begin{tikzpicture}\node at (0,0)[]{\begin{tikzpicture} \filldraw[white, line width=4mm] (0,0) to [bend right=40](1.4,1.5)[rounded corners=3] -- (1.4,2.4)[rounded corners=5] -- (1.6,2.6)[rounded corners=5] to [bend right=10](0,2.9)[rounded corners=5] to [bend right=10](-1.6,2.6)[rounded corners=3] -- (-1.4,2.4) -- (-1.4,1.5) to [bend right=40](0,0);\end{tikzpicture}}; \node at (0,0)[scale=0.9]{\begin{tikzpicture} \draw[line width=2mm, fill=black] (0,0) to [bend right=40](1.4,1.5)[rounded corners=3] -- (1.4,2.4)[rounded corners=5] -- (1.6,2.6)[rounded corners=0] to [bend right=10](0,2.9) -- (0,0); \draw[line width=2mm] (0,0) to [bend left=40](-1.4,1.5)[rounded corners=3] -- (-1.4,2.4)[rounded corners=5] -- (-1.6,2.6)[rounded corners=0] to [bend left=10](0,2.9) -- (0,0); \end{tikzpicture}}; \node at (1,1.3)[scale=0.8]{\begin{tikzpicture} \fill[white] (0,0) circle(1.5cm); \fill[black] (0,0) circle(1.25cm); \node at (0,0)[]{\begin{tikzpicture} \draw[line width=3mm, white] (-0.5,0.5) -- (0,0) -- (0.8,1.2); \end{tikzpicture}}; \end{tikzpicture}}; \end{tikzpicture}}; \node at (0.85,3.5)[]{\begin{tikzpicture} \fill[white] (0,0) circle(8pt); \draw[white, line width=2mm] (0,0) -- (0,1.25); \end{tikzpicture}}; \node at (-0.8,3.2)[]{\begin{tikzpicture} \draw[line width=2mm, white] (-0.25,0.1)[rounded corners=3] -- (-0.4,0.15)[rounded corners=10]  -- (-0.4,1) -- (0.4,1)[rounded corners=3] -- (0.4,0.15) -- (0.25,0.1); \draw[line width=2mm, white] (0,1) -- (0,1.6); \end{tikzpicture}}; \end{tikzpicture}}; \node at (3.3,-1.3)[scale=0.8]{\begin{tikzpicture} \node at (0,-0.2)[scale=0.5]{\begin{tikzpicture} \fill[black!70, rounded corners=5] (0,0) -- (1.5,0) to [bend right=15](1.3,1.8) to [bend right=10](0.6,2) to [bend left](-0.6,2) to [bend right=10](-1.3,1.8) to [bend right=15](-1.5,0) -- (0,0); \fill[black!70] (0,3) circle(0.86cm); \end{tikzpicture}};	\node at (1.5,0)[]{\begin{tikzpicture}\node at (0,0)[]{\begin{tikzpicture}\fill[white] (0,0)[rounded corners=13] -- (1.3,0)[rounded corners=20] -- (1.3,1.65)[rounded corners=0] -- (0.5,1.65) -- (0,0.7) -- (-0.5,1.65)[rounded corners=20] -- (-1.3,1.65)[rounded corners=13] -- (-1.3,0) -- (0,0);\fill[white, rounded corners=3] (0.12,0.5) -- (0.05,1.25) -- (0.2,1.65) -- (-0.2,1.65) -- (-0.05,1.25) -- (-0.12,0.5); \fill[white] (0,2.7) circle(0.88cm and 0.85cm); \end{tikzpicture}}; \node at (0,-0.05)[scale=0.95]{\begin{tikzpicture}\fill[black!70] (0,0)[rounded corners=13] -- (1.3,0)[rounded corners=20] -- (1.3,1.65)[rounded corners=0] -- (0.5,1.65) -- (0,0.7) -- (-0.5,1.65)[rounded corners=20] -- (-1.3,1.65)[rounded corners=13] -- (-1.3,0) -- (0,0);\fill[black!70, rounded corners=3] (0.12,0.5) -- (0.05,1.25) -- (0.2,1.65) -- (-0.2,1.65) -- (-0.05,1.25) -- (-0.12,0.5); \fill[black!70] (0,2.7) circle(0.88cm and 0.85cm); \end{tikzpicture}}; \end{tikzpicture}};\end{tikzpicture}}; \end{tikzpicture}}; \end{tikzpicture}}

\def\ContactIconeSecondxx#1#2{
	\begin{tikzpicture}
		\node at (0,0)[scale=#1]{
			\begin{tikzpicture}
				\draw[line width=0.6mm, rounded corners=35, fill=white] (-4.5,-4.4) rectangle (6,3);	\node at (0.75,-3.5)[scale=#2]{\Huge \sf Researchers};	\node at (4.7,1.5)[scale=0.35]{\begin{tikzpicture} \draw[line width=0.9mm] (0,0) circle(2cm); \node at (0,0.35)[scale=1.1]{\begin{tikzpicture} \node at (0,0)[scale=0.65, yscale=1.3, xscale=1]{\begin{tikzpicture}\fill[black!80, rounded corners=4] (0,0) -- (2,0) -- (2,1) -- (0.3,1.5) -- (0,0.8) -- (-0.3,1.5) -- (-2,1) -- (-2,0) -- (0,0); \end{tikzpicture}}; \node at (0,1.35)[scale=1.18]{\begin{tikzpicture} \fill[black!80] (0.4,1) to [bend right=70](-0.4,1)[rounded corners=5] to [bend right=40](0,0) to [bend right=40](0.4,1); \end{tikzpicture}}; \end{tikzpicture}}; \end{tikzpicture}}; \node at (1.7,1.5)[scale=0.35]{\begin{tikzpicture} \draw[line width=0.9mm] (0,0) circle(2cm); \node at (0,0.35)[scale=1.1]{\begin{tikzpicture} \node at (0,0)[scale=0.65, yscale=1.3, xscale=1]{\begin{tikzpicture}\fill[black!80, rounded corners=4] (0,0) -- (2,0) -- (2,1) -- (0.3,1.5) -- (0,0.8) -- (-0.3,1.5) -- (-2,1) -- (-2,0) -- (0,0); \end{tikzpicture}}; \node at (0,1.35)[scale=1.18]{\begin{tikzpicture} \fill[black!80] (0.4,1) to [bend right=70](-0.4,1)[rounded corners=5] to [bend right=40](0,0) to [bend right=40](0.4,1); \end{tikzpicture}}; \end{tikzpicture}}; \end{tikzpicture}}; \node at (3.7,-0.7)[scale=0.85]{\begin{tikzpicture}[scale=1.2] \node at (-0.04,-0.04)[rotate=38]{\begin{tikzpicture} \draw[line width=2.5mm] (0,0) circle(1.2cm); \draw[line width=2.5mm] (0,-1.2) -- (0,-2); \draw[line width=4.5mm] (0,-1.6) -- (0,-2); \draw[line width=7.5mm] (0,-1.85) -- (0,-3); \end{tikzpicture}}; \draw[line width=0.8mm] (-2.03,0) -- (0,0); \draw[line width=0.8mm] (-2.03,-0.15) -- (0,-0.15); \draw[line width=3mm] (-0.1,0.1) -- (-0.1,1); \draw[line width=3mm] (-0.55,0.1) -- (-0.55,1.4); \draw[line width=3mm] (-1,0.1) -- (-1,0.65); \draw[line width=3mm, white] (-1.45,0.1) -- (-1.45,0.9); \draw[line width=1mm] (-1.35,0.1) -- (-1.35,0.9); \draw[line width=3mm] (-1.9,0.1) -- (-1.9,0.55);\end{tikzpicture}}; \node at (1.7,-1.8)[scale=0.35]{\begin{tikzpicture} \draw[line width=0.9mm] (0,0) circle(2cm); \node at (0,0.35)[scale=1.1]{\begin{tikzpicture} \node at (0,0)[scale=0.65, yscale=1.3, xscale=1]{\begin{tikzpicture}\fill[black!80, rounded corners=4] (0,0) -- (2,0) -- (2,1) -- (0.3,1.5) -- (0,0.8) -- (-0.3,1.5) -- (-2,1) -- (-2,0) -- (0,0); \end{tikzpicture}}; \node at (0,1.35)[scale=1.18]{\begin{tikzpicture} \fill[black!80] (0.4,1) to [bend right=70](-0.4,1)[rounded corners=5] to [bend right=40](0,0) to [bend right=40](0.4,1); \end{tikzpicture}}; \end{tikzpicture}}; \end{tikzpicture}};	\node at (-0.8,1.5)[scale=0.95]{\begin{tikzpicture}\draw[line width=1.2mm, black!80] (0.12,0.2) -- (0.64,0.2) (0.76,0.2) -- (0.88,0.2); \draw[line width=1.2mm, black!80] (0.12,0.45) -- (0.88,0.45); \draw[line width=1.2mm, black!80] (0.12,0.7) -- (0.88,0.7); \draw[line width=1.2mm, black!80] (0.12,0.95) -- (0.88,0.95); \draw[line width=1.2mm, black!80] (-0.7,0.15) -- (0.16,0.96);\end{tikzpicture}};	\node at (-3,1.5)[scale=0.7]{\begin{tikzpicture} \draw[line width=1.2mm, black!80] (0,0) -- (2.25,0); \draw[line width=3mm, black!80] (0.4,0) -- (0.4,1.2); \draw[line width=3mm, black!80] (1.1,0) -- (1.1,0.8); \draw[line width=3mm, black!80] (1.8,0) -- (1.8,0.5);	 \end{tikzpicture}};	\node at (-0.4,-0.3)[scale=0.8]{\begin{tikzpicture} \node at (0,0)[]{\begin{tikzpicture} \draw[line width=1.5mm] (0,0) -- (0,1); \end{tikzpicture}}; \node at (0,0)[rotate=45]{\begin{tikzpicture} \draw[line width=1.5mm] (0,0) -- (0,1); \end{tikzpicture}}; \node at (0,0)[rotate=90]{\begin{tikzpicture} \draw[line width=1.5mm] (0,0) -- (0,1); \end{tikzpicture}}; \node at (0,0)[rotate=135]{\begin{tikzpicture} \draw[line width=1.5mm] (0,0) -- (0,1); \end{tikzpicture}}; \fill[white] (0,0) circle(0.27cm); \draw[line width=1.5mm] (0,-0.7) -- (0,-1.3) (0,-1.45) -- (0,-1.65); \end{tikzpicture}};	\node at (-1.8,-1)[scale=0.7]{\begin{tikzpicture}\fill[black!80] (0,0)[rounded corners=10] -- (1.3,0) -- (1.3,1.65)[rounded corners=0] -- (0.5,1.65) -- (0,0.7) -- (-0.5,1.65)[rounded corners=10] -- (-1.3,1.65) -- (-1.3,0) -- (0,0);\fill[black!80, rounded corners=3] (0.12,0.5) -- (0.05,1.25) -- (0.2,1.65) -- (-0.2,1.65) -- (-0.05,1.25) -- (-0.12,0.5); \fill[black!80] (0,2.7) circle(0.88cm and 0.85cm); \end{tikzpicture}};	\node at (-3.35,-0.6)[scale=0.9]{\begin{tikzpicture} \fill[black!80] (0,0) rectangle (1,1.2); \draw[line width=1.2mm, white] (0.12,0.2) -- (0.88,0.2); \draw[line width=1.2mm, white] (0.12,0.45) -- (0.88,0.45); \draw[line width=1.2mm, white] (0.12,0.7) -- (0.88,0.7); \draw[line width=1.2mm, white] (0.12,0.95) -- (0.88,0.95); \end{tikzpicture}}; \end{tikzpicture}};\end{tikzpicture}}

\def\ContactIconeThirdxx#1#2{
	\begin{tikzpicture}
		\node at (0,0)[scale=#1]{
			\begin{tikzpicture}
				\draw[line width=0.6mm, rounded corners=35, fill=white] (-4.5,-4.4) rectangle (6,3);
				\node at (-2,-2.1)[scale=0.7]{\begin{tikzpicture} \draw[line width=0.5mm, fill=black] (0,0) to [bend right=15](1.5,0.5) to [bend right](0.5,2.25) -- (0,1.4) -- (-0.5,2.25) to [bend right](-1.5,0.5) to [bend right=15](0,0);
						\node at (0.4,0.75)[]{\begin{tikzpicture}\draw[line width=1mm, white] (0,0.2) -- (0,-0.2); \draw[line width=1mm, white] (-0.2,0) -- (0.2,0); \end{tikzpicture}};
						\node at (-0.85,1.4)[scale=0.7]{\begin{tikzpicture} \draw[line width=0.8mm, white, rounded corners=4] (-0.4,0)node[color=white]{\Large $\bullet$} -- (-0.6,0.15) to [bend left](0,1.25) to [bend left](0.6,0.15) -- (0.4,0)node[color=white]{\Large $\bullet$}; \draw[line width=0.8mm, white] (0,1.2) to [bend left=15](0.3,2.25); \end{tikzpicture}};
						\node at (0.8,1.65)[]{\begin{tikzpicture} \draw[line width=1mm, white] (0,0) to [bend right=10](-0.2,1); \draw[line width=1mm, white, fill=black] (0,0) circle(8pt); \end{tikzpicture}}; \draw[thick, fill=black] (0,3) circle(0.7cm); \end{tikzpicture}}; \node at (-2,-3.75)[scale=#2]{\Huge \sf Doctor}; \node at (-2,1.65)[scale=0.6]{\begin{tikzpicture} \draw[line width=0.2mm, fill=black] (0,0) -- (1.5,0) arc(0:180:1.5 and 1.4) -- (0,0); \draw[thick, fill=black] (0,2.25) circle(0.7cm);
						\draw[thick, fill=black] (0,3.13) to [bend left=15](0.6,2.9) -- (1,3.5) to [bend right=50](-1,3.4) -- (-0.6,2.9) to [bend left=15](0,3.13);
						\node at (0,3.5)[]{\begin{tikzpicture} \draw[line width=1.5mm, white] (-0.25,0) -- (0.25,0); \draw[line width=1.5mm, white] (0,0.25) -- (0,-0.25); \end{tikzpicture}};\end{tikzpicture}}; \node at (-2,-0.15)[scale=#2]{\Huge \sf Nurse};
				\node at (2.5,-2.1)[scale=0.7]{\begin{tikzpicture} \draw[thick, fill=black] (0,0) -- (2,0) to [bend right=10](1.8,1.25) to [bend right=5](0.85,1.8) -- (0.6,1.65) -- (0,0.5) (0,0) -- (-2,0) to [bend left=10](-1.8,1.25) to [bend left=5](-0.85,1.8) -- (-0.6,1.65) -- (0,0.5); \draw[thick, fill=black] (0,1.8) -- (-0.2,1.5) -- (-0.08,1.25) -- (-0.2,0.4) -- (0.2,0.4) -- (0.08,1.25) -- (0.2,1.5) -- (0,1.8); \draw[fill=black] (0,2.9) circle(0.8cm and 0.85cm); \node at (-1.5,1.2)[scale=0.8]{\begin{tikzpicture} \node at (0.1,-0.1)[scale=1.5]{\begin{tikzpicture} \draw[line width=1mm, white, fill=white] (0,0)[rounded corners=10] -- (1,0)[rounded corners=0] -- (0.2,1.6) -- (0.2,2.2)[rounded corners=2] -- (0.35,2.2) -- (0.35,2.4) -- (-0.35,2.4) -- (-0.35,2.2)[rounded corners=0] -- (-0.2,2.2) -- (-0.2,1.6)[rounded corners=10] -- (-1,0) -- (0,0); \end{tikzpicture}};	\node at (0,0)[scale=1.2]{\begin{tikzpicture} \draw[line width=1mm, fill=black] (0,0)[rounded corners=10] -- (1,0)[rounded corners=0] -- (0.2,1.6) -- (0.2,2.2)[rounded corners=2] -- (0.35,2.2) -- (0.35,2.4) -- (-0.35,2.4) -- (-0.35,2.2)[rounded corners=0] -- (-0.2,2.2) -- (-0.2,1.6)[rounded corners=10] -- (-1,0) -- (0,0); \end{tikzpicture}}; \node at (0,-0.75)[scale=1]{\begin{tikzpicture} \draw[line width=1mm, fill=white] (0,0)[rounded corners=5] -- (1,0)[rounded corners=0] -- (0.4,1.1) -- (-0.4,1.1)[rounded corners=5] -- (-1,0) -- (0,0); \end{tikzpicture}}; \node at (-0.15,1.8)[scale=2]{$\bullet$}; \node at (-0.3,2.2)[scale=2]{$\bullet$}; \node at (-0.8,1.8)[scale=3]{$\bullet$}; \end{tikzpicture}};  \end{tikzpicture}}; \node at (2.9,-3.75)[scale=#2]{\Huge \sf Lab Worker}; \node at (2.5,1.6)[scale=0.7]{\begin{tikzpicture} \draw[line width=0.2mm, fill=black] (0,0) -- (1.5,0) arc(0:180:1.5 and 1.4) -- (0,0); \draw[thick, fill=black] (0,2.25) circle(0.7cm); \node at (-0.8,0.5)[]{\begin{tikzpicture} \draw[line width=2mm, white] (-0.35,0) -- (0.35,0); \draw[line width=2mm, white] (0,0.35) -- (0,-0.35); \end{tikzpicture}}; \node at (0.8,0.6)[scale=0.7]{\begin{tikzpicture} \fill[white] (0,0) circle(1.3cm); \node at (0,0.2)[rotate=50, yscale=1.4, xscale=1.1]{\begin{tikzpicture} \fill[black] (0.1,0)[rounded corners=7] -- (1,0) -- (1,0.5)[rounded corners=0] -- (0.1,0.5) -- (0.1,0); \fill[black] (-0.1,0)[rounded corners=7] -- (-1,0) -- (-1,0.5)[rounded corners=0] -- (-0.1,0.5) -- (-0.1,0); \end{tikzpicture}}; \end{tikzpicture}}; \end{tikzpicture}}; \node at (2.9,0.1)[scale=#2]{\Huge \sf Pharmacist};\end{tikzpicture}};\end{tikzpicture}}

\def\SecondHomexx#1{\begin{tikzpicture} \node at (0,0)[scale=#1, xscale=1.3]{\begin{tikzpicture} \draw[line width=2mm] (0,0) -- (0,6.08); \fill[bluef!20] (-1.7,0) rectangle (1.7,6); \fill[bluef!40] (1.7,0.05) rectangle (4.25,4.25); \draw[line width=1.2mm, fill=bluef!60, rounded corners=1] (-0.5,0) rectangle (0.5,1.2); \draw[line width=1.2mm, fill=bluef!40, rounded corners=1] (0.5,0) rectangle (1,1.2); \draw[line width=1.2mm, fill=bluef!40, rounded corners=1] (-0.5,0) rectangle (-1,1.2); \draw[line width=1.2mm] (-1.4,0) -- (1.7,0) -- (1.7,4); \node at (-1,2)[scale=0.55]{\begin{tikzpicture} \draw[line width=1.2mm, fill=bluef!37, rounded corners=1] (0,0) rectangle (1,1);\end{tikzpicture}}; \node at (-1,3)[scale=0.55]{\begin{tikzpicture} \draw[line width=1.2mm, fill=bluef!37, rounded corners=1] (0,0) rectangle (1,1);\end{tikzpicture}}; \node at (-1,4)[scale=0.55]{\begin{tikzpicture} \draw[line width=1.2mm, fill=bluef!37, rounded corners=1] (0,0) rectangle (1,1);\end{tikzpicture}}; \node at (0,2)[scale=0.55]{\begin{tikzpicture} \draw[line width=1.2mm, fill=bluef!37, rounded corners=1] (0,0) rectangle (1,1);\end{tikzpicture}}; \node at (0,3)[scale=0.55]{\begin{tikzpicture} \draw[line width=1.2mm, fill=bluef!37, rounded corners=1] (0,0) rectangle (1,1);\end{tikzpicture}}; \node at (0,4)[scale=0.55]{\begin{tikzpicture} \draw[line width=1.2mm, fill=bluef!37, rounded corners=1] (0,0) rectangle (1,1);\end{tikzpicture}}; \node at (1,2)[scale=0.55]{\begin{tikzpicture} \draw[line width=1.2mm, fill=bluef!37, rounded corners=1] (0,0) rectangle (1,1);\end{tikzpicture}}; \node at (1,3)[scale=0.55]{\begin{tikzpicture} \draw[line width=1.2mm, fill=bluef!37, rounded corners=1] (0,0) rectangle (1,1);\end{tikzpicture}}; \node at (1,4)[scale=0.55]{\begin{tikzpicture} \draw[line width=1.2mm, fill=bluef!37, rounded corners=1] (0,0) rectangle (1,1);\end{tikzpicture}}; \node at (2.9,1)[]{\begin{tikzpicture} \draw[line width=1.2mm, fill=bluef!60, rounded corners=1] (0,0) rectangle (1.8,0.6); \end{tikzpicture}}; \node at (2.9,2)[]{\begin{tikzpicture} \draw[line width=1.2mm, fill=bluef!60, rounded corners=1] (0,0) rectangle (1.8,0.6); \end{tikzpicture}}; \node at (2.9,3)[]{\begin{tikzpicture} \draw[line width=1.2mm, fill=bluef!60, rounded corners=1] (0,0) rectangle (1.8,0.6); \end{tikzpicture}}; \node at (0,5.2)[scale=0.4]{\begin{tikzpicture} \draw[line width=3mm, fill=green!70] (-1,1.7)[rounded corners=2] -- (-1,1.5)[rounded corners=4] to [bend right](0,0)[rounded corners=2] to [bend right](1,1.5)[rounded corners=3] -- (1,2.25)[rounded corners=3] to [bend left=20](0,2.25)[rounded corners=2] to [bend left=20](-1,2.25) -- (-1,1.7); \end{tikzpicture}}; \end{tikzpicture}};\end{tikzpicture}}

\def\Hospitalxx#1{
	\begin{tikzpicture}
		\node at (0,0)[scale=#1]{\begin{tikzpicture}
				\fill[black!30] (0,0) circle(4.2cm and 0.8cm);
				\node at (0,2.7)[scale=0.5]{\begin{tikzpicture}
						\draw[black!25, fill=black!10] (-2.5,0) rectangle (2.5,9.8);
						\draw[black!25, fill=black!3] (-2.4,2.3) rectangle (2.4,9.7);
						\draw[line width=0.8mm, fill=bluef!40, rounded corners=1] (-1,0) rectangle (1,2);
						\draw[line width=0.8mm] (0,0) -- (0,2);
						\draw[line width=0.8mm, rounded corners=1, fill=bluef!40] (1.25,0) rectangle (2.25,2);
						\draw[line width=0.8mm, rounded corners=1, fill=bluef!40] (-1.25,0) rectangle (-2.25,2);
						\node at (0,3)[xscale=1.1]{\begin{tikzpicture} \fill[bluef!50] (0,0) rectangle (0.8,1); \fill[bluef!20] (0.8,0) rectangle (1.6,1); \fill[bluef!50] (1.6,0) rectangle (2.4,1); \fill[bluef!20] (2.4,0) rectangle (3.2,1); \fill[bluef!50] (3.2,0) rectangle (4,1); \end{tikzpicture}};
						\node at (0,4.5)[xscale=1.1]{\begin{tikzpicture} \fill[bluef!50] (0,0) rectangle (0.8,1); \fill[bluef!20] (0.8,0) rectangle (1.6,1); \fill[bluef!50] (1.6,0) rectangle (2.4,1); \fill[bluef!20] (2.4,0) rectangle (3.2,1); \fill[bluef!50] (3.2,0) rectangle (4,1); \end{tikzpicture}};
						\node at (0,6)[xscale=1.1]{\begin{tikzpicture} \fill[bluef!50] (0,0) rectangle (0.8,1); \fill[bluef!20] (0.8,0) rectangle (1.6,1); \fill[bluef!50] (1.6,0) rectangle (2.4,1); \fill[bluef!20] (2.4,0) rectangle (3.2,1); \fill[bluef!50] (3.2,0) rectangle (4,1); \end{tikzpicture}};
						\node at (0,7.5)[xscale=1.1]{\begin{tikzpicture} \fill[bluef!50] (0,0) rectangle (0.8,1); \fill[bluef!20] (0.8,0) rectangle (1.6,1); \fill[bluef!50] (1.6,0) rectangle (2.4,1); \fill[bluef!20] (2.4,0) rectangle (3.2,1); \fill[bluef!50] (3.2,0) rectangle (4,1); \end{tikzpicture}};
						\node at (0,9)[xscale=1.1]{\begin{tikzpicture} \fill[bluef!50] (0,0) rectangle (0.8,1); \fill[bluef!20] (0.8,0) rectangle (1.6,1); \fill[bluef!50] (1.6,0) rectangle (2.4,1); \fill[bluef!20] (2.4,0) rectangle (3.2,1); \fill[bluef!50] (3.2,0) rectangle (4,1); \end{tikzpicture}};
						\fill[fill=red] (-3,9.8) rectangle (3,11);
						\node at (0,10.4)[color=white]{\Huge \sf HOSPITAL};
						\fill[fill=black!5] (2.5,0) rectangle (6.55,5.1);
						\draw[line width=1.4mm, black!40] (2.5,0) -- (6.55,0);
						\fill[fill=black!5] (-2.5,0) rectangle (-6.55,5.1);
						\draw[line width=1.4mm, black!40] (-2.5,0) -- (-6.55,0);
						\fill[black!20, opacity=0.8] (2.5,0) -- (2.5,5.1) -- (5,0);
						\fill[black!20, opacity=0.8] (-2.5,0) -- (-2.5,5.1) -- (-5,0);
						\draw[line width=5mm, black!65] (-2.5,5) -- (-6.55,5);
						\draw[line width=5mm, black!65] (2.5,5) -- (6.55,5);
						\node at (3,0.8)[]{\begin{tikzpicture}\draw[line width=0.4mm, fill=blue!30] (0,0) rectangle (0.4,1); \end{tikzpicture}};
						\node at (3.6,0.8)[]{\begin{tikzpicture}\draw[line width=0.4mm, fill=blue!30] (0,0) rectangle (0.4,1); \end{tikzpicture}};
						\node at (4.2,0.8)[]{\begin{tikzpicture}\draw[line width=0.4mm, fill=blue!30] (0,0) rectangle (0.4,1); \end{tikzpicture}};
						\node at (4.8,0.8)[]{\begin{tikzpicture}\draw[line width=0.4mm, fill=blue!30] (0,0) rectangle (0.4,1); \end{tikzpicture}};
						\node at (5.4,0.8)[]{\begin{tikzpicture}\draw[line width=0.4mm, fill=blue!30] (0,0) rectangle (0.4,1); \end{tikzpicture}};
						\node at (6,0.8)[]{\begin{tikzpicture}\draw[line width=0.4mm, fill=blue!30] (0,0) rectangle (0.4,1); \end{tikzpicture}};
						\node at (3,2.3)[]{\begin{tikzpicture}\draw[line width=0.4mm, fill=blue!30] (0,0) rectangle (0.4,1); \end{tikzpicture}};
						\node at (3.6,2.3)[]{\begin{tikzpicture}\draw[line width=0.4mm, fill=blue!30] (0,0) rectangle (0.4,1); \end{tikzpicture}};
						\node at (4.2,2.3)[]{\begin{tikzpicture}\draw[line width=0.4mm, fill=blue!30] (0,0) rectangle (0.4,1); \end{tikzpicture}};
						\node at (4.8,2.3)[]{\begin{tikzpicture}\draw[line width=0.4mm, fill=blue!30] (0,0) rectangle (0.4,1); \end{tikzpicture}};
						\node at (5.4,2.3)[]{\begin{tikzpicture}\draw[line width=0.4mm, fill=blue!30] (0,0) rectangle (0.4,1); \end{tikzpicture}};
						\node at (6,2.3)[]{\begin{tikzpicture}\draw[line width=0.4mm, fill=blue!30] (0,0) rectangle (0.4,1); \end{tikzpicture}};
						\node at (3,3.8)[]{\begin{tikzpicture}\draw[line width=0.4mm, fill=blue!30] (0,0) rectangle (0.4,1); \end{tikzpicture}};
						\node at (3.6,3.8)[]{\begin{tikzpicture}\draw[line width=0.4mm, fill=blue!30] (0,0) rectangle (0.4,1); \end{tikzpicture}};
						\node at (4.2,3.8)[]{\begin{tikzpicture}\draw[line width=0.4mm, fill=blue!30] (0,0) rectangle (0.4,1); \end{tikzpicture}};
						\node at (4.8,3.8)[]{\begin{tikzpicture}\draw[line width=0.4mm, fill=blue!30] (0,0) rectangle (0.4,1); \end{tikzpicture}};
						\node at (5.4,3.8)[]{\begin{tikzpicture}\draw[line width=0.4mm, fill=blue!30] (0,0) rectangle (0.4,1); \end{tikzpicture}};
						\node at (6,3.8)[]{\begin{tikzpicture}\draw[line width=0.4mm, fill=blue!30] (0,0) rectangle (0.4,1); \end{tikzpicture}};
						\node at (-3,0.8)[]{\begin{tikzpicture}\draw[line width=0.4mm, fill=blue!30] (0,0) rectangle (0.4,1); \end{tikzpicture}};
						\node at (-3.6,0.8)[]{\begin{tikzpicture}\draw[line width=0.4mm, fill=blue!30] (0,0) rectangle (0.4,1); \end{tikzpicture}};
						\node at (-4.2,0.8)[]{\begin{tikzpicture}\draw[line width=0.4mm, fill=blue!30] (0,0) rectangle (0.4,1); \end{tikzpicture}};
						\node at (-4.8,0.8)[]{\begin{tikzpicture}\draw[line width=0.4mm, fill=blue!30] (0,0) rectangle (0.4,1); \end{tikzpicture}};
						\node at (-5.4,0.8)[]{\begin{tikzpicture}\draw[line width=0.4mm, fill=blue!30] (0,0) rectangle (0.4,1); \end{tikzpicture}};
						\node at (-6,0.8)[]{\begin{tikzpicture}\draw[line width=0.4mm, fill=blue!30] (0,0) rectangle (0.4,1); \end{tikzpicture}};
						\node at (-3,2.3)[]{\begin{tikzpicture}\draw[line width=0.4mm, fill=blue!30] (0,0) rectangle (0.4,1); \end{tikzpicture}};
						\node at (-3.6,2.3)[]{\begin{tikzpicture}\draw[line width=0.4mm, fill=blue!30] (0,0) rectangle (0.4,1); \end{tikzpicture}};
						\node at (-4.2,2.3)[]{\begin{tikzpicture}\draw[line width=0.4mm, fill=blue!30] (0,0) rectangle (0.4,1); \end{tikzpicture}};
						\node at (-4.8,2.3)[]{\begin{tikzpicture}\draw[line width=0.4mm, fill=blue!30] (0,0) rectangle (0.4,1); \end{tikzpicture}};
						\node at (-5.4,2.3)[]{\begin{tikzpicture}\draw[line width=0.4mm, fill=blue!30] (0,0) rectangle (0.4,1); \end{tikzpicture}};
						\node at (-6,2.3)[]{\begin{tikzpicture}\draw[line width=0.4mm, fill=blue!30] (0,0) rectangle (0.4,1); \end{tikzpicture}};
						\node at (-3,3.8)[]{\begin{tikzpicture}\draw[line width=0.4mm, fill=blue!30] (0,0) rectangle (0.4,1); \end{tikzpicture}};
						\node at (-3.6,3.8)[]{\begin{tikzpicture}\draw[line width=0.4mm, fill=blue!30] (0,0) rectangle (0.4,1); \end{tikzpicture}};
						\node at (-4.2,3.8)[]{\begin{tikzpicture}\draw[line width=0.4mm, fill=blue!30] (0,0) rectangle (0.4,1); \end{tikzpicture}};
						\node at (-4.8,3.8)[]{\begin{tikzpicture}\draw[line width=0.4mm, fill=blue!30] (0,0) rectangle (0.4,1); \end{tikzpicture}};
						\node at (-5.4,3.8)[]{\begin{tikzpicture}\draw[line width=0.4mm, fill=blue!30] (0,0) rectangle (0.4,1); \end{tikzpicture}};
						\node at (-6,3.8)[]{\begin{tikzpicture}\draw[line width=0.4mm, fill=blue!30] (0,0) rectangle (0.4,1); \end{tikzpicture}}; \end{tikzpicture}}; \end{tikzpicture}};\end{tikzpicture}}

\def\SPCPServeurFunctionxx#1#2#3{
	\begin{tikzpicture}
		\node at (0,0)[scale=#1]{
			\begin{tikzpicture}
				\node at (0,3.3)[]{\Nuage{\Huge \textbf{#2}}{0.8}};
				\draw[line width=2mm, Triangle-Triangle] (0,1.5) -- (0,-1.5);
				\node at (0.4,-3.4)[]{\Serveur{0.55}};
				\node at (2,-3.1)[]{\Huge \sf \textbf{#3}};
				\node at (0.4,-3.4)[]{};
				\node at (2,-3.1)[]{};
				\node at (2,-5)[]{};
			\end{tikzpicture}
		};
	\end{tikzpicture}
}

\def\Nuage#1#2{
	\begin{tikzpicture}
		\node at (0,0)[scale=#2]{
			\begin{tikzpicture}[scale=1.5]
				\draw[black!40,line width=0.8mm, fill=nuageColor] (-1,1)[rounded corners=21] -- (-1.5,1) -- (-1.5,0) -- (1.5,0) -- (1.5,1)[rounded corners=0] -- (1,1) .. controls(1,1.8) and (-0.25,2) .. (-0.5,1.4) to [bend right=70](-1,1);
				\draw[line width=0.8mm, black!40] (-0.5,1.4) to [bend left](-0.22,1.1); \draw[line width=0.8mm, black!40] (1,1) -- (0.7,1);
				\node at (0,0.15)[scale=1.1]{
					\begin{tikzpicture}
						\node at (0.75,1.1)[]{
							\begin{tikzpicture}
								\draw[line width=1.5mm, rounded corners=13, white] (0,0) -- (0,1.2) -- (0.9,1.2) -- (0.9,0);
						\end{tikzpicture}};
						\fill[blue!20] (0,0) rectangle (1.5,1.15);
						\node at (0.75,0.575)[]{
							\begin{tikzpicture}
								\fill[black!70, rounded corners=2] (0,0) rectangle (0.15,0.5); \fill[black!70] (0.075,0.4) circle(0.2cm);
						\end{tikzpicture}};
				\end{tikzpicture}};
				\node at (0.25,1.1)[scale=1.2]{\sf #1};
		\end{tikzpicture}};
\end{tikzpicture}}

\def\Bande{
	\begin{tikzpicture}
		\fill[thick, rounded corners=3, blue!60] (3,0) arc(0:-180:3cm and 0.7cm) -- (-3,1) arc(-180:0:3cm and 0.7cm) -- (3,0);
		\filldraw[white, line width=1.3mm, xshift=0.5cm] (0,0.1) -- (0,-0.5);
		\filldraw[white, line width=1.3mm, xshift=1.5cm, yshift=0.05cm] (0,0.1) -- (0,-0.5);
		\filldraw[white, line width=1.3mm, xshift=2.5cm, yshift=0.25cm] (0,0.1) -- (0,-0.5);
		\filldraw[white, line width=1.3mm, xshift=-0.5cm] (0,0.1) -- (0,-0.5);
		\filldraw[white, line width=1.3mm, xshift=-1.5cm, yshift=0.08cm] (0,0.1) -- (0,-0.5);
		\filldraw[white, line width=1.3mm, xshift=-2.5cm, yshift=0.25cm] (0,0.1) -- (0,-0.5);
	\end{tikzpicture}
}

\def\Parametre{
	\begin{tikzpicture}
		\def\BandeR{
			\begin{tikzpicture}
				\filldraw[white, rounded corners=2] (-0.1,-0.1) rectangle (2.9, 0.5);
				\fill[black!60, rounded corners=1] (0,0) rectangle (2.8, 0.4);
			\end{tikzpicture}
		}
		\filldraw[white] (0,0) circle(1.15cm);
		\node at (0,0)[]{\BandeR};
		\node at (0,0)[rotate=45]{\BandeR};
		\node at (0,0)[rotate=90]{\BandeR};
		\node at (0,0)[rotate=135]{\BandeR};
		\filldraw[black!60] (0,0) circle(1.05cm);
		\filldraw[white] (0,0) circle(0.55cm);
	\end{tikzpicture}
}

\def\Serveur#1{
	\begin{tikzpicture}
		\node at (0,)[scale=#1]{
			\begin{tikzpicture}
				\node at (0,0)[xscale=0.7]{
					\begin{tikzpicture}
						\filldraw[blue!60] (3,0) arc(0:360:3cm and 0.7cm);
						\node at (0,-1.2)[]{\Bande};
						\node at (0,-2.55)[]{\Bande};
						\node at (0,-3.9)[]{\Bande};
					\end{tikzpicture}
				};
				\node at (1.8,-1.8)[]{\Parametre};
			\end{tikzpicture}
		};
	\end{tikzpicture}
}

\def\MyFinalDiagramxx#1#2{\begin{tikzpicture}
		\node at (0,0)[scale=#1]{
			\begin{tikzpicture}[xscale=1.05]
				\node at (0,10)[scale=0.4]{\ManAvatarxx}; \node at (2,10)[scale=0.4]{\RoundAssociatorxx{\MagnificClockxx{0.6}}{\VitesseBulletxx{0.8}}}; \node at (4.5,10)[scale=0.26]{\phonexx};
				
				\node at (0,5)[scale=0.4]{\ManAvatarxx}; \node at (2,5)[scale=0.4]{\RoundAssociatorxx{\VitesseBulletxx{0.8}}{\Tishortxx{0.48}}}; \node at (4.5,5)[scale=0.26]{\phonexx};
				
				\node at (0,0)[scale=0.4]{\ManAvatarxx}; \node at (2,0)[scale=0.4]{\RoundAssociatorxx{\Tishortxx{0.48}}{\MagnificClockxx{0.6}}}; \node at (4.5,0)[scale=0.26]{\phonexx};
				
				\node at (9,5.05)[]{\SPCPServeurFunctionxx{0.7}{SP}{CP}};

				\draw[line width=0.5mm, Stealth-Stealth] (5.2, 0)--(8,6.8);
				\draw[line width=0.5mm, Stealth-Stealth] (5.2, 5)--(7.5,7);
				\draw[line width=0.5mm, Stealth-Stealth] (5.2, 10)--(7.5,7.6);
				
				\draw[line width=0.4mm, rounded corners=10, Stealth-] (0,12.4) -- (0,13.2) -- (8.75,13.2) -- (8.75,14);
				\draw[line width=0.4mm, rounded corners=10, -Stealth] (8.75,14) -- (8.75,13.2) -- (17.5,13.2) -- (17.6,12.3);
				
				\node at (0,14)[scale=#2]{\Titledefxx{\large \textbf{DO}}{0.5}};
				\node at (4.5,13.15)[rotate=40]{\Keyxx{0.3}};
				\node at (8.75,13.1)[below, scale=#2]{\textcolor{black}{\large \sf \textbf{KA}}};
				\node at (8.75,14.3)[]{\FirstHomexx{0.22}};
				\draw[line width=0.5mm, Stealth-Stealth] (8.75,8.5) -- (8.75,12.2);
				\node at (8,10.2)[rotate=40]{\Keyxx{0.3}};
				
				\node at (12.75,13.15)[rotate=40]{\Keyxx{0.3}};
				\node at (17.5,14)[scale=#2]{\Titledefxx{\large \textbf{DR}}{0.5}};
				\draw[line width=0.8mm, dashed, red, rounded corners=30] (14.5,-2.3) rectangle (20.5,12.25);
				\draw[line width=0.6mm, Stealth-Stealth] (10,6.9) -- (14.8,0);
				\draw[line width=0.6mm, Stealth-Stealth] (10,7.3) -- (14.8,5);
				\draw[line width=0.6mm, Stealth-Stealth] (10,7.6) -- (14.8,10);

				\node at (17.53,10)[]{\ContactIconeFirstxx{0.5}{1.3}};
				
				\node at (17.53,5)[]{\ContactIconeSecondxx{0.5}{1.3}};
				
				\node at (17.53,0)[]{\ContactIconeThirdxx{0.5}{1.2}};
				
				
		\end{tikzpicture}};
\end{tikzpicture}}

\def\boxText#1#2#3#4{
	\begin{tikzpicture}
		\draw[line width=0.5mm, fill=white] (-0.1,-0.4) rectangle (3.4-#4,1.5);
		\node at (0,0.9)[right, scale=#3+0.3]{\sf \textbf{#1}};
		\node at (0,0.15)[right, scale=#3+.3]{\sf \textbf{#2}};
	\end{tikzpicture}
}
\def\newFigureOne#1{
	\begin{tikzpicture}
		\node at (0,0)[scale=#1]{
			\begin{tikzpicture}
				\draw[line width=0.5mm] (-2,-3) -- (0,0) -- (2,-3);
				\draw[line width=0.5mm] (-6,-6)-- (-2,-3) -- (-1.5,-6);
				\draw[line width=0.5mm] (6,-6)-- (2,-3) -- (1.5,-6);
				\node at (0,0)[]{\begin{tikzpicture}
						\draw[line width=0.8mm, fill=white] (0,0) circle(0.9cm);
						\node at (0,0)[]{\Large \sf \textbf{AND}};
				\end{tikzpicture}};
				
				\node at (-2,-3)[]{\begin{tikzpicture}
						\draw[line width=0.8mm, fill=white] (0,0) circle(0.8cm);
						\node at (0,0)[]{\Large \sf \textbf{OR}};
				\end{tikzpicture}};
				
				\node at (2,-3)[]{\begin{tikzpicture}
						\draw[line width=0.8mm, fill=white] (0,0) circle(0.8cm);
						\node at (0,0)[]{\Large \sf \textbf{OR}};
				\end{tikzpicture}};
				
				\node at (-5.5,-6.2)[]{\boxText{Research}{Institutes}{1.2}{0.5}};
				\node at (-2,-6.2)[]{\boxText{Government}{Regulators}{1.2}{0}};
				\node at (2,-6.2)[]{\boxText{International}{Autorities}{1.15}{0}};
				\node at (5.5,-6.2)[]{\boxText{Local city}{regulators}{1.2}{0.5}};
			\end{tikzpicture}
		};
	\end{tikzpicture}
}

\def\newFigureTwo#1{
	\begin{tikzpicture}
		\node at (0,0)[scale=#1]{
			\begin{tikzpicture}
				\draw[line width=0.5mm] (-2,-3) -- (0,0) -- (2,-3);
				\draw[line width=0.5mm] (-6,-6)-- (-2,-3) -- (-1.5,-6);
				\draw[line width=0.5mm] (6,-6)-- (2,-3) -- (1.5,-6);
				\node at (0,0)[]{\begin{tikzpicture}
						\draw[line width=0.8mm, fill=white] (0,0) circle(0.9cm);
						\node at (0,0)[]{\Large \sf \textbf{AND}};
				\end{tikzpicture}};
				
				\node at (-2,-3)[]{\begin{tikzpicture}
						\draw[line width=0.8mm, fill=white] (0,0) circle(0.8cm);
						\node at (0,0)[]{\Large \sf \textbf{OR}};
				\end{tikzpicture}};
				
				\node at (2,-3)[]{\begin{tikzpicture}
						\draw[line width=0.8mm, fill=white] (0,0) circle(0.8cm);
						\node at (0,0)[]{\Large \sf \textbf{OR}};
				\end{tikzpicture}};
				
				\node at (-5.4,-6.2)[]{\boxText{Family}{members}{1.2}{0.5}};
				\node at (-2,-6.2)[]{\boxText{Specific}{Doctor}{1.2}{0.6}};
				\node at (1.6,-6.2)[]{\boxText{Specific}{Friends}{1.15}{0.7}};
				\node at (5,-6.2)[]{\boxText{Specific}{Hospital}{1.2}{0.6}};
			\end{tikzpicture}
		};
	\end{tikzpicture}
}
\definecolor{bluef}{HTML}{132380}
\definecolor{titleColor}{rgb}{1.0,0.86,0.6}
\definecolor{nuageColor}{rgb}{0.57,0.73,0.88}

\def\ManAvatarRoundWhite{\begin{tikzpicture} \node at (0,0)[scale=0.97]{\begin{tikzpicture} \fill[white] (0,4.5)[rounded corners=35] -- (-2,4.5)[rounded corners=11] -- (-2,0.1) -- (-1.2,0.1)[rounded corners=0] -- (-1.2,3) -- (-1,3)[rounded corners=11] -- (-1,-3.5) -- (-0.1,-3.5)[rounded corners=0] -- (-0.1,0) -- (0,0)[rounded corners=0] -- (0.1,0)[rounded corners=11] -- (0.1,-3.5) -- (1,-3.5)[rounded corners=0] -- (1,3) -- (1.2,3)[rounded corners=11] -- (1.2,0.1) -- (2,0.1)[rounded corners=35] -- (2,4.5) -- (0,4.5); \fill[fill=white] (0,5.75) circle(1cm); \end{tikzpicture}}; \node at (0,0)[scale=0.9]{\begin{tikzpicture}[fill=black!80] \fill[black!85] (0,4.5)[rounded corners=35] -- (-2,4.5)[rounded corners=11] -- (-2,0.1) -- (-1.2,0.1)[rounded corners=0] -- (-1.2,3) -- (-1,3)[rounded corners=11] -- (-1,-3.5) -- (-0.1,-3.5)[rounded corners=0] -- (-0.1,0) -- (0,0)[rounded corners=0] -- (0.1,0)[rounded corners=11] -- (0.1,-3.5) -- (1,-3.5)[rounded corners=0] -- (1,3) -- (1.2,3)[rounded corners=11] -- (1.2,0.1) -- (2,0.1)[rounded corners=35] -- (2,4.5) -- (0,4.5); \fill[fill=black!85] (0,5.75) circle(1cm); \end{tikzpicture}}; \end{tikzpicture}}

\def\ThreeManAvatar{\begin{tikzpicture} \node at (0,0)[xscale=0.5]{\begin{tikzpicture} \node at (0,0)[yscale=0.8]{\ManAvatarRoundWhite}; \node at (2.9,0)[yscale=0.8]{\ManAvatarRoundWhite}; \node at (-2.9,0)[yscale=0.8]{\ManAvatarRoundWhite}; \node at (5.2,1.2)[yscale=0.5]{\begin{tikzpicture} \fill[black] (0,0) arc(-90:90:0.7 and 0.7) -- (-0.05,0.7); \end{tikzpicture}}; \node at (-5.2,1.2)[xscale=-1,yscale=0.5]{\begin{tikzpicture} \fill[black] (0,0) arc(-90:90:0.7 and 0.7) -- (-0.05,0.7); \end{tikzpicture}}; \end{tikzpicture}}; \end{tikzpicture}}

\def\UsersAvatar{\begin{tikzpicture} \node at (0,0)[yscale=0.8]{\ManAvatarRoundWhite}; \end{tikzpicture}}

\def\NewNuage{\begin{tikzpicture}\node at (0,0)[scale=1]{\begin{tikzpicture} \draw[black!40,line width=0.8mm, fill=nuageColor] (-1,1)[rounded corners=15] -- (-1.5,1) -- (-1.5,0) -- (1.5,0) -- (1.5,1)[rounded corners=0] -- (1,1) .. controls(1,1.8) and (-0.25,2) .. (-0.5,1.4) to [bend right=70](-1,1); \draw[line width=0.8mm, black!40] (-0.5,1.4) to [bend left](-0.22,1.1); \draw[line width=0.8mm, black!40] (1,1) -- (0.7,1); \node at (0,0.15)[scale=0.65]{\begin{tikzpicture} \node at (0.75,1.1)[]{\begin{tikzpicture} \draw[line width=1.5mm, rounded corners=13, white] (0,0) -- (0,1.2) -- (0.9,1.2) -- (0.9,0); \end{tikzpicture}}; \fill[blue!30] (0,0) rectangle (1.5,1.15); \node at (0.75,0.575)[]{\begin{tikzpicture} \fill[black!70, rounded corners=2] (0,0) rectangle (0.15,0.5); \fill[black!70] (0.075,0.4) circle(0.2cm); \end{tikzpicture}}; \end{tikzpicture}}; \end{tikzpicture}};\end{tikzpicture}}

\def\OtherTitle#1#2{\begin{tikzpicture}\fill[titleColor] (-#2,-0.4) rectangle (#2,0.4); \node at (0,0)[color=black]{\huge \sf \textbf{#1}}; \end{tikzpicture}}
\def\TitledefOne#1#2{\begin{tikzpicture}\fill[blue!70] (-#2,-0.4) rectangle (#2,0.4); \node at (0,-0.05)[color=black]{\Large \sf \textbf{#1}}; \end{tikzpicture}}

\def\DiagramOne{\begin{tikzpicture}
		\draw[line width=0.8mm, dashed] (-7,-20/2) -- (-7,20/2);
		\draw[line width=0.8mm, dashed] (7,-20/2) -- (7,20/2);
		\draw[line width=0.8mm, dashed] (0,-20/2) -- (0,20/2);
		\node at (0,{(20/2) - (20/3)})[yscale=1.2]{\OtherTitle{User access policy setting}{14}};
		\node at (0,{(20/2) - (2*20/3)})[yscale=1.2]{\OtherTitle{Data uploading}{14}};
		\node at (-10.5,6.5)[scale=0.4, xscale=0.7]{\UsersAvatar};
		\node at (-10.5,9)[]{\LARGE \sf \textbf{\textcolor{red}{Users}}};
		\node at (-3.5,7)[scale=1]{\NewNuage};
		\node at (-3.5,9)[]{\LARGE \sf \textbf{\textcolor{red}{CSP$_{\sf \textbf{A}}$}}};
		\node at (3.5,7)[scale=1]{\NewNuage};
		\node at (3.5,9)[]{\LARGE \sf \textbf{\textcolor{red}{{CSP$_{\sf \textbf{B}}$}}}};
		\node at (10.5,6.5)[scale=0.4, xscale=1.5]{\ThreeManAvatar};
		\node at (10.5,9)[]{\LARGE \sf \textbf{\textcolor{red}{Recipients}}};
		\node at (-14,2)[right, text width=6.7cm, draw, color=black,line width=0.8mm]{\textcolor{black}{\textbf{\large \sf \textbf{User defines single and multiple user(s) access policy (AP\_S, AP\_M) and sends to CSP$_{\sf A}$}}}};
		\node at (-14,-4.8)[right, text width=6.7cm, draw, color=black,line width=0.8mm]{\textcolor{black}{\textbf{\large \sf \textbf{User encrypts wearable data and sends it to CSP$_{\sf A}$}}}};
		\draw[line width=0.8mm, black, yshift=0.7cm] (-7,-0.4) rectangle (0,0.4);
		\node at (-3.5,0)[yshift=0.7cm]{\textcolor{black}{\large \sf \textbf{AP Storage}}};
		\draw[line width=0.8mm, black, yshift=-5.8cm] (-7,-0.4) rectangle (0,0.4);
		\node at (-3.5,0)[yshift=-5.8cm]{\textcolor{black}{\large \sf \textbf{Data Storage}}};
\end{tikzpicture}}

\def\DiagramTwo{\begin{tikzpicture}
		\draw[line width=0.8mm, dashed] (-7,-20/2) -- (-7,20/2);
		\draw[line width=0.8mm, dashed] (7,-20/2) -- (7,20/2);
		\draw[line width=0.8mm, dashed] (0,-20/2) -- (0,20/2);
		\node at (0,{(20/2) - (20/3)})[yscale=1.2]{\OtherTitle{Data access request and processing}{14}};
		\node at (0,{(20/2) - (20/3)-0.85})[yscale=1]{\TitledefOne{\textcolor{white}{User Request case}}{14}};
		\node at (-10.5,6.2)[scale=0.4, xscale=0.7]{\UsersAvatar};
		\node at (-10.5,9)[]{\LARGE \sf \textbf{\textcolor{red}{Users}}};
		\node at (-3.5,7)[scale=1]{\NewNuage};
		\node at (-3.5,9)[]{\LARGE \sf \textbf{\textcolor{red}{CSP$_{\sf \textbf{A}}$}}};
		\node at (3.5,7)[scale=1]{\NewNuage};
		\node at (3.5,9)[]{\LARGE \sf \textbf{\textcolor{red}{{CSP$_{\sf \textbf{B}}$}}}};
		\node at (10.5,6.2)[scale=0.4, xscale=1.5]{\ThreeManAvatar};
		\node at (10.5,9)[]{\LARGE \sf \textbf{\textcolor{red}{Recipients}}};
		\node at (-14,1.3)[right, text width=6.7cm]{\textcolor{black}{\Large \sf \textbf{User rquest single Processing}}};
		\draw[line width=0.8mm, blue, -Stealth] (-14,0.5) -- (-7,0.5);
		\node at (-14,-1.3)[right, text width=6.7cm, draw, color=blue,line width=0.8mm]{\textcolor{black}{\Large \sf \textbf{User decrypts the processing results}}};
		\node at (-7,-0.1)[right, text width=6.7cm, draw, color=red,line width=0.8mm]{\textcolor{black}{\Large \sf \textbf{Performs additive homomorphie encryption}}};
\end{tikzpicture}}

\def\DiagramThree{\begin{tikzpicture}
		\draw[line width=0.8mm, dashed] (-7,-20/2-2) -- (-7,20/2);
		\draw[line width=0.8mm, dashed] (7,-20/2-2) -- (7,20/2);
		\draw[line width=0.8mm, dashed] (0,-20/2-2) -- (0,20/2);
		\node at (0,{(20/2) - (20/3)})[yscale=1.2]{\OtherTitle{Data access request and processing}{14}};
		\node at (0,{(20/2) - (20/3)-0.85})[yscale=1]{\TitledefOne{\textcolor{white}{DR request case - Single processing}}{14}};
		\node at (-10.5,6.2)[scale=0.4, xscale=0.7]{\UsersAvatar};
		\node at (-10.5,9)[]{\LARGE \sf \textbf{\textcolor{red}{Users}}};
		\node at (-3.5,7)[scale=1]{\NewNuage};
		\node at (-3.5,9)[]{\LARGE \sf \textbf{\textcolor{red}{CSP$_{\sf \textbf{A}}$}}};
		\node at (3.5,7)[scale=1]{\NewNuage};
		\node at (3.5,9)[]{\LARGE \sf \textbf{\textcolor{red}{{CSP$_{\sf \textbf{B}}$}}}};
		\node at (10.5,6.2)[scale=0.4, xscale=1.5]{\ThreeManAvatar};
		\node at (10.5,9)[]{\LARGE \sf \textbf{\textcolor{red}{Recipients}}};
		\node at (14,1.3)[left, text width=6.7cm]{\textcolor{black}{\Large \sf \textbf{DR request single processing}}};
		\draw[line width=0.8mm, blue, -Stealth] (14,0.7) -- (0,0.7);
		\node at (-7,-0)[right, text width=6.7cm, draw, color=red,line width=0.8mm]{\textcolor{black}{\Large \sf \textbf{Masking data and sends it to CSP$_{\sf B}$}}};
		\node at (0,-1.6)[right, text width=6.7cm, draw, color=green,line width=0.8mm]{\textcolor{black}{\Large \sf \textbf{CSP$_{\sf B}$ decrypts the received ciphertexts to obtains the masked data}}};
		\node at (0,-4)[right, text width=6.7cm, draw, color=green,line width=0.8mm]{\textcolor{black}{\Large \sf \textbf{Re-encrypts the masked data with new ppk and encrypts wsk with CP-ABE using AP, and send both ciphertexts to CSP$_{\sf A}$}}};
		\node at (-7,-5.9)[right, text width=6.7cm, draw, color=red,line width=0.8mm]{\textcolor{black}{\Large \sf \textbf{Store both ciphertext}}};
		\node at (-7,-7.3)[right, text width=6.7cm, draw, color=red,line width=0.8mm]{\textcolor{black}{\Large \sf \textbf{De-masking: remove random numbers and sends it to the DR}}};
		\draw[line width=0.7mm, -Stealth] (0,-8) -- (7,-8); 
		\node at (7,-8.6)[right, text width=6.7cm, draw, color=red,line width=0.8mm]{\textcolor{black}{\Large \sf \textbf{Decrypt CP-ABE(wsk) to decrypt and get the processing result}}};
\end{tikzpicture}}

\def\DiagramFour{\begin{tikzpicture}
		\draw[line width=0.8mm, dashed] (-7,-20/2) -- (-7,20/2);
		\draw[line width=0.8mm, dashed] (7,-20/2) -- (7,20/2);
		\draw[line width=0.8mm, dashed] (0,-20/2) -- (0,20/2);
		\node at (0,{(20/2) - (20/3)})[yscale=1.2]{\OtherTitle{Data access request and processing}{14}};
		\node at (0,{(20/2) - (20/3)-0.85})[yscale=1]{\TitledefOne{\textcolor{white}{DR request case - Multiple processing}}{14}};
		\node at (-10.5,6.2)[scale=0.4, xscale=0.7]{\UsersAvatar};
		\node at (-10.5,9)[]{\LARGE \sf \textbf{\textcolor{red}{Users}}};
		\node at (-3.5,7)[scale=1]{\NewNuage};
		\node at (-3.5,9)[]{\LARGE \sf \textbf{\textcolor{red}{CSP$_{\sf \textbf{A}}$}}};
		\node at (3.5,7)[scale=1]{\NewNuage};
		\node at (3.5,9)[]{\LARGE \sf \textbf{\textcolor{red}{{CSP$_{\sf \textbf{B}}$}}}};
		\node at (10.5,6.2)[scale=0.4, xscale=1.5]{\ThreeManAvatar};
		\node at (10.5,9)[]{\LARGE \sf \textbf{\textcolor{red}{Recipients}}};
		\node at (14,1.3)[left, text width=6.7cm]{\textcolor{black}{\Large \sf \textbf{DR request multiple processing}}};
		\draw[line width=0.8mm, blue, -Stealth] (14,0.7) -- (0,0.7);
		\node at (-7,0)[right, text width=6.7cm, draw, color=red,line width=0.8mm]{\textcolor{black}{\Large \sf \textbf{Group users based on matching AP\_M with DR request}}};
		\node at (-7,-1.5)[right, text width=6.7cm, draw, color=red,line width=0.8mm]{\textcolor{black}{\Large \sf \textbf{Masking data and sends it to CSP$_{\sf B}$}}};
		\node at (0,-1.6)[right, text width=6.7cm, draw, color=green,line width=0.8mm]{\textcolor{black}{\Large \sf \textbf{CSP$_{\sf B}$ decrypts the received ciphertexts using (ssk) to obtains the masked data}}};
		\node at (0,-4)[right, text width=6.7cm, draw, color=green,line width=0.8mm]{\textcolor{black}{\Large \sf \textbf{Encrypts the masked data with new ppk and encrypts wsk with CP-ABE using AP, and send both ciphertexts to CSP$_{\sf A}$}}};
		\node at (-7,-5.9)[right, text width=6.7cm, draw, color=red,line width=0.8mm]{\textcolor{black}{\Large \sf \textbf{Store both ciphertext}}};
		\node at (-7,-7.1)[right, text width=6.7cm, draw, color=red,line width=0.8mm]{\textcolor{black}{\Large \sf \textbf{De-masking data and sends it to the DR}}};
		\draw[line width=0.7mm, -Stealth] (0,-7.7) -- (7,-7.7); 
		\node at (7,-8.4)[right, text width=6.7cm, draw, color=red,line width=0.8mm]{\textcolor{black}{\Large \sf \textbf{Decrypt CP-ABE(wsk) and use it to decrypt and get the processing result}}};
\end{tikzpicture}}

\def\TitleBarAndRow#1#2{\begin{tikzpicture} \node at (0,-0.8)[]{\begin{tikzpicture}[scale=0.7] \fill[titleColor] (0.5,0) -- (1,0) -- (1,-1.5) -- (1.5,-1.5) -- (0.5,-2.5) -- (-0.5,-1.5) -- (0,-1.5) -- (0,0) -- (0.5,0); \end{tikzpicture}}; \node at (0,0)[]{\begin{tikzpicture}[yscale=1.3] \fill[titleColor, rounded corners=6] (0,0) rectangle (#2,1); \node at (#2/2,0.5)[]{\LARGE \sf \textbf{#1}};\end{tikzpicture}}; \end{tikzpicture}}

\def\TitleBarAndRowClean#1#2{\begin{tikzpicture} \node at (0,0)[]{\begin{tikzpicture}[yscale=1.3] \fill[titleColor, rounded corners=6] (0,0) rectangle (#2,1); \node at (#2/2,0.5)[]{\LARGE \sf \textbf{#1}};\end{tikzpicture}}; \end{tikzpicture}}

\def\TitleBarDashed#1#2{\begin{tikzpicture}[yscale=1.2] \draw[line width=0.7mm, blue, dashed, fill=white] (0,0) rectangle (#2,1); \node at (#2/2,0.5)[color=blue]{\LARGE \sf \textbf{#1}}; \end{tikzpicture}}

\def\DRRTitleBarDashed#1#2{\begin{tikzpicture}[yscale=1.2] \draw[line width=0.7mm, red, dashed, fill=white] (0,0) rectangle (#2,1); \node at (#2/2,0.5)[color=red]{\LARGE \sf \textbf{#1}}; \end{tikzpicture}}

\def\TitleRoundSimple#1#2#3#4{\begin{tikzpicture} \draw[orange, rounded corners=8, fill=titleColor] (0,0) rectangle (#3,2); \node at (#3/2,1.5)[]{\Large \sf \textbf{#1}}; \node at (#3/2,1)[]{\Large \sf \textbf{#4}}; \node at (#3/2,0.5)[]{\Large \sf \textbf{#2}}; \end{tikzpicture}}

\def\DiagramFiveCh{
\begin{tikzpicture}
	\node at (0,2.35*2)[]{\TitleBarAndRow{User access policy setting}{13}}; \node at (0,2.35)[]{\TitleBarAndRow{Data uploading}{13}};
	\node at (0,0.5)[]{\TitleBarAndRowClean{Data access request and processing}{13}};
	\draw[line width=0.5mm] (-6.5,-2.8) rectangle (6.5,-0.2);
	\node at (0,-1.5)[scale=0.8]{\begin{tikzpicture} \node at (4.5,-5.15)[scale=1]{ \begin{tikzpicture}[xscale=0.7]  \node at (-4.5,-1.7)[]{\TitleRoundSimple{Single User}{Processing}{4}{data}}; \node at (4.5,-1.7)[]{\TitleRoundSimple{Multiple Users}{Processing}{4}{data}}; \node at (-0.5,-1.7)[scale=0.35, xscale=0.8]{\UsersAvatar}; \node at (9.6,-1.7)[scale=0.35, xscale=1.4, yscale=0.9]{\ThreeManAvatar}; \end{tikzpicture}}; \end{tikzpicture}};
	\draw[line width=0.6mm, -Stealth, blue] (-3.75,-5) -- (-3.75,-2.35);
	\draw[line width=0.6mm, -Stealth, red] (3.5,-5) -- (3.5,-2.8);
	\node at (-3.75,-5)[scale=0.9]{\TitleBarDashed{User Request case}{6}};
	\node at (3.5,-5)[scale=0.9]{\DRRTitleBarDashed{DR Request case}{6}};
\end{tikzpicture}}

\def\DiagramSixCh{
	\begin{tikzpicture}
	\fill[titleColor] (-8,7) rectangle (8,8);
	\node at (0,7.5)[]{\Large \sf \textbf{User access policy setting}};
	\draw[] (-8,7) rectangle (8,8);
	\draw[] (-8,2) rectangle (8,7);
	\fill[titleColor] (-8,0) rectangle (8,1);
	\node at (0,0.5)[]{\Large \sf \textbf{Data uploading}};
	\draw[] (-8,0) rectangle (8,1);
	\draw[] (-8,-5) rectangle (8,0);
	\draw[thick, dashed] (-4,7) -- (-4,1);
	\draw[thick, dashed] (4,7) -- (4,-5);
	\draw[thick, dashed] (-4,0) -- (-4,-5);
	\node at (-6.5,5.6)[right, text width=5cm, draw, black, line width=0.5mm, fill=white]{\large \sf \textbf{\textcolor{black}{User defines single and multiple user(s) access policy ($\sf \textbf{AP}_\textbf{S}, \textbf{AP}_\textbf{M}$)}}};
	\draw[line width=0.5mm, -Stealth] (-4,3.75) -- (4,3.75)node[midway, above]{\large \sf {$\sf {AP}_{S}, {AP}_{M}$}};
	\node at (4,2.75)[text width=5cm, draw, black, line width=0.5mm, fill=white]{\large \sf \textcolor{white}{storage} \textbf{\textcolor{black}{AP storage}}};
	\node at (-6.5,-1.25)[right, text width=4.0cm, draw, black, line width=0.5mm, fill=white]{\large \sf \textbf{\textcolor{black}{Users encrypt wearable data using $vpk_i$}}};
	\draw[line width=0.5mm, -Stealth] (-4,-2.5) -- (4,-2.5)node[midway, above]{\large \sf {Encrypted data}};
	\node at (4,-3.75)[text width=5cm, draw, black, line width=0.5mm, fill=white]{\large \sf \textcolor{white}{storage} \textbf{\textcolor{black}{Data storage}}};
	\node at (-4,9.1)[scale=0.25]{\UsersAvatar};
	\node at (-4,10.5)[]{\large \sf \text{\textcolor{black}{Data Owners}}};
	\node at (4,10.4)[]{\large \sf \text{\textcolor{black}{SP$_{\mbox{SP}}$}}};
	\node at (4,9.1)[scale=0.65]{\NewNuage};
\end{tikzpicture}}

\def\DiagramSevenCh{\begin{tikzpicture}
		\fill[titleColor] (-8,7) rectangle (8,8);
		\node at (0,7.5)[]{\Large \sf \textbf{Data access request and processing \textcolor{black}{(User Request case)}}};
		\draw[] (-8,7) rectangle (8,8);
		\draw[] (-8,0) rectangle (8,7);
		\draw[thick, dashed] (-4,7) -- (-4,0);
		\draw[thick, dashed] (4,7) -- (4,0);
		\node at (-6.0,1.5)[right, text width=4.0cm, draw, black, line width=0.5mm, fill=white]{\large \sf \textbf{\textcolor{black}{Decrypts the\\ processing \\ results using \large \sf{$\textbf{wsk}_\textbf{i}$}}}};
		\draw[line width=0.6mm, -Stealth] (-4,6) -- (4,6)node[midway, above]{\large \sf \textbf{Requests own Data}};
		\draw[line width=0.5mm, Stealth-] (-4,3) -- (4,3)node[midway, above]{\large \sf {Processing results}};
		\node at (2.4,4.5)[right, text width=3.7cm, draw, black, line width=0.5mm, fill=white]{\large \sf \textbf{\textcolor{black}{Performs additive \\ homomorphic \\ encryption}}};
		\node at (-4,9.1)[scale=0.25, xscale=0.8]{\UsersAvatar};
		\node at (-4,10.5)[]{\large \sf \text{\textcolor{black}{Users}}};
		\node at (4,10.4)[]{\large \sf \text{\textcolor{black}{SP}}};
		\node at (4,9.1)[scale=0.65]{\NewNuage};
\end{tikzpicture}}

\def\DiagramEightCh{
\begin{tikzpicture}
	\fill[titleColor] (-10,7) rectangle (10,8);
	\node at (0,7.5)[]{\Large \sf \textbf{Data access request and processing \textcolor{black}{(DR request case)}}};
	\draw[] (-10,7) rectangle (10,8);
	\draw[] (-10,-5.5) rectangle (10,7);
	
	\draw[thick, dashed] (-7,7) -- (-7,-5.5);
	\draw[thick, dashed] (-3,7) -- (-3,-5.5);
	
	\draw[thick, dashed] (3,7) -- (3,-5.5);
	\draw[thick, dashed] (7,7) -- (7,-5.5);
	
	\draw[line width=0.6mm, -Stealth, black] (7,6) -- (-3,6)node[midway, above, fill=white]{\large \sf \textbf{Request single / multiple user(s) data}};
	
	\node at (-4.5,5.25)[right, text width=3.5cm, draw, black, line width=0.3mm, fill=white]{\large \sf \textbf{\textcolor{black}{Select N users*}}};
	
	\node at (-5.7,4.45)[right, text width=5.7cm, draw, black, line width=0.3mm, fill=white]{\large \sf \textbf{\textcolor{black}{Aggregate encrypted data**}}};
	
	\node at (-4.5,3.70)[right, text width=2.5cm, draw, black, line width=0.3mm, fill=white]{\large \sf \textbf{\textcolor{black}{Mask data}}};
	
	\draw[line width=0.6mm, -Stealth] (-3,2.9) -- (3,2.9)node[midway, above]{\large \sf {Masked Data}};
	
	\node at (0.3,2.07)[right, text width=5.5cm, draw, gray, line width=0.3mm, fill=white]{\large \sf \textbf{\textcolor{black}{Decrypts and obtains the masked data}}};
	
	\node at (0.3,1.07)[right, text width=5.5cm, draw, black, line width=0.3mm, fill=white]{\large \sf \textbf{\textcolor{black}{Aggregate masked data*}}};

	\draw[line width=0.5mm, -Stealth, black, dashed] (-3,1.5) -- (-7,1.5)node[midway, above]{\large \sf {Regular notifications}};
	
	\node at (-0.2,-0.2)[right, text width=6.5cm, draw, gray, line width=0.3mm, fill=white]{\large \sf \textbf{\textcolor{black}{Encrypts the masked data with $\textbf{ppk}_\textbf{j}$ and encrypts $\textbf{psk}_\textbf{j}$ with CP-ABE using $ \textbf{AP}_\textbf{S} / \textbf{AP}_\textbf{M}$}}};

	\draw[line width=0.6mm, Stealth-] (-3,-1.73) -- (3,-1.73)node[midway, above]{\large \sf {Masked Data}};
	
	\node at (-5.5,-2.43)[right, text width=5.0cm, draw, black, line width=0.3mm, fill=white]{\large \sf \textbf{\textcolor{black}{De-mask encrypted data}}};
	
	\draw[line width=0.6mm, Stealth-] (7,-3.0) -- (-3,-3.0)node[midway, above, fill=white]{\large \sf {Data}};
	
	\node at (4.0,-4.3)[right, text width=5.5cm, draw, black, line width=0.3mm, fill=white]{\large \sf \textbf{\textcolor{black}{Decrypt CP-ABE($\textbf{psk}_\textbf{j}$) and use $\textbf{psk}_\textbf{j}$ to decrypt and get the processing result}}};
	
	\node at (-9.5,-4.6)[right, fill=white]{\sf \textbf{*Only for multiple users data processing request}}; 
	\node at (-9.5,-5)[right, fill=white]{\sf \textbf{\newline **Only for single user data processing request}};
	
	\node at (7,9.1)[scale=0.25, xscale=1.5]{\ThreeManAvatar};
	\node at (7,10.5)[]{\large \sf \text{\textcolor{black}{DR}}};
	
	\node at (-7,9.1)[scale=0.25, xscale=0.8]{\UsersAvatar};
	\node at (-7,10.5)[]{\large \sf \text{\textcolor{black}{Users}}};
	
	\node at (-3,10.4)[]{\large \sf \text{\textcolor{black}{SP}}};
	\node at (-3,9.1)[scale=0.65]{\NewNuage};
	
	\node at (3,10.4)[]{\large \sf \text{\textcolor{black}{CP}}};
	\node at (3,9.1)[scale=0.65]{\NewNuage};
\end{tikzpicture}}
\definecolor{bluef}{HTML}{132380}
\definecolor{titleColor}{rgb}{1.0,0.86,0.6}
\definecolor{nuageColor}{rgb}{0.57,0.73,0.88}

\def\OtherNge#1{\begin{tikzpicture}\node at (0,0)[scale=#1]{\begin{tikzpicture} \draw[black!40,line width=0.8mm, fill=nuageColor] (-1,1)[rounded corners=15] -- (-1.5,1) -- (-1.5,0) -- (1.5,0) -- (1.5,1)[rounded corners=0] -- (1,1) .. controls(1,1.8) and (-0.25,2) .. (-0.5,1.4) to [bend right=70](-1,1); \draw[line width=0.8mm, black!40] (-0.5,1.4) to [bend left](-0.22,1.1); \draw[line width=0.8mm, black!40] (1,1) -- (0.7,1); \node at (0,0.15)[scale=0.65]{\begin{tikzpicture} \node at (0.75,1.1)[]{\begin{tikzpicture} \draw[line width=1.5mm, rounded corners=13, white] (0,0) -- (0,1.2) -- (0.9,1.2) -- (0.9,0); \end{tikzpicture}}; \fill[blue!20] (0,0) rectangle (1.5,1.15); \node at (0.75,0.575)[]{\begin{tikzpicture} \fill[black!70, rounded corners=2] (0,0) rectangle (0.15,0.5); \fill[black!70] (0.075,0.4) circle(0.2cm); \end{tikzpicture}}; \end{tikzpicture}}; \end{tikzpicture}};\end{tikzpicture}}

\def\ManAvatar{\begin{tikzpicture}[fill=black!80] \fill[black!85] (0,4.5)[rounded corners=35] -- (-2,4.5)[rounded corners=11] -- (-2,0.1) -- (-1.2,0.1)[rounded corners=0] -- (-1.2,3) -- (-1,3)[rounded corners=11] -- (-1,-3.5) -- (-0.1,-3.5)[rounded corners=0] -- (-0.1,0) -- (0,0)[rounded corners=0] -- (0.1,0)[rounded corners=11] -- (0.1,-3.5) -- (1,-3.5)[rounded corners=0] -- (1,3) -- (1.2,3)[rounded corners=11] -- (1.2,0.1) -- (2,0.1)[rounded corners=35] -- (2,4.5) -- (0,4.5); \fill[fill=black!85] (0,5.75) circle(1cm); \end{tikzpicture}}

\def\ManAvatarRoundWhite{\begin{tikzpicture} \node at (0,0)[scale=0.97]{\begin{tikzpicture} \fill[white] (0,4.5)[rounded corners=35] -- (-2,4.5)[rounded corners=11] -- (-2,0.1) -- (-1.2,0.1)[rounded corners=0] -- (-1.2,3) -- (-1,3)[rounded corners=11] -- (-1,-3.5) -- (-0.1,-3.5)[rounded corners=0] -- (-0.1,0) -- (0,0)[rounded corners=0] -- (0.1,0)[rounded corners=11] -- (0.1,-3.5) -- (1,-3.5)[rounded corners=0] -- (1,3) -- (1.2,3)[rounded corners=11] -- (1.2,0.1) -- (2,0.1)[rounded corners=35] -- (2,4.5) -- (0,4.5); \fill[fill=white] (0,5.75) circle(1cm); \end{tikzpicture}}; \node at (0,0)[scale=0.9]{\begin{tikzpicture}[fill=black!80] \fill[black!85] (0,4.5)[rounded corners=35] -- (-2,4.5)[rounded corners=11] -- (-2,0.1) -- (-1.2,0.1)[rounded corners=0] -- (-1.2,3) -- (-1,3)[rounded corners=11] -- (-1,-3.5) -- (-0.1,-3.5)[rounded corners=0] -- (-0.1,0) -- (0,0)[rounded corners=0] -- (0.1,0)[rounded corners=11] -- (0.1,-3.5) -- (1,-3.5)[rounded corners=0] -- (1,3) -- (1.2,3)[rounded corners=11] -- (1.2,0.1) -- (2,0.1)[rounded corners=35] -- (2,4.5) -- (0,4.5); \fill[fill=black!85] (0,5.75) circle(1cm); \end{tikzpicture}}; \end{tikzpicture}}

\def\ThreeManAvatar{\begin{tikzpicture} \node at (0,0)[xscale=0.5]{\begin{tikzpicture} \node at (0,0)[yscale=0.8]{\ManAvatarRoundWhite}; \node at (2.9,0)[yscale=0.8]{\ManAvatarRoundWhite}; \node at (-2.9,0)[yscale=0.8]{\ManAvatarRoundWhite}; \node at (5.2,1.2)[yscale=0.5]{\begin{tikzpicture} \fill[black] (0,0) arc(-90:90:0.7 and 0.7) -- (-0.05,0.7); \end{tikzpicture}}; \node at (-4.0,1.2)[xscale=-1,yscale=0.5]{\begin{tikzpicture} \fill[black] (0,0) arc(-90:90:0.7 and 0.7) -- (-0.05,0.7); \end{tikzpicture}}; \end{tikzpicture}}; \end{tikzpicture}}

\def\maskingAvatar{\begin{tikzpicture} \node at (0,0)[yscale=0.8]{\ManAvatarRoundWhite}; \end{tikzpicture}}

\def\Extrafigure#1#2{

\begin{tikzpicture}
	\node at (0,0)[scale=#1]{
	\begin{tikzpicture}[xscale=1, yscale=0.75]
			\fill[color=titleColor] (0,-4.0) rectangle (3,20);
			\draw[line width=0.3mm] (0,-4.0) rectangle (20,20);
			
			\draw[line width=0.3mm] (0,17) -- (20,17);
			\draw[line width=0.3mm] (1.25,9.4) -- (20,9.4);
			\draw[line width=0.3mm] (0,14) -- (20,14);
			\draw[line width=0.3mm] (1.25,-4.0) rectangle (1.25,14);
			\draw[line width=0.3mm] (3,-4.0) -- (3,20);
			
        	\node at (1.5,19.5)[]{#2 \sf{\textbf{DO }}};
			\node at (3/2,18.8)[]{#2 \sf{\textbf{access}}};
			\node at (3/2,18.2)[]{#2 \sf{\textbf{ policy}}};
			\node at (3/2,17.6)[]{#2 \sf{\textbf{ setting}}};
			
			\node at (3/2,16)[]{#2 \sf{\textbf{Data}}};
			\node at (3/2,15.2)[]{#2 \sf{\textbf{uploading}}};
			
			\node at (1.25+1.75/2, 11.6)[rotate=90]{
				\begin{tikzpicture}
					\node at (1.5+1.75/2, 9.8)[]{\large \sf{\textcolor{red!80!green!90}{#2 \textbf{DO}}}};
					\node at (1.75+1.75/2, 9.2)[]{\large \sf{\textcolor{red!80!green!90}{#2 \textbf{Request case}}}};
				\end{tikzpicture}
			};
			
			\node at (1.25+1.75/2, 2.5)[rotate=90]{
				\begin{tikzpicture}
					\node at (1.25+1.75/2, 9.8)[]{\large \sf{\textcolor{red!80!green!90}{#2 \textbf{DR}}}};
					\node at (1.25+1.75/2, 9.2)[]{\large \sf{\textcolor{red!80!green!90}{#2 \textbf{Request case}}}};
				\end{tikzpicture}
			};
			
		\node at (1.25/2,5)[rotate=90]{
				\begin{tikzpicture}
					\node at (1.75+1.75/2, 9.8)[]{\large \sf{\textcolor{black}{\textbf{#2 Data access request and processing}}}};
				\end{tikzpicture}
			};
			
			\draw[line width=0.2mm, dashed] (6,-4.0) -- (6,20);
			\draw[line width=0.2mm, dashed] (9.75,-4.0) -- (9.75,20);
			\draw[line width=0.2mm, dashed] (13.75,-4.0) -- (13.75,20);
			\draw[line width=0.2mm, dashed] (17.75,-4.0) -- (17.75,20);
			
			\node at (6,19.3)[draw, black, fill=white]{#2 \textbf{Set $AP_S$ and $AP_M$}};
			\draw[thick, -Stealth] (6,18.5) -- (9.75,18.5);
			\node at (9.75,17.7)[draw, black, fill=white]{#2 \textbf{Store $AP_S$ and $AP_M$}};
			
			\node at (6,16.3)[draw, black, fill=white]{#2 \textbf{Encrypt data}};
			\draw[thick, -Stealth] (6,15.5) -- (9.75,15.5);
			\node at (9.75,14.7)[draw, black, fill=white]{#2 \textbf{Store encrypted data}};
			
			\node at (6,13.3)[draw, color=black, fill=white, line width=0.2mm]{#2 \textbf{Request aggregated data}};
			\draw[thick, -Stealth] (6,12.5) -- (9.75,12.5);
			\node at (9.75,11.7)[draw, color=black, fill=white, line width=0.2mm]{ \textbf{#2 Aggregate encrypted data}};
			\draw[thick, Stealth-] (6,10.9) -- (9.75,10.9);
			\node at (6,10.1)[draw, color=black, fill=white, line width=0.2mm]{#2 \textbf{Decrypt aggregated data}};
			
			\node at (17.1,8.7)[draw, black, fill=white, line width=0.2mm]{#2\textbf{Request aggregated data}};
			
			\draw[thick, -Stealth] (17.75,7.9) -- (9.75,7.9);
			
			\node at (9.75,7.1)[draw, black, fill=white, line width=0.2mm]{#2\textbf{\textcolor{black}{Select} {{#2\textcolor{black}{N}}}\textcolor{black}{\large users*}}};
			
			\node at (9.75,6.0)[draw, black, fill=white, line width=0.2mm]{#2\textbf{\textcolor{black}{Aggregate encrypted data**}}};
			
			\node at (9.75,4.9)[draw, black, fill=white, line width=0.2mm, text width=3.0cm]{#2 \textbf{Mask data}};
			
			\draw[thick, Stealth-] (13.75,4.2) -- (9.75,4.2);
			
			\node at (13.75,3.4)[draw, black, fill=white, line width=0.2mm]{#2\textbf{Decrypt masked data}};
			
			\node at (13.75,2.2)[draw, black, fill=white, line width=0.2mm]{#2 \textbf{\textcolor{black}{Aggregate masked data*}}};
			
			\node at (13.75,0.7)[draw, black, fill=white, line width=0.2mm, text width=6.7cm]{#2 \textbf{Encrypt masked data by \textcolor{black}{$ppk_j$}\newline Encrypt \textcolor{black}{$psk_j$} by \textcolor{black}{$AP_S$} or \textcolor{black}{$AP_M$}}};
			
			\draw[thick, -Stealth] (13.75,-0.4) -- (9.75,-0.4);
			
			\node at (9.8,-1.1)[draw, black, fill=white, line width=0.2mm, text width=3.5cm]{#2 \textbf{\textcolor{black}{De-mask data}}};
			
			\draw[thick, Stealth-] (17.75,-1.7) -- (9.75,-1.7);
			
			\node at (16.3,-2.9)[draw, black, fill=white, line width=0.2mm, text width=6.9cm]{#2 \textbf{\textcolor{black}{Decrypt encrypted \textcolor{black}{$psk_j$} by $\sk_j$ \newline Decrypt aggregated data by $psk_j$}}};
			
			\node at (0.2,-2.9)[right, draw, black, dashed, fill=white, text width=6cm]{ \textcolor{red!80!green!90}{#2 \textbf{*Only for DRs-DOs case}} \newline \textcolor{red!80!green!90}{#2 \textbf{ **Only for DRs-DO case}}};
			
			\node at (6,21.6)[scale=0.2, xscale=0.75, yscale=0.9]{\maskingAvatar};
			\node at (6,20.40)[black]{\Large \sf{\textbf{DO}}};
			\node at (9.75,21.8)[]{\OtherNge{0.55}};
			\node at (9.75,20.4)[black]{\Large \sf{{$\textbf{SP}$}}};
			\node at (13.75,21.8)[]{\OtherNge{0.55}};
			\node at (13.75,20.4)[black]{\Large \sf{\textbf{$\textbf{CP}$}}};
			\node at (17.75,21.6)[scale=0.2, xscale=1.4, yscale=0.85]{\ThreeManAvatar};
			\node at (17.75,20.4)[black]{\Large \sf{\textbf{DRs}}};
	\end{tikzpicture}};
\end{tikzpicture}}

\definecolor{bluef}{HTML}{132380}
\definecolor{titleColor}{rgb}{1.0,0.86,0.6}
\definecolor{nuageColor}{rgb}{0.57,0.73,0.88}

\def\phone{\begin{tikzpicture} \fill[black!15, rounded corners=20] (0,0) rectangle (5,10); \fill[fill=bluef!75] (0,1.25) rectangle (5,8.75); \node at (1.7,0.65)[]{\begin{tikzpicture}[scale=1.15] \fill[black!35] (0,0) circle(0.115cm and 0.15cm); \end{tikzpicture}}; \node at (2.1,0.65)[]{\begin{tikzpicture}[scale=1.15] \fill[black!35] (0,0) circle(0.115cm and 0.15cm); \end{tikzpicture}}; \node at (2.5,0.65)[]{\begin{tikzpicture}[scale=1.15] \fill[black!35] (0,0) circle(0.115cm and 0.15cm); \end{tikzpicture}}; \node at (2.9,0.65)[]{\begin{tikzpicture}[scale=1.15] \fill[black!35] (0,0) circle(0.115cm and 0.15cm); \end{tikzpicture}}; \node at (3.3,0.65)[]{\begin{tikzpicture}[scale=1.15] \fill[black!35] (0,0) circle(0.115cm and 0.15cm); \end{tikzpicture}}; \node at (1.7,9.4)[]{\begin{tikzpicture}[scale=1.15] \fill[black!35] (0,0) circle(0.115cm and 0.15cm); \end{tikzpicture}}; \node at (2.1,9.4)[]{\begin{tikzpicture}[scale=1.15] \fill[black!35] (0,0) circle(0.115cm and 0.15cm); \end{tikzpicture}}; \node at (2.5,9.4)[]{\begin{tikzpicture}[scale=1.15] \fill[black!35] (0,0) circle(0.115cm and 0.15cm); \end{tikzpicture}}; \node at (2.9,9.4)[]{\begin{tikzpicture}[scale=1.15] \fill[black!35] (0,0) circle(0.115cm and 0.15cm); \end{tikzpicture}}; \node at (3.3,9.4)[]{\begin{tikzpicture}[scale=1.15] \fill[black!35] (0,0) circle(0.115cm and 0.15cm); \end{tikzpicture}}; \node at (0.5,9.4)[]{\begin{tikzpicture}[scale=1.15] \fill[black!35] (0,0) circle(0.115cm and 0.15cm); \end{tikzpicture}}; \node at (4.5,9.4)[]{\begin{tikzpicture}[scale=1.15] \fill[black!35] (0,0) circle(0.115cm and 0.15cm); \end{tikzpicture}}; \node at (0.7,7.9)[rotate=50, scale=0.6]{\begin{tikzpicture} \fill[blue!60, rounded corners=7] (0,0) rectangle (2,0.5);	\end{tikzpicture}}; \node at (1.1,7.5)[rotate=50, scale=0.6]{\begin{tikzpicture} \fill[blue!60, rounded corners=7] (0,0) rectangle (2,0.5); \fill[blue!60, rounded corners=7] (2.2,0) rectangle (2.7,0.5); \fill[blue!60, rounded corners=7] (2.9,0) rectangle (4.5,0.5);	 \end{tikzpicture}}; \node at (1.55,7.05)[rotate=50, scale=0.65]{\begin{tikzpicture} \fill[blue!60, rounded corners=7] (0,0) rectangle (3.5,0.5); \fill[blue!60, rounded corners=7] (3.7,0) rectangle (4.2,0.5); \fill[blue!60, rounded corners=7] (4.5,0) rectangle (5.5,0.5); \end{tikzpicture}}; \end{tikzpicture}}

\def\ManAvatar{\begin{tikzpicture}[fill=black!80] \fill[black!85] (0,4.5)[rounded corners=35] -- (-2,4.5)[rounded corners=11] -- (-2,0.1) -- (-1.2,0.1)[rounded corners=0] -- (-1.2,3) -- (-1,3)[rounded corners=11] -- (-1,-3.5) -- (-0.1,-3.5)[rounded corners=0] -- (-0.1,0) -- (0,0)[rounded corners=0] -- (0.1,0)[rounded corners=11] -- (0.1,-3.5) -- (1,-3.5)[rounded corners=0] -- (1,3) -- (1.2,3)[rounded corners=11] -- (1.2,0.1) -- (2,0.1)[rounded corners=35] -- (2,4.5) -- (0,4.5); \fill[fill=black!85] (0,5.75) circle(1cm); \end{tikzpicture}}

\def\RoundAssociator#1#2{\begin{tikzpicture} \draw[line width=1mm, rounded corners=25, dashed, red] (0,0) rectangle (3,3); \node at (1.5,1.5)[]{#1}; \draw[line width=1mm, rounded corners=25, dashed, red] (0,4) rectangle (3,7); \node at (1.5,5.5)[]{#2}; \draw[line width=1mm, red, dashed] (0,1.5) -- (-3,3.5) -- (0,5.5); \node at (-0.25,1.75)[rotate=60]{\begin{tikzpicture} \fill[black!50] (0,0) circle(0.35cm and 0.4cm); \node at (0,0)[]{\begin{tikzpicture}[scale=0.7] \draw[line width=1mm, red] (0,0) circle(0.35cm and 0.4cm); \end{tikzpicture}}; \node at (0,0)[]{\begin{tikzpicture}[scale=0.4] \fill[white] (0,0) circle(0.35cm and 0.4cm); \end{tikzpicture}}; \end{tikzpicture}}; \node at (-0.25,5.25)[rotate=-60]{\begin{tikzpicture} \fill[black!50] (0,0) circle(0.35cm and 0.4cm); \node at (0,0)[]{\begin{tikzpicture}[scale=0.7] \draw[line width=1mm, red] (0,0) circle(0.35cm and 0.4cm); \end{tikzpicture}};\node at (0,0)[]{\begin{tikzpicture}[scale=0.4] \fill[white] (0,0) circle(0.35cm and 0.4cm); \end{tikzpicture}}; \end{tikzpicture}}; \draw[line width=0.mm, -Stealth] (3.2,1.5) -- (6,3.2); \draw[line width=0.6mm, -Stealth] (3.2,5.5) -- (6,3.8);\end{tikzpicture}}

\def\MagnificClock#1{\begin{tikzpicture} \node at (0,0)[scale=#1]{\begin{tikzpicture} \node at (0,0)[]{\begin{tikzpicture} \fill[black!45, rounded corners=5] (0,0) rectangle (1.5,4); \end{tikzpicture}}; \node at (0.2,0.5)[]{\begin{tikzpicture} \fill[black!35, rounded corners=2] (0,0) rectangle (2,0.5); \end{tikzpicture}}; \node at (0,0)[]{\begin{tikzpicture} \fill[black!65, rounded corners=12] (0,0) rectangle (2.2,2.5); \end{tikzpicture}}; \node at (0,0)[scale=0.9]{\begin{tikzpicture} \fill[black!80, rounded corners=11] (0,0) rectangle (2.2,2.5); \end{tikzpicture}}; \node at (0,0)[scale=0.7, xscale=0.85]{\begin{tikzpicture} \fill[white] (0,0) circle(0.54cm); \draw[black!80] (0.9,0) arc(0:{7*360/8}:0.9cm)node[]{\begin{tikzpicture} \fill[green!70] (0,0) circle(0.25cm); \end{tikzpicture}}; \draw[black!80] (0.9,0) arc(0:{6*360/8}:0.9cm)node[]{\begin{tikzpicture} \fill[red!70] (0,0) circle(0.25cm); \end{tikzpicture}}; \draw[black!80] (0.9,0) arc(0:{5*360/8}:0.9cm)node[]{\begin{tikzpicture} \fill[blue!70] (0,0) circle(0.25cm); \end{tikzpicture}}; \draw[black!80] (0.9,0) arc(0:{4*360/8}:0.9cm)node[]{\begin{tikzpicture} \fill[green!70] (0,0) circle(0.25cm); \end{tikzpicture}}; \draw[black!80] (0.9,0) arc(0:{3*360/8}:0.9cm)node[]{\begin{tikzpicture} \fill[blue!70] (0,0) circle(0.25cm); \end{tikzpicture}}; \draw[black!80] (0.9,0) arc(0:{2*360/8}:0.9cm)node[]{\begin{tikzpicture} \fill[red!50] (0,0) circle(0.25cm); \end{tikzpicture}}; \draw[black!80] (0.9,0)node[]{\begin{tikzpicture} \fill[blue!70] (0,0) circle(0.25cm); \end{tikzpicture}} arc(0:{360/8}:0.9cm)node[]{\begin{tikzpicture} \fill[green!70] (0,0) circle(0.25cm); \end{tikzpicture}}; \end{tikzpicture}}; \end{tikzpicture}}; \end{tikzpicture}}

\def\VitesseBullet#1{\begin{tikzpicture} \node at (0,0)[scale=#1]{\begin{tikzpicture} \draw[black!75, line width=0.3mm] (0,0) circle(1.2cm and 0.35cm); \fill[black!45] (0,0) circle(1.2cm and 0.35cm); \node at (0,0.22)[]{\begin{tikzpicture} \fill[blue!45] (0,0) rectangle (1.8,0.9); \end{tikzpicture}}; \fill[black!60] (-1.2,0) arc(180:0:1.2cm and 0.35cm) -- (1.2,0.85) arc(0:180:1.2cm and 0.35cm) -- (-1.2,0); \draw[thick, line width=0.5mm] (-1,0.3) -- (-1,0.9); \draw[thick, line width=0.5mm] (1,0.3) -- (1,0.9); \node at (-0.85,0.6)[color=white]{\scriptsize $\bullet$}; \node at (-0.5,0.7)[color=red, scale=1.6]{\ding{170}}; \node at (0.05,0.73)[color=red, scale=1.5, xscale=0.7]{$\sf 120$}; \node at (0.68,0.7)[scale=0.3]{\begin{tikzpicture} \draw[line width=1.5mm, rounded corners=2, red] (-0.06,0) -- (-0.35,-0.3) -- (-0.4,-0.75); \draw[line width=1.5mm, rounded corners=2, red] (0.06,0) -- (0.25,-0.4) -- (0.7,-0.4); \draw[line width=1.5mm, rounded corners=2, red] (-0.06,0.68) -- (-0.35,0.5) -- (-0.7,0.8); \draw[line width=1.5mm, rounded corners=2, red] (0.06,0.7) -- (0.45,0.65) -- (0.65,0.3); \fill[red] (0,1.03) circle(5pt); \fill[red, rounded corners=2] (-0.15,-0.06) rectangle (0.15,0.8); \end{tikzpicture}}; \draw[line width=0.3mm] (1.2,0) arc(0:180:1.2cm and 0.35cm); \end{tikzpicture}}; \end{tikzpicture}}

\def\Tishort#1{\begin{tikzpicture} \node at (0,0)[scale=#1]{\begin{tikzpicture} \fill[black!10, rounded corners=3] (-1.45,1.4) -- (-1.5,0) to [bend right=20](1.5,0) -- (1.4,2.8) to [bend left=25](1,3.4) -- (0.95,4.2) -- (0.5,4.2) .. controls(0.2,3.7) and (-0.2,3.7) .. (-0.5,4.2) -- (-0.95,4.2) -- (-1,3.4) to [bend left](-1.4,2.8) -- (-1.45,1.4); \draw[line width=0.5mm, rounded corners=3] (-1.45,1.4) -- (-1.5,0) to [bend right=20](1.5,0) -- (1.4,2.8) to [bend left=25](1,3.4) -- (0.95,4.2) -- (0.5,4.2) .. controls(0.2,3.7) and (-0.2,3.7) .. (-0.5,4.2) -- (-0.95,4.2) -- (-1,3.4) to [bend left](-1.4,2.8) -- (-1.45,1.4); \fill[black!75] (0.5,4.2) .. controls(0.2,3.7) and (-0.2,3.7) .. (-0.5,4.2) -- (-0.75,4.2) .. controls(-0.6,3) and (0.6,3) .. (0.75,4.2); \draw[line width=0.5mm] (0.15,-0.3) -- (0.15,2);\node at (0.39,1.85)[scale=0.7]{\begin{tikzpicture} \fill[black!75, rounded corners=2] (0,0.2) -- (0,0) to [bend right=22](0.7,0) -- (0.7,0.4) to [bend left=22](0,0.4) -- (0,0.2); \end{tikzpicture}}; \draw[line width=0.5mm] (-0.15,-0.3) -- (-0.15,2); \node at (-0.39,1.85)[scale=0.7]{\begin{tikzpicture} \fill[black!75, rounded corners=2] (0,0.2) -- (0,0) to [bend right=22](0.7,0) -- (0.7,0.4) to [bend left=22](0,0.4) -- (0,0.2); \end{tikzpicture}}; \draw[line width=0.5mm, rounded corners=2] (0.35,-0.28) -- (0.35,1.21) -- (0.725,1.21) -- (0.8,2.75); \node at (0.9,2.75)[scale=0.7]{\begin{tikzpicture} \fill[black!75, rounded corners=2] (0,0.2) -- (0,0) to [bend right=22](0.7,0) -- (0.7,0.4) to [bend left=22](0,0.4) -- (0,0.2); \end{tikzpicture}}; \draw[line width=0.5mm, rounded corners=2] (-0.35,-0.28) -- (-0.35,1.21) -- (-0.725,1.21) -- (-0.8,2.75); \node at (-0.9,2.75)[scale=0.7]{\begin{tikzpicture} \fill[black!75, rounded corners=2] (0,0.2) -- (0,0) to [bend right=22](0.7,0) -- (0.7,0.4) to [bend left=22](0,0.4) -- (0,0.2); \end{tikzpicture}}; \draw[line width=0.5mm] (-0.5,-0.26) -- (-0.5,0.3) -- (-0.85,0.3) -- (-0.85,0.7); \node at (-1.08,0.7)[scale=0.7]{\begin{tikzpicture} \fill[black!75, rounded corners=2] (0,0.2) -- (0,0) to [bend right=22](0.7,0) -- (0.7,0.4) to [bend left=22](0,0.4) -- (0,0.2); \end{tikzpicture}}; \node at (1.2,0.6)[scale=0.5]{\begin{tikzpicture} \fill[black!10] (0,0) circle(1.15cm); \draw[thick] (0,0) circle(1cm); \draw[thick] (0,0) circle(0.9cm); \draw[thick] (0,0) circle(0.8cm); \fill[black!10] (-0.1,-1.1) rectangle (0.1,1.1); \fill[black!10] (-1.1,-0.1) rectangle (1.1,0.1); \draw[thick, rounded corners=2] (-0.1,-0.7) rectangle (0.1,0.7); \draw[thick, rounded corners=2] (-0.5,0.3) rectangle (0.5,0.5); \end{tikzpicture}}; \end{tikzpicture}}; \end{tikzpicture}}

\def\Nuage#1#2{\begin{tikzpicture}\node at (0,0)[scale=#2]{\begin{tikzpicture} \draw[black!40,line width=0.8mm, fill=nuageColor] (-1,1)[rounded corners=15] -- (-1.5,1) -- (-1.5,0) -- (1.5,0) -- (1.5,1)[rounded corners=0] -- (1,1) .. controls(1,1.8) and (-0.25,2) .. (-0.5,1.4) to [bend right=70](-1,1); \draw[line width=0.8mm, black!40] (-0.5,1.4) to [bend left](-0.22,1.1); \draw[line width=0.8mm, black!40] (1,1) -- (0.7,1); \node at (0,0.15)[scale=0.65]{\begin{tikzpicture} \node at (0.75,1.1)[]{\begin{tikzpicture} \draw[line width=1.5mm, rounded corners=13, white] (0,0) -- (0,1.2) -- (0.9,1.2) -- (0.9,0); \end{tikzpicture}}; \fill[blue!20] (0,0) rectangle (1.5,1.15); \node at (0.75,0.575)[]{\begin{tikzpicture} \fill[black!70, rounded corners=2] (0,0) rectangle (0.15,0.5); \fill[black!70] (0.075,0.4) circle(0.2cm); \end{tikzpicture}}; \end{tikzpicture}}; \node at (0,0.9)[scale=1.2]{\Large $\mbox{CSP}_{\mbox{\scriptsize #1}}$}; \end{tikzpicture}};\end{tikzpicture}}

\def\Titledef#1#2{\begin{tikzpicture}[xscale=#2, yscale=1.3] \fill[titleColor, rounded corners=5] (0,0) rectangle (3.5,0.5); \node at (1.75,0.25)[color=black]{\sf #1}; \end{tikzpicture}}

\def\Key#1{\begin{tikzpicture} \node at (0,0)[scale=#1]{\begin{tikzpicture}[xscale=1.15, yscale=1.1] \filldraw[line width=0.8mm, fill=titleColor] (1,-0.25) -- (2.5,-0.25) arc(-155:0:0.9cm and 0.8cm) arc(0:165:0.9cm and 0.8cm) -- (2.2,0.35) -- (2,0.2) -- (1.8,0.4) -- (1.6,0.2) -- (1.4,0.45) -- (1.2,0.2) -- (1,0.45) -- (0.75,0.1) -- (1,-0.25); \filldraw[line width=0.8mm, fill=white] (3.7,0.1) circle(6pt and 5pt); \end{tikzpicture}}; \end{tikzpicture}}

\def\FirstHome#1{\begin{tikzpicture} \node at (0,0)[scale=#1]{\begin{tikzpicture} \draw[line width=1mm, fill=bluef!55] (4,0) rectangle (8,4.5); \draw[line width=0.8mm] (4,3.9) -- (6.1,3.9) (6.3,3.9) -- (6.6,3.9) (6.9,3.9) -- (8,3.9); \node at (4.8,0.8)[scale=0.65]{\begin{tikzpicture} \draw[line width=0.5mm, fill=white] (0.5,1.5) rectangle (1.5,2.25); \end{tikzpicture}}; \node at (4.8,1.6)[scale=0.65]{\begin{tikzpicture} \draw[line width=0.5mm, fill=white] (0.5,1.5) rectangle (1.5,2.25); \end{tikzpicture}}; \node at (4.8,2.4)[scale=0.65]{\begin{tikzpicture} \draw[line width=0.5mm, fill=white] (0.5,1.5) rectangle (1.5,2.25); \end{tikzpicture}}; \node at (4.8,3.2)[scale=0.65]{\begin{tikzpicture} \draw[line width=0.5mm, fill=white] (0.5,1.5) rectangle (1.5,2.25); \end{tikzpicture}}; \node at (6,0.8)[scale=0.65]{\begin{tikzpicture} \draw[line width=0.5mm, fill=white] (0.5,1.5) rectangle (1.5,2.25); \end{tikzpicture}}; \node at (6,1.6)[scale=0.65]{\begin{tikzpicture} \draw[line width=0.5mm, fill=white] (0.5,1.5) rectangle (1.5,2.25); \end{tikzpicture}}; \node at (6,2.4)[scale=0.65]{\begin{tikzpicture} \draw[line width=0.5mm, fill=white] (0.5,1.5) rectangle (1.5,2.25); \end{tikzpicture}}; \node at (6,3.2)[scale=0.65]{\begin{tikzpicture} \draw[line width=0.5mm, fill=white] (0.5,1.5) rectangle (1.5,2.25); \end{tikzpicture}}; \node at (7.2,0.8)[scale=0.65]{\begin{tikzpicture} \draw[line width=0.5mm, fill=white] (0.5,1.5) rectangle (1.5,2.25); \end{tikzpicture}}; \node at (7.2,1.6)[scale=0.65]{\begin{tikzpicture} \draw[line width=0.5mm, fill=white] (0.5,1.5) rectangle (1.5,2.25); \end{tikzpicture}}; \node at (7.2,2.4)[scale=0.65]{\begin{tikzpicture} \draw[line width=0.5mm, fill=white] (0.5,1.5) rectangle (1.5,2.25); \end{tikzpicture}}; \node at (7.2,3.2)[scale=0.65]{\begin{tikzpicture} \draw[line width=0.5mm, fill=white] (0.5,1.5) rectangle (1.5,2.25); \end{tikzpicture}}; \node at (5.9,4.5)[]{\begin{tikzpicture} \filldraw[line width=1mm, fill=black!25, rounded corners=8] (-0.3,0) rectangle (4.6,0.5); \end{tikzpicture}}; \fill[black!25] (0,0) rectangle (4,0.85); \fill[bluef!55] (0,0.85) rectangle (4,6); \draw[line width=0.6mm] (0,0.85) -- (0.4,0.85) (0.8,0.85) -- (3.2,0.85) (3.6,0.85) -- (4,0.85); \draw[line width=0.6mm] (0,5.5) -- (2,5.5) (2.4,5.5) -- (3,5.5) (3.4,5.5) -- (4,5.5); \node at (1,1.75)[scale=0.85]{\begin{tikzpicture} \draw[line width=0.5mm, fill=white] (0.5,1.5) rectangle (1.5,2.25); \end{tikzpicture}}; \node at (1,2.75)[scale=0.85]{\begin{tikzpicture} \draw[line width=0.5mm, fill=white] (0.5,1.5) rectangle (1.5,2.25); \end{tikzpicture}}; \node at (1,3.75)[scale=0.85]{\begin{tikzpicture} \draw[line width=0.5mm, fill=white] (0.5,1.5) rectangle (1.5,2.25); \end{tikzpicture}}; \node at (1,4.75)[scale=0.85]{\begin{tikzpicture} \draw[line width=0.5mm, fill=white] (0.5,1.5) rectangle (1.5,2.25); \end{tikzpicture}}; \node at (3,1.75)[scale=0.85]{\begin{tikzpicture} \draw[line width=0.5mm, fill=white] (0.5,1.5) rectangle (1.5,2.25); \end{tikzpicture}}; \node at (3,2.75)[scale=0.85]{\begin{tikzpicture} \draw[line width=0.5mm, fill=white] (0.5,1.5) rectangle (1.5,2.25); \end{tikzpicture}}; \node at (3,3.75)[scale=0.85]{\begin{tikzpicture} \draw[line width=0.5mm, fill=white] (0.5,1.5) rectangle (1.5,2.25); \end{tikzpicture}}; \node at (3,4.75)[scale=0.85]{\begin{tikzpicture} \draw[line width=0.5mm, fill=white] (0.5,1.5) rectangle (1.5,2.25); \end{tikzpicture}}; \draw[line width=1mm] (0,0) rectangle (4,6); \node at (2,6.15)[]{\begin{tikzpicture} \filldraw[line width=1mm, fill=black!25, rounded corners=8] (-0.3,0) rectangle (4.6,0.5); \end{tikzpicture}}; \draw[line width=1.1mm, fill=white, rounded corners=20] (1.35,0) -- (1.35,1.5) -- (2.65,1.5) -- (2.65,0); \draw[line width=1mm] (-0.8,0) -- (8.8,0); \end{tikzpicture}}; \end{tikzpicture}}

\def\ContactIconeFirst#1#2{
    \begin{tikzpicture}
        \node at (0,0)[scale=#1]{
            \begin{tikzpicture}
                \draw[line width=0.6mm, rounded corners=35, fill=white] (-4.5,-4.4) rectangle (6,3); \node at (0.75,-3.5)[scale=#2]{\Huge \sf \textbf{Insurance Staff}}; \node at (-2,-1)[scale=0.55]{\begin{tikzpicture} \fill[black] (0,0)[rounded corners=15] -- (2.3,0)[rounded corners=2] to [bend right](0.7,3)[rounded corners=2] -- (0,1.75)[rounded corners=2] -- (-0.7,3)[rounded corners=15] to [bend right](-2.3,0)[rounded corners=2] -- (0,0); \fill[black] (0,2.95) -- (-0.17,2.45) -- (0,2.15) -- (0.17,2.45) -- (0,2.95); \fill[white] (0,0.15) -- (1.4,0.15) -- (1.6,1.5) -- (-1.6,1.5) -- (-1.4,0.15) -- (0,0.15); \fill[black] (0,4) circle(0.9cm); \draw[line width=0.4cm] (-2.8,-0.4) -- (2.8,-0.4); \end{tikzpicture}}; \node at (0.8,0.7)[scale=0.55]{\begin{tikzpicture} \fill[black] (0,0)[rounded corners=3] -- (1.2,0)[rounded corners=3] -- (1.2,0.7)[rounded corners=3] to [bend right](1.65,0.9)[rounded corners=8] -- (1.65,3.25)[rounded corners=1] to [bend right=20](0.5,4) -- (0,3.5) -- (-0.5,4)[rounded corners=8] to [bend right=20](-1.65,3.25)[rounded corners=3] -- (-1.65,0.9)[rounded corners=3] to [bend right](-1.2,0.7) -- (-1.2,0) -- (0,0); \fill[black] (0,5) circle(0.85cm); \node at (1.2,1)[scale=0.7]{\begin{tikzpicture}\node at (0,0)[]{\begin{tikzpicture} \filldraw[white, line width=4mm] (0,0) to [bend right=40](1.4,1.5)[rounded corners=3] -- (1.4,2.4)[rounded corners=5] -- (1.6,2.6)[rounded corners=5] to [bend right=10](0,2.9)[rounded corners=5] to [bend right=10](-1.6,2.6)[rounded corners=3] -- (-1.4,2.4) -- (-1.4,1.5) to [bend right=40](0,0);\end{tikzpicture}}; \node at (0,0)[scale=0.9]{\begin{tikzpicture} \draw[line width=2mm, fill=black] (0,0) to [bend right=40](1.4,1.5)[rounded corners=3] -- (1.4,2.4)[rounded corners=5] -- (1.6,2.6)[rounded corners=0] to [bend right=10](0,2.9) -- (0,0); \draw[line width=2mm] (0,0) to [bend left=40](-1.4,1.5)[rounded corners=3] -- (-1.4,2.4)[rounded corners=5] -- (-1.6,2.6)[rounded corners=0] to [bend left=10](0,2.9) -- (0,0); \end{tikzpicture}}; \node at (1,1.3)[scale=0.8]{\begin{tikzpicture} \fill[white] (0,0) circle(1.5cm); \fill[black] (0,0) circle(1.25cm); \node at (0,0)[]{\begin{tikzpicture} \draw[line width=3mm, white] (-0.5,0.5) -- (0,0) -- (0.8,1.2); \end{tikzpicture}}; \end{tikzpicture}}; \end{tikzpicture}}; \node at (0.85,3.5)[]{\begin{tikzpicture} \fill[white] (0,0) circle(8pt); \draw[white, line width=2mm] (0,0) -- (0,1.25); \end{tikzpicture}}; \node at (-0.8,3.2)[]{\begin{tikzpicture} \draw[line width=2mm, white] (-0.25,0.1)[rounded corners=3] -- (-0.4,0.15)[rounded corners=10]  -- (-0.4,1) -- (0.4,1)[rounded corners=3] -- (0.4,0.15) -- (0.25,0.1); \draw[line width=2mm, white] (0,1) -- (0,1.6); \end{tikzpicture}}; \end{tikzpicture}}; \node at (3.3,-1.3)[scale=0.8]{\begin{tikzpicture} \node at (0,-0.2)[scale=0.5]{\begin{tikzpicture} \fill[black!70, rounded corners=5] (0,0) -- (1.5,0) to [bend right=15](1.3,1.8) to [bend right=10](0.6,2) to [bend left](-0.6,2) to [bend right=10](-1.3,1.8) to [bend right=15](-1.5,0) -- (0,0); \fill[black!70] (0,3) circle(0.86cm); \end{tikzpicture}};	\node at (1.5,0)[]{\begin{tikzpicture}\node at (0,0)[]{\begin{tikzpicture}\fill[white] (0,0)[rounded corners=13] -- (1.3,0)[rounded corners=20] -- (1.3,1.65)[rounded corners=0] -- (0.5,1.65) -- (0,0.7) -- (-0.5,1.65)[rounded corners=20] -- (-1.3,1.65)[rounded corners=13] -- (-1.3,0) -- (0,0);\fill[white, rounded corners=3] (0.12,0.5) -- (0.05,1.25) -- (0.2,1.65) -- (-0.2,1.65) -- (-0.05,1.25) -- (-0.12,0.5); \fill[white] (0,2.7) circle(0.88cm and 0.85cm); \end{tikzpicture}}; \node at (0,-0.05)[scale=0.95]{\begin{tikzpicture}\fill[black!70] (0,0)[rounded corners=13] -- (1.3,0)[rounded corners=20] -- (1.3,1.65)[rounded corners=0] -- (0.5,1.65) -- (0,0.7) -- (-0.5,1.65)[rounded corners=20] -- (-1.3,1.65)[rounded corners=13] -- (-1.3,0) -- (0,0);\fill[black!70, rounded corners=3] (0.12,0.5) -- (0.05,1.25) -- (0.2,1.65) -- (-0.2,1.65) -- (-0.05,1.25) -- (-0.12,0.5); \fill[black!70] (0,2.7) circle(0.88cm and 0.85cm); \end{tikzpicture}}; \end{tikzpicture}};\end{tikzpicture}}; \end{tikzpicture}}; \end{tikzpicture}}

\def\ContactIconeSecond#1#2{
    \begin{tikzpicture}
        \node at (0,0)[scale=#1]{
            \begin{tikzpicture}
                \draw[line width=0.6mm, rounded corners=35, fill=white] (-4.5,-4.4) rectangle (6,3);	\node at (0.75,-3.5)[scale=#2]{\Huge \sf \textbf{Researchers}};	\node at (4.7,1.5)[scale=0.35]{\begin{tikzpicture} \draw[line width=0.9mm] (0,0) circle(2cm); \node at (0,0.35)[scale=1.1]{\begin{tikzpicture} \node at (0,0)[scale=0.65, yscale=1.3, xscale=1]{\begin{tikzpicture}\fill[black!80, rounded corners=4] (0,0) -- (2,0) -- (2,1) -- (0.3,1.5) -- (0,0.8) -- (-0.3,1.5) -- (-2,1) -- (-2,0) -- (0,0); \end{tikzpicture}}; \node at (0,1.35)[scale=1.18]{\begin{tikzpicture} \fill[black!80] (0.4,1) to [bend right=70](-0.4,1)[rounded corners=5] to [bend right=40](0,0) to [bend right=40](0.4,1); \end{tikzpicture}}; \end{tikzpicture}}; \end{tikzpicture}}; \node at (1.7,1.5)[scale=0.35]{\begin{tikzpicture} \draw[line width=0.9mm] (0,0) circle(2cm); \node at (0,0.35)[scale=1.1]{\begin{tikzpicture} \node at (0,0)[scale=0.65, yscale=1.3, xscale=1]{\begin{tikzpicture}\fill[black!80, rounded corners=4] (0,0) -- (2,0) -- (2,1) -- (0.3,1.5) -- (0,0.8) -- (-0.3,1.5) -- (-2,1) -- (-2,0) -- (0,0); \end{tikzpicture}}; \node at (0,1.35)[scale=1.18]{\begin{tikzpicture} \fill[black!80] (0.4,1) to [bend right=70](-0.4,1)[rounded corners=5] to [bend right=40](0,0) to [bend right=40](0.4,1); \end{tikzpicture}}; \end{tikzpicture}}; \end{tikzpicture}}; \node at (3.7,-0.7)[scale=0.85]{\begin{tikzpicture}[scale=1.2] \node at (-0.04,-0.04)[rotate=38]{\begin{tikzpicture} \draw[line width=2.5mm] (0,0) circle(1.2cm); \draw[line width=2.5mm] (0,-1.2) -- (0,-2); \draw[line width=4.5mm] (0,-1.6) -- (0,-2); \draw[line width=7.5mm] (0,-1.85) -- (0,-3); \end{tikzpicture}}; \draw[line width=0.8mm] (-2.03,0) -- (0,0); \draw[line width=0.8mm] (-2.03,-0.15) -- (0,-0.15); \draw[line width=3mm] (-0.1,0.1) -- (-0.1,1); \draw[line width=3mm] (-0.55,0.1) -- (-0.55,1.4); \draw[line width=3mm] (-1,0.1) -- (-1,0.65); \draw[line width=3mm, white] (-1.45,0.1) -- (-1.45,0.9); \draw[line width=1mm] (-1.35,0.1) -- (-1.35,0.9); \draw[line width=3mm] (-1.9,0.1) -- (-1.9,0.55);\end{tikzpicture}}; \node at (1.7,-1.8)[scale=0.35]{\begin{tikzpicture} \draw[line width=0.9mm] (0,0) circle(2cm); \node at (0,0.35)[scale=1.1]{\begin{tikzpicture} \node at (0,0)[scale=0.65, yscale=1.3, xscale=1]{\begin{tikzpicture}\fill[black!80, rounded corners=4] (0,0) -- (2,0) -- (2,1) -- (0.3,1.5) -- (0,0.8) -- (-0.3,1.5) -- (-2,1) -- (-2,0) -- (0,0); \end{tikzpicture}}; \node at (0,1.35)[scale=1.18]{\begin{tikzpicture} \fill[black!80] (0.4,1) to [bend right=70](-0.4,1)[rounded corners=5] to [bend right=40](0,0) to [bend right=40](0.4,1); \end{tikzpicture}}; \end{tikzpicture}}; \end{tikzpicture}};	\node at (-0.8,1.5)[scale=0.95]{\begin{tikzpicture}\draw[line width=1.2mm, black!80] (0.12,0.2) -- (0.64,0.2) (0.76,0.2) -- (0.88,0.2); \draw[line width=1.2mm, black!80] (0.12,0.45) -- (0.88,0.45); \draw[line width=1.2mm, black!80] (0.12,0.7) -- (0.88,0.7); \draw[line width=1.2mm, black!80] (0.12,0.95) -- (0.88,0.95); \draw[line width=1.2mm, black!80] (-0.7,0.15) -- (0.16,0.96);\end{tikzpicture}};	\node at (-3,1.5)[scale=0.7]{\begin{tikzpicture} \draw[line width=1.2mm, black!80] (0,0) -- (2.25,0); \draw[line width=3mm, black!80] (0.4,0) -- (0.4,1.2); \draw[line width=3mm, black!80] (1.1,0) -- (1.1,0.8); \draw[line width=3mm, black!80] (1.8,0) -- (1.8,0.5);	 \end{tikzpicture}};	\node at (-0.4,-0.3)[scale=0.8]{\begin{tikzpicture} \node at (0,0)[]{\begin{tikzpicture} \draw[line width=1.5mm] (0,0) -- (0,1); \end{tikzpicture}}; \node at (0,0)[rotate=45]{\begin{tikzpicture} \draw[line width=1.5mm] (0,0) -- (0,1); \end{tikzpicture}}; \node at (0,0)[rotate=90]{\begin{tikzpicture} \draw[line width=1.5mm] (0,0) -- (0,1); \end{tikzpicture}}; \node at (0,0)[rotate=135]{\begin{tikzpicture} \draw[line width=1.5mm] (0,0) -- (0,1); \end{tikzpicture}}; \fill[white] (0,0) circle(0.27cm); \draw[line width=1.5mm] (0,-0.7) -- (0,-1.3) (0,-1.45) -- (0,-1.65); \end{tikzpicture}};	\node at (-1.8,-1)[scale=0.7]{\begin{tikzpicture}\fill[black!80] (0,0)[rounded corners=10] -- (1.3,0) -- (1.3,1.65)[rounded corners=0] -- (0.5,1.65) -- (0,0.7) -- (-0.5,1.65)[rounded corners=10] -- (-1.3,1.65) -- (-1.3,0) -- (0,0);\fill[black!80, rounded corners=3] (0.12,0.5) -- (0.05,1.25) -- (0.2,1.65) -- (-0.2,1.65) -- (-0.05,1.25) -- (-0.12,0.5); \fill[black!80] (0,2.7) circle(0.88cm and 0.85cm); \end{tikzpicture}};	\node at (-3.35,-0.6)[scale=0.9]{\begin{tikzpicture} \fill[black!80] (0,0) rectangle (1,1.2); \draw[line width=1.2mm, white] (0.12,0.2) -- (0.88,0.2); \draw[line width=1.2mm, white] (0.12,0.45) -- (0.88,0.45); \draw[line width=1.2mm, white] (0.12,0.7) -- (0.88,0.7); \draw[line width=1.2mm, white] (0.12,0.95) -- (0.88,0.95); \end{tikzpicture}}; \end{tikzpicture}};\end{tikzpicture}}

\def\ContactIconeThird#1#2{
    \begin{tikzpicture}
        \node at (0,0)[scale=#1]{
            \begin{tikzpicture}
                \draw[line width=0.6mm, rounded corners=35, fill=white] (-4.5,-4.4) rectangle (6,3);
                \node at (-2,-2.1)[scale=0.7]{\begin{tikzpicture} \draw[line width=0.5mm, fill=black] (0,0) to [bend right=15](1.5,0.5) to [bend right](0.5,2.25) -- (0,1.4) -- (-0.5,2.25) to [bend right](-1.5,0.5) to [bend right=15](0,0);
                \node at (0.4,0.75)[]{\begin{tikzpicture}\draw[line width=1mm, white] (0,0.2) -- (0,-0.2); \draw[line width=1mm, white] (-0.2,0) -- (0.2,0); \end{tikzpicture}};
                \node at (-0.85,1.4)[scale=0.7]{\begin{tikzpicture} \draw[line width=0.8mm, white, rounded corners=4] (-0.4,0)node[color=white]{\Large $\bullet$} -- (-0.6,0.15) to [bend left](0,1.25) to [bend left](0.6,0.15) -- (0.4,0)node[color=white]{\Large $\bullet$}; \draw[line width=0.8mm, white] (0,1.2) to [bend left=15](0.3,2.25); \end{tikzpicture}};
                \node at (0.8,1.65)[]{\begin{tikzpicture} \draw[line width=1mm, white] (0,0) to [bend right=10](-0.2,1); \draw[line width=1mm, white, fill=black] (0,0) circle(8pt); \end{tikzpicture}}; \draw[thick, fill=black] (0,3) circle(0.7cm); \end{tikzpicture}}; \node at (-2,-3.75)[scale=#2]{\Huge \sf \textbf{Doctor}}; \node at (-2,1.65)[scale=0.6]{\begin{tikzpicture} \draw[line width=0.2mm, fill=black] (0,0) -- (1.5,0) arc(0:180:1.5 and 1.4) -- (0,0); \draw[thick, fill=black] (0,2.25) circle(0.7cm);
                \draw[thick, fill=black] (0,3.13) to [bend left=15](0.6,2.9) -- (1,3.5) to [bend right=50](-1,3.4) -- (-0.6,2.9) to [bend left=15](0,3.13);
                \node at (0,3.5)[]{\begin{tikzpicture} \draw[line width=1.5mm, white] (-0.25,0) -- (0.25,0); \draw[line width=1.5mm, white] (0,0.25) -- (0,-0.25); \end{tikzpicture}};\end{tikzpicture}}; \node at (-2,-0.15)[scale=#2]{\Huge \sf \textbf{Nurse}};
                \node at (2.5,-2.1)[scale=0.7]{\begin{tikzpicture} \draw[thick, fill=black] (0,0) -- (2,0) to [bend right=10](1.8,1.25) to [bend right=5](0.85,1.8) -- (0.6,1.65) -- (0,0.5) (0,0) -- (-2,0) to [bend left=10](-1.8,1.25) to [bend left=5](-0.85,1.8) -- (-0.6,1.65) -- (0,0.5); \draw[thick, fill=black] (0,1.8) -- (-0.2,1.5) -- (-0.08,1.25) -- (-0.2,0.4) -- (0.2,0.4) -- (0.08,1.25) -- (0.2,1.5) -- (0,1.8); \draw[fill=black] (0,2.9) circle(0.8cm and 0.85cm); \node at (-1.5,1.2)[scale=0.8]{\begin{tikzpicture} \node at (0.1,-0.1)[scale=1.5]{\begin{tikzpicture} \draw[line width=1mm, white, fill=white] (0,0)[rounded corners=10] -- (1,0)[rounded corners=0] -- (0.2,1.6) -- (0.2,2.2)[rounded corners=2] -- (0.35,2.2) -- (0.35,2.4) -- (-0.35,2.4) -- (-0.35,2.2)[rounded corners=0] -- (-0.2,2.2) -- (-0.2,1.6)[rounded corners=10] -- (-1,0) -- (0,0); \end{tikzpicture}};	\node at (0,0)[scale=1.2]{\begin{tikzpicture} \draw[line width=1mm, fill=black] (0,0)[rounded corners=10] -- (1,0)[rounded corners=0] -- (0.2,1.6) -- (0.2,2.2)[rounded corners=2] -- (0.35,2.2) -- (0.35,2.4) -- (-0.35,2.4) -- (-0.35,2.2)[rounded corners=0] -- (-0.2,2.2) -- (-0.2,1.6)[rounded corners=10] -- (-1,0) -- (0,0); \end{tikzpicture}}; \node at (0,-0.75)[scale=1]{\begin{tikzpicture} \draw[line width=1mm, fill=white] (0,0)[rounded corners=5] -- (1,0)[rounded corners=0] -- (0.4,1.1) -- (-0.4,1.1)[rounded corners=5] -- (-1,0) -- (0,0); \end{tikzpicture}}; \node at (-0.15,1.8)[scale=2]{$\bullet$}; \node at (-0.3,2.2)[scale=2]{$\bullet$}; \node at (-0.8,1.8)[scale=3]{$\bullet$}; \end{tikzpicture}};  \end{tikzpicture}}; \node at (2.9,-3.75)[scale=#2]{\Huge \sf \textbf{Lab Worker}}; \node at (2.5,1.6)[scale=0.7]{\begin{tikzpicture} \draw[line width=0.2mm, fill=black] (0,0) -- (1.5,0) arc(0:180:1.5 and 1.4) -- (0,0); \draw[thick, fill=black] (0,2.25) circle(0.7cm); \node at (-0.8,0.5)[]{\begin{tikzpicture} \draw[line width=2mm, white] (-0.35,0) -- (0.35,0); \draw[line width=2mm, white] (0,0.35) -- (0,-0.35); \end{tikzpicture}}; \node at (0.8,0.6)[scale=0.7]{\begin{tikzpicture} \fill[white] (0,0) circle(1.3cm); \node at (0,0.2)[rotate=50, yscale=1.4, xscale=1.1]{\begin{tikzpicture} \fill[black] (0.1,0)[rounded corners=7] -- (1,0) -- (1,0.5)[rounded corners=0] -- (0.1,0.5) -- (0.1,0); \fill[black] (-0.1,0)[rounded corners=7] -- (-1,0) -- (-1,0.5)[rounded corners=0] -- (-0.1,0.5) -- (-0.1,0); \end{tikzpicture}}; \end{tikzpicture}}; \end{tikzpicture}}; \node at (2.9,0.1)[scale=#2]{\Huge \sf \textbf{Pharmacist}};\end{tikzpicture}};\end{tikzpicture}}

\def\SecondHome#1{\begin{tikzpicture} \node at (0,0)[scale=#1, xscale=1.3]{\begin{tikzpicture} \draw[line width=2mm] (0,0) -- (0,6.08); \fill[bluef!20] (-1.7,0) rectangle (1.7,6); \fill[bluef!40] (1.7,0.05) rectangle (4.25,4.25); \draw[line width=1.2mm, fill=bluef!60, rounded corners=1] (-0.5,0) rectangle (0.5,1.2); \draw[line width=1.2mm, fill=bluef!40, rounded corners=1] (0.5,0) rectangle (1,1.2); \draw[line width=1.2mm, fill=bluef!40, rounded corners=1] (-0.5,0) rectangle (-1,1.2); \draw[line width=1.2mm] (-1.4,0) -- (1.7,0) -- (1.7,4); \node at (-1,2)[scale=0.55]{\begin{tikzpicture} \draw[line width=1.2mm, fill=bluef!37, rounded corners=1] (0,0) rectangle (1,1);\end{tikzpicture}}; \node at (-1,3)[scale=0.55]{\begin{tikzpicture} \draw[line width=1.2mm, fill=bluef!37, rounded corners=1] (0,0) rectangle (1,1);\end{tikzpicture}}; \node at (-1,4)[scale=0.55]{\begin{tikzpicture} \draw[line width=1.2mm, fill=bluef!37, rounded corners=1] (0,0) rectangle (1,1);\end{tikzpicture}}; \node at (0,2)[scale=0.55]{\begin{tikzpicture} \draw[line width=1.2mm, fill=bluef!37, rounded corners=1] (0,0) rectangle (1,1);\end{tikzpicture}}; \node at (0,3)[scale=0.55]{\begin{tikzpicture} \draw[line width=1.2mm, fill=bluef!37, rounded corners=1] (0,0) rectangle (1,1);\end{tikzpicture}}; \node at (0,4)[scale=0.55]{\begin{tikzpicture} \draw[line width=1.2mm, fill=bluef!37, rounded corners=1] (0,0) rectangle (1,1);\end{tikzpicture}}; \node at (1,2)[scale=0.55]{\begin{tikzpicture} \draw[line width=1.2mm, fill=bluef!37, rounded corners=1] (0,0) rectangle (1,1);\end{tikzpicture}}; \node at (1,3)[scale=0.55]{\begin{tikzpicture} \draw[line width=1.2mm, fill=bluef!37, rounded corners=1] (0,0) rectangle (1,1);\end{tikzpicture}}; \node at (1,4)[scale=0.55]{\begin{tikzpicture} \draw[line width=1.2mm, fill=bluef!37, rounded corners=1] (0,0) rectangle (1,1);\end{tikzpicture}}; \node at (2.9,1)[]{\begin{tikzpicture} \draw[line width=1.2mm, fill=bluef!60, rounded corners=1] (0,0) rectangle (1.8,0.6); \end{tikzpicture}}; \node at (2.9,2)[]{\begin{tikzpicture} \draw[line width=1.2mm, fill=bluef!60, rounded corners=1] (0,0) rectangle (1.8,0.6); \end{tikzpicture}}; \node at (2.9,3)[]{\begin{tikzpicture} \draw[line width=1.2mm, fill=bluef!60, rounded corners=1] (0,0) rectangle (1.8,0.6); \end{tikzpicture}}; \node at (0,5.2)[scale=0.4]{\begin{tikzpicture} \draw[line width=3mm, fill=green!70] (-1,1.7)[rounded corners=2] -- (-1,1.5)[rounded corners=4] to [bend right](0,0)[rounded corners=2] to [bend right](1,1.5)[rounded corners=3] -- (1,2.25)[rounded corners=3] to [bend left=20](0,2.25)[rounded corners=2] to [bend left=20](-1,2.25) -- (-1,1.7); \end{tikzpicture}}; \end{tikzpicture}};\end{tikzpicture}}

\def\Hospital#1{
    \begin{tikzpicture}
    \node at (0,0)[scale=#1]{\begin{tikzpicture}
    \fill[black!30] (0,0) circle(4.2cm and 0.8cm);
    \node at (0,2.7)[scale=0.5]{\begin{tikzpicture}
    \draw[black!25, fill=black!10] (-2.5,0) rectangle (2.5,9.8);
    \draw[black!25, fill=black!3] (-2.4,2.3) rectangle (2.4,9.7);
    \draw[line width=0.8mm, fill=bluef!40, rounded corners=1] (-1,0) rectangle (1,2);
    \draw[line width=0.8mm] (0,0) -- (0,2);
    \draw[line width=0.8mm, rounded corners=1, fill=bluef!40] (1.25,0) rectangle (2.25,2);
    \draw[line width=0.8mm, rounded corners=1, fill=bluef!40] (-1.25,0) rectangle (-2.25,2);
    \node at (0,3)[xscale=1.1]{\begin{tikzpicture} \fill[bluef!50] (0,0) rectangle (0.8,1); \fill[bluef!20] (0.8,0) rectangle (1.6,1); \fill[bluef!50] (1.6,0) rectangle (2.4,1); \fill[bluef!20] (2.4,0) rectangle (3.2,1); \fill[bluef!50] (3.2,0) rectangle (4,1); \end{tikzpicture}};
    \node at (0,4.5)[xscale=1.1]{\begin{tikzpicture} \fill[bluef!50] (0,0) rectangle (0.8,1); \fill[bluef!20] (0.8,0) rectangle (1.6,1); \fill[bluef!50] (1.6,0) rectangle (2.4,1); \fill[bluef!20] (2.4,0) rectangle (3.2,1); \fill[bluef!50] (3.2,0) rectangle (4,1); \end{tikzpicture}};
    \node at (0,6)[xscale=1.1]{\begin{tikzpicture} \fill[bluef!50] (0,0) rectangle (0.8,1); \fill[bluef!20] (0.8,0) rectangle (1.6,1); \fill[bluef!50] (1.6,0) rectangle (2.4,1); \fill[bluef!20] (2.4,0) rectangle (3.2,1); \fill[bluef!50] (3.2,0) rectangle (4,1); \end{tikzpicture}};
    \node at (0,7.5)[xscale=1.1]{\begin{tikzpicture} \fill[bluef!50] (0,0) rectangle (0.8,1); \fill[bluef!20] (0.8,0) rectangle (1.6,1); \fill[bluef!50] (1.6,0) rectangle (2.4,1); \fill[bluef!20] (2.4,0) rectangle (3.2,1); \fill[bluef!50] (3.2,0) rectangle (4,1); \end{tikzpicture}};
    \node at (0,9)[xscale=1.1]{\begin{tikzpicture} \fill[bluef!50] (0,0) rectangle (0.8,1); \fill[bluef!20] (0.8,0) rectangle (1.6,1); \fill[bluef!50] (1.6,0) rectangle (2.4,1); \fill[bluef!20] (2.4,0) rectangle (3.2,1); \fill[bluef!50] (3.2,0) rectangle (4,1); \end{tikzpicture}};
    \fill[fill=red] (-3,9.8) rectangle (3,11);
    \node at (0,10.4)[color=white]{\Huge \sf HOSPITAL};
    \fill[fill=black!5] (2.5,0) rectangle (6.55,5.1);
    \draw[line width=1.4mm, black!40] (2.5,0) -- (6.55,0);
    \fill[fill=black!5] (-2.5,0) rectangle (-6.55,5.1);
    \draw[line width=1.4mm, black!40] (-2.5,0) -- (-6.55,0);
    \fill[black!20, opacity=0.8] (2.5,0) -- (2.5,5.1) -- (5,0);
    \fill[black!20, opacity=0.8] (-2.5,0) -- (-2.5,5.1) -- (-5,0);
    \draw[line width=5mm, black!65] (-2.5,5) -- (-6.55,5);
    \draw[line width=5mm, black!65] (2.5,5) -- (6.55,5);
    \node at (3,0.8)[]{\begin{tikzpicture}\draw[line width=0.4mm, fill=blue!30] (0,0) rectangle (0.4,1); \end{tikzpicture}};
    \node at (3.6,0.8)[]{\begin{tikzpicture}\draw[line width=0.4mm, fill=blue!30] (0,0) rectangle (0.4,1); \end{tikzpicture}};
    \node at (4.2,0.8)[]{\begin{tikzpicture}\draw[line width=0.4mm, fill=blue!30] (0,0) rectangle (0.4,1); \end{tikzpicture}};
    \node at (4.8,0.8)[]{\begin{tikzpicture}\draw[line width=0.4mm, fill=blue!30] (0,0) rectangle (0.4,1); \end{tikzpicture}};
    \node at (5.4,0.8)[]{\begin{tikzpicture}\draw[line width=0.4mm, fill=blue!30] (0,0) rectangle (0.4,1); \end{tikzpicture}};
    \node at (6,0.8)[]{\begin{tikzpicture}\draw[line width=0.4mm, fill=blue!30] (0,0) rectangle (0.4,1); \end{tikzpicture}};
    \node at (3,2.3)[]{\begin{tikzpicture}\draw[line width=0.4mm, fill=blue!30] (0,0) rectangle (0.4,1); \end{tikzpicture}};
    \node at (3.6,2.3)[]{\begin{tikzpicture}\draw[line width=0.4mm, fill=blue!30] (0,0) rectangle (0.4,1); \end{tikzpicture}};
    \node at (4.2,2.3)[]{\begin{tikzpicture}\draw[line width=0.4mm, fill=blue!30] (0,0) rectangle (0.4,1); \end{tikzpicture}};
    \node at (4.8,2.3)[]{\begin{tikzpicture}\draw[line width=0.4mm, fill=blue!30] (0,0) rectangle (0.4,1); \end{tikzpicture}};
    \node at (5.4,2.3)[]{\begin{tikzpicture}\draw[line width=0.4mm, fill=blue!30] (0,0) rectangle (0.4,1); \end{tikzpicture}};
    \node at (6,2.3)[]{\begin{tikzpicture}\draw[line width=0.4mm, fill=blue!30] (0,0) rectangle (0.4,1); \end{tikzpicture}};
    \node at (3,3.8)[]{\begin{tikzpicture}\draw[line width=0.4mm, fill=blue!30] (0,0) rectangle (0.4,1); \end{tikzpicture}};
    \node at (3.6,3.8)[]{\begin{tikzpicture}\draw[line width=0.4mm, fill=blue!30] (0,0) rectangle (0.4,1); \end{tikzpicture}};
    \node at (4.2,3.8)[]{\begin{tikzpicture}\draw[line width=0.4mm, fill=blue!30] (0,0) rectangle (0.4,1); \end{tikzpicture}};
    \node at (4.8,3.8)[]{\begin{tikzpicture}\draw[line width=0.4mm, fill=blue!30] (0,0) rectangle (0.4,1); \end{tikzpicture}};
    \node at (5.4,3.8)[]{\begin{tikzpicture}\draw[line width=0.4mm, fill=blue!30] (0,0) rectangle (0.4,1); \end{tikzpicture}};
    \node at (6,3.8)[]{\begin{tikzpicture}\draw[line width=0.4mm, fill=blue!30] (0,0) rectangle (0.4,1); \end{tikzpicture}};
    \node at (-3,0.8)[]{\begin{tikzpicture}\draw[line width=0.4mm, fill=blue!30] (0,0) rectangle (0.4,1); \end{tikzpicture}};
    \node at (-3.6,0.8)[]{\begin{tikzpicture}\draw[line width=0.4mm, fill=blue!30] (0,0) rectangle (0.4,1); \end{tikzpicture}};
    \node at (-4.2,0.8)[]{\begin{tikzpicture}\draw[line width=0.4mm, fill=blue!30] (0,0) rectangle (0.4,1); \end{tikzpicture}};
    \node at (-4.8,0.8)[]{\begin{tikzpicture}\draw[line width=0.4mm, fill=blue!30] (0,0) rectangle (0.4,1); \end{tikzpicture}};
    \node at (-5.4,0.8)[]{\begin{tikzpicture}\draw[line width=0.4mm, fill=blue!30] (0,0) rectangle (0.4,1); \end{tikzpicture}};
    \node at (-6,0.8)[]{\begin{tikzpicture}\draw[line width=0.4mm, fill=blue!30] (0,0) rectangle (0.4,1); \end{tikzpicture}};
    \node at (-3,2.3)[]{\begin{tikzpicture}\draw[line width=0.4mm, fill=blue!30] (0,0) rectangle (0.4,1); \end{tikzpicture}};
    \node at (-3.6,2.3)[]{\begin{tikzpicture}\draw[line width=0.4mm, fill=blue!30] (0,0) rectangle (0.4,1); \end{tikzpicture}};
    \node at (-4.2,2.3)[]{\begin{tikzpicture}\draw[line width=0.4mm, fill=blue!30] (0,0) rectangle (0.4,1); \end{tikzpicture}};
    \node at (-4.8,2.3)[]{\begin{tikzpicture}\draw[line width=0.4mm, fill=blue!30] (0,0) rectangle (0.4,1); \end{tikzpicture}};
    \node at (-5.4,2.3)[]{\begin{tikzpicture}\draw[line width=0.4mm, fill=blue!30] (0,0) rectangle (0.4,1); \end{tikzpicture}};
    \node at (-6,2.3)[]{\begin{tikzpicture}\draw[line width=0.4mm, fill=blue!30] (0,0) rectangle (0.4,1); \end{tikzpicture}};
    \node at (-3,3.8)[]{\begin{tikzpicture}\draw[line width=0.4mm, fill=blue!30] (0,0) rectangle (0.4,1); \end{tikzpicture}};
    \node at (-3.6,3.8)[]{\begin{tikzpicture}\draw[line width=0.4mm, fill=blue!30] (0,0) rectangle (0.4,1); \end{tikzpicture}};
    \node at (-4.2,3.8)[]{\begin{tikzpicture}\draw[line width=0.4mm, fill=blue!30] (0,0) rectangle (0.4,1); \end{tikzpicture}};
    \node at (-4.8,3.8)[]{\begin{tikzpicture}\draw[line width=0.4mm, fill=blue!30] (0,0) rectangle (0.4,1); \end{tikzpicture}};
    \node at (-5.4,3.8)[]{\begin{tikzpicture}\draw[line width=0.4mm, fill=blue!30] (0,0) rectangle (0.4,1); \end{tikzpicture}};
    \node at (-6,3.8)[]{\begin{tikzpicture}\draw[line width=0.4mm, fill=blue!30] (0,0) rectangle (0.4,1); \end{tikzpicture}}; \end{tikzpicture}}; \end{tikzpicture}};\end{tikzpicture}}

\def\MyFinalDiagram#1#2{\begin{tikzpicture}
		\node at (0,0)[scale=#1]{
			\begin{tikzpicture}[xscale=1.05]
				\node at (0,10)[scale=0.4]{\ManAvatar}; \node at (2,10)[scale=0.4]{\RoundAssociator{\MagnificClock{0.6}}{\VitesseBullet{0.8}}}; \node at (4.1,10)[scale=0.15]{\phone};
				
				\node at (0,5)[scale=0.4]{\ManAvatar}; \node at (2,5)[scale=0.4]{\RoundAssociator{\VitesseBullet{0.8}}{\Tishort{0.48}}}; \node at (4.1,5)[scale=0.15]{\phone};
				
				\node at (0,0)[scale=0.4]{\ManAvatar}; \node at (2,0)[scale=0.4]{\RoundAssociator{\Tishort{0.48}}{\MagnificClock{0.6}}}; \node at (4.1,0)[scale=0.15]{\phone};
				
				\node at (9,4.8)[]{\SPCPServeurFunction{0.75}{SP}{CP}};
				\draw[line width=0.12cm, Triangle-Triangle] (8.75,3.5) -- (8.75,6.2);
				
				\draw[line width=0.5mm, Stealth-Stealth] (4.55, 0)--(8,6.8);
                \draw[line width=0.5mm, Stealth-Stealth] (4.55, 5)--(7.5,7);
                \draw[line width=0.5mm, Stealth-Stealth] (4.55, 10)--(7.5,7.6);
				
				\draw[line width=0.5mm, rounded corners=10, Stealth-] (0,12.4) -- (0,13.2) -- (8.75,13.2) -- (8.75,14);
				\draw[line width=0.4mm, rounded corners=10, -Stealth] (8.75,14) -- (8.75,13.2) -- (17.5,13.2) -- (17.6,12.3);
				
				\node at (0,14)[scale=#2]{\Titledef{\Large\textbf{DOs}}{0.5}};
				\node at (4.5,13.15)[rotate=40]{\Key{0.3}};
				\node at (8.75,13.1)[below, scale=#2]{\textcolor{black}{\sf \Large \textbf{KA}}};
				\node at (8.75,14.3)[]{\FirstHome{0.22}};
				\draw[line width=0.5mm, Stealth-Stealth] (8.75,8.5) -- (8.75,12.3);
				\node at (8,10.2)[rotate=40]{\Key{0.3}};
				
				\node at (12.75,13.15)[rotate=40]{\Key{0.3}};
				\node at (17.5,14)[scale=#2]{\Titledef{\large \textbf{DRs}}{0.5}};
				\draw[line width=0.8mm, dashed, red, rounded corners=30] (14.5,-2.3) rectangle (20.5,12.25);
				\draw[line width=0.6mm, Stealth-Stealth] (10,6.9) -- (14.8,0);
				\draw[line width=0.6mm, Stealth-Stealth] (10,7.3) -- (14.8,5);
				\draw[line width=0.6mm, Stealth-Stealth] (10,7.6) -- (14.8,10);

				\node at (17.53,10)[]{\ContactIconeFirst{0.5}{1.3}};
				
				\node at (17.53,5)[]{\ContactIconeSecond{0.5}{1.3}};
				
				\node at (17.53,0)[]{\ContactIconeThird{0.5}{1.2}};
				
				
		\end{tikzpicture}};
\end{tikzpicture}}

\def\boxText#1#2#3#4{
\begin{tikzpicture}
\draw[line width=0.5mm, fill=white] (-0.1,-0.4) rectangle (3.4-#4,1.5);
\node at (0,0.9)[right, scale=#3+0.3]{\sf \textbf{#1}};
\node at (0,0.15)[right, scale=#3+.3]{\sf \textbf{#2}};
\end{tikzpicture}
}
\def\newFigureOne#1{
    \begin{tikzpicture}
	\node at (0,0)[scale=#1]{
		\begin{tikzpicture}
			\draw[line width=0.5mm] (-2,-3) -- (0,0) -- (2,-3);
			\draw[line width=0.5mm] (-6,-6)-- (-2,-3) -- (-1.5,-6);
			\draw[line width=0.5mm] (6,-6)-- (2,-3) -- (1.5,-6);
			\node at (0,0)[]{\begin{tikzpicture}
					\draw[line width=0.8mm, fill=white] (0,0) circle(0.9cm);
					\node at (0,0)[]{\Large \sf \textbf{AND}};
			\end{tikzpicture}};
			
			\node at (-2,-3)[]{\begin{tikzpicture}
					\draw[line width=0.8mm, fill=white] (0,0) circle(0.8cm);
					\node at (0,0)[]{\Large \sf \textbf{OR}};
			\end{tikzpicture}};
			
			\node at (2,-3)[]{\begin{tikzpicture}
					\draw[line width=0.8mm, fill=white] (0,0) circle(0.8cm);
					\node at (0,0)[]{\Large \sf \textbf{OR}};
			\end{tikzpicture}};
			
			\node at (-5.5,-6.2)[]{\boxText{Research}{Institutes}{1.2}{0.5}};
			\node at (-2,-6.2)[]{\boxText{Government}{Regulators}{1.2}{0}};
			\node at (2,-6.2)[]{\boxText{International}{Autorities}{1.15}{0}};
			\node at (5.5,-6.2)[]{\boxText{Local city}{regulators}{1.2}{0.5}};
		\end{tikzpicture}
	};
    \end{tikzpicture}
}

\def\newFigureTwo#1{
    \begin{tikzpicture}
	\node at (0,0)[scale=#1]{
		\begin{tikzpicture}
			\draw[line width=0.5mm] (-2,-3) -- (0,0) -- (2,-3);
			\draw[line width=0.5mm] (-6,-6)-- (-2,-3) -- (-1.5,-6);
			\draw[line width=0.5mm] (6,-6)-- (2,-3) -- (1.5,-6);
			\node at (0,0)[]{\begin{tikzpicture}
					\draw[line width=0.8mm, fill=white] (0,0) circle(0.9cm);
					\node at (0,0)[]{\Large \sf \textbf{AND}};
			\end{tikzpicture}};
			
			\node at (-2,-3)[]{\begin{tikzpicture}
					\draw[line width=0.8mm, fill=white] (0,0) circle(0.8cm);
					\node at (0,0)[]{\Large \sf \textbf{OR}};
			\end{tikzpicture}};
			
			\node at (2,-3)[]{\begin{tikzpicture}
					\draw[line width=0.8mm, fill=white] (0,0) circle(0.8cm);
					\node at (0,0)[]{\Large \sf \textbf{OR}};
			\end{tikzpicture}};
			
			\node at (-5.4,-6.2)[]{\boxText{Family}{members}{1.2}{0.5}};
			\node at (-2,-6.2)[]{\boxText{Specific}{Doctor}{1.2}{0.6}};
			\node at (1.6,-6.2)[]{\boxText{Specific}{Friends}{1.15}{0.7}};
			\node at (5,-6.2)[]{\boxText{Specific}{Hospital}{1.2}{0.6}};
		\end{tikzpicture}
	};
    \end{tikzpicture}
}
	


%
\title{Efficient User-Centric Privacy-Friendly and Flexible Wearable Data Aggregation and Sharing}
\def\Bande{
	\begin{tikzpicture}
		\fill[thick, rounded corners=3, blue!60] (3,0) arc(0:-180:3cm and 0.7cm) -- (-3,1) arc(-180:0:3cm and 0.7cm) -- (3,0);
		\filldraw[white, line width=1.3mm, xshift=0.5cm] (0,0.1) -- (0,-0.5);
		\filldraw[white, line width=1.3mm, xshift=1.5cm, yshift=0.05cm] (0,0.1) -- (0,-0.5);
		\filldraw[white, line width=1.3mm, xshift=2.5cm, yshift=0.25cm] (0,0.1) -- (0,-0.5);
		\filldraw[white, line width=1.3mm, xshift=-0.5cm] (0,0.1) -- (0,-0.5);
		\filldraw[white, line width=1.3mm, xshift=-1.5cm, yshift=0.08cm] (0,0.1) -- (0,-0.5);
		\filldraw[white, line width=1.3mm, xshift=-2.5cm, yshift=0.25cm] (0,0.1) -- (0,-0.5);
	\end{tikzpicture}
}

\def\Parametre{
	\begin{tikzpicture}
		\def\BandeR{
			\begin{tikzpicture}
				\filldraw[white, rounded corners=2] (-0.1,-0.1) rectangle (2.9, 0.5);
				\fill[black!60, rounded corners=1] (0,0) rectangle (2.8, 0.4);
			\end{tikzpicture}
		}
		\filldraw[white] (0,0) circle(1.15cm);
		\node at (0,0)[]{\BandeR};
		\node at (0,0)[rotate=45]{\BandeR};
		\node at (0,0)[rotate=90]{\BandeR};
		\node at (0,0)[rotate=135]{\BandeR};
		\filldraw[black!60] (0,0) circle(1.05cm);
		\filldraw[white] (0,0) circle(0.55cm);
	\end{tikzpicture}
}

\def\Serveur#1{
	\begin{tikzpicture}
		\node at (0,)[scale=#1]{
			\begin{tikzpicture}
				\node at (0,0)[xscale=0.7]{
					\begin{tikzpicture}
						\filldraw[blue!60] (3,0) arc(0:360:3cm and 0.7cm);
						\node at (0,-1.2)[]{\Bande};
						\node at (0,-2.55)[]{\Bande};
						\node at (0,-3.9)[]{\Bande};
					\end{tikzpicture}
				};
				\node at (1.8,-1.8)[]{\Parametre};
			\end{tikzpicture}
		};
	\end{tikzpicture}
}


\def\phonex#1{
	\begin{tikzpicture}
		\node at (0,)[scale=#1]{\begin{tikzpicture}
				\fill[black!15, rounded corners=20] (0,0) rectangle (5,10);
				\fill[fill=bluef!75] (0,1.25) rectangle (5,8.75);
				\node at (1.7,0.65)[]{
					\begin{tikzpicture}[scale=1.15]
						\fill[black!35] (0,0) circle(0.115cm and 0.15cm);
				\end{tikzpicture}};
				\node at (2.1,0.65)[]{
					\begin{tikzpicture}[scale=1.15]
						\fill[black!35] (0,0) circle(0.115cm and 0.15cm);
				\end{tikzpicture}};
				\node at (2.5,0.65)[]{
					\begin{tikzpicture}[scale=1.15]
						\fill[black!35] (0,0) circle(0.115cm and 0.15cm);
				\end{tikzpicture}};
				\node at (2.9,0.65)[]{
					\begin{tikzpicture}[scale=1.15]
						\fill[black!35] (0,0) circle(0.115cm and 0.15cm);
				\end{tikzpicture}};
				\node at (3.3,0.65)[]{
					\begin{tikzpicture}[scale=1.15]
						\fill[black!35] (0,0) circle(0.115cm and 0.15cm);
				\end{tikzpicture}};
				\node at (1.7,9.4)[]{
					\begin{tikzpicture}[scale=1.15]
						\fill[black!35] (0,0) circle(0.115cm and 0.15cm);
				\end{tikzpicture}};
				\node at (2.1,9.4)[]{
					\begin{tikzpicture}[scale=1.15]
						\fill[black!35] (0,0) circle(0.115cm and 0.15cm);
				\end{tikzpicture}};
				\node at (2.5,9.4)[]{
					\begin{tikzpicture}[scale=1.15]
						\fill[black!35] (0,0) circle(0.115cm and 0.15cm);
				\end{tikzpicture}};
				\node at (2.9,9.4)[]{
					\begin{tikzpicture}[scale=1.15] 
						\fill[black!35] (0,0) circle(0.115cm and 0.15cm); 
				\end{tikzpicture}};
				\node at (3.3,9.4)[]{\begin{tikzpicture}[scale=1.15]
						\fill[black!35] (0,0) circle(0.115cm and 0.15cm); 
				\end{tikzpicture}}; 
				\node at (0.5,9.4)[]{
					\begin{tikzpicture}[scale=1.15] 
						\fill[black!35] (0,0) circle(0.115cm and 0.15cm); 
				\end{tikzpicture}}; 
				\node at (4.5,9.4)[]{
					\begin{tikzpicture}[scale=1.15] 
						\fill[black!35] (0,0) circle(0.115cm and 0.15cm); 
				\end{tikzpicture}}; 
				\node at (0.7,7.9)[rotate=50, scale=0.6]{
					\begin{tikzpicture} 
						\fill[blue!60, rounded corners=7] (0,0) rectangle (2,0.5);	
				\end{tikzpicture}}; 
				\node at (1.1,7.5)[rotate=50, scale=0.6]{
					\begin{tikzpicture} 
						\fill[blue!60, rounded corners=7] (0,0) rectangle (2,0.5); \fill[blue!60, rounded corners=7] (2.2,0) rectangle (2.7,0.5); \fill[blue!60, rounded corners=7] (2.9,0) rectangle (4.5,0.5);	 
				\end{tikzpicture}}; 
				\node at (1.55,7.05)[rotate=50, scale=0.65]{
					\begin{tikzpicture} 
						\fill[blue!60, rounded corners=7] (0,0) rectangle (3.5,0.5); \fill[blue!60, rounded corners=7] (3.7,0) rectangle (4.2,0.5); \fill[blue!60, rounded corners=7] (4.5,0) rectangle (5.5,0.5); 
				\end{tikzpicture}};
		\end{tikzpicture}};
	\end{tikzpicture}
}

\def\ManAvatarx#1{
	\begin{tikzpicture}
		\node at (0,0)[scale=#1]{\begin{tikzpicture}[fill=black!80]
				\fill[black!85] (0,4.5)[rounded corners=35] -- (-2,4.5)[rounded corners=11] -- (-2,0.1) -- (-1.2,0.1)[rounded corners=0] -- (-1.2,3) -- (-1,3)[rounded corners=11] -- (-1,-3.5) -- (-0.1,-3.5)[rounded corners=0] -- (-0.1,0) -- (0,0)[rounded corners=0] -- (0.1,0)[rounded corners=11] -- (0.1,-3.5) -- (1,-3.5)[rounded corners=0] -- (1,3) -- (1.2,3)[rounded corners=11] -- (1.2,0.1) -- (2,0.1)[rounded corners=35] -- (2,4.5) -- (0,4.5); \fill[fill=black!85] (0,5.75) circle(1cm);
		\end{tikzpicture}};
	\end{tikzpicture}
}

\def\ManAvatarHandUp#1{
	\begin{tikzpicture}
		\node at (0,0)[scale=#1]{
			\begin{tikzpicture}[fill=black!80]
				\fill[black!85] (0,4.5)[rounded corners=0] -- (-2,4.5)[rounded corners=11] -- (-3.8,6) -- (-4.3,5.5)[rounded corners=5] -- (-2,3.7) -- (-1,3.6)[rounded corners=11] -- (-1,-3.5) -- (-0.1,-3.5)[rounded corners=0] -- (-0.1,0) -- (0,0)[rounded corners=0] -- (0.1,0)[rounded corners=11] -- (0.1,-3.5) -- (1,-3.5)[rounded corners=0] -- (1,3.3) -- (1.2,3.3)[rounded corners=8] -- (1.2,0.1) -- (1.8,0.1)[rounded corners=35] -- (1.8,4.5) -- (0,4.5); \fill[fill=black!85] (0,5.75) circle(1cm);
			\end{tikzpicture}
		};
\end{tikzpicture}}

\def\MagnificClockx#1{
	\begin{tikzpicture} 
		\node at (0,0)[scale=#1]{
			\begin{tikzpicture} 
				\node at (0,0)[]{
					\begin{tikzpicture} 
						\fill[black!45, rounded corners=5] (0,0) rectangle (1.5,4); 
				\end{tikzpicture}}; 
				\node at (0.2,0.5)[]{
					\begin{tikzpicture} 
						\fill[black!35, rounded corners=2] (0,0) rectangle (2,0.5); 
				\end{tikzpicture}}; 
				\node at (0,0)[]{
					\begin{tikzpicture} 
						\fill[black!65, rounded corners=12] (0,0) rectangle (2.2,2.5); \end{tikzpicture}}; 
				\node at (0,0)[scale=0.9]{
					\begin{tikzpicture} 
						\fill[black!80, rounded corners=11] (0,0) rectangle (2.2,2.5); 
				\end{tikzpicture}}; 
				\node at (0,0)[scale=0.7, xscale=0.85]{
					\begin{tikzpicture} 
						\fill[white] (0,0) circle(0.54cm); 
						\draw[black!80] (0.9,0) arc(0:{7*360/8}:0.9cm)node[]{
							\begin{tikzpicture} 
								\fill[green!70] (0,0) circle(0.25cm); 
						\end{tikzpicture}}; 
						\draw[black!80] (0.9,0) arc(0:{6*360/8}:0.9cm)node[]{
							\begin{tikzpicture} 
								\fill[red!70] (0,0) circle(0.25cm); 
						\end{tikzpicture}}; 
						\draw[black!80] (0.9,0) arc(0:{5*360/8}:0.9cm)node[]{
							\begin{tikzpicture} 
								\fill[blue!70] (0,0) circle(0.25cm); 
						\end{tikzpicture}}; 
						\draw[black!80] (0.9,0) arc(0:{4*360/8}:0.9cm)node[]{
							\begin{tikzpicture} 
								\fill[green!70] (0,0) circle(0.25cm); 
						\end{tikzpicture}};
						\draw[black!80] (0.9,0) arc(0:{3*360/8}:0.9cm)node[]{
							\begin{tikzpicture} 
								\fill[blue!70] (0,0) circle(0.25cm); 
						\end{tikzpicture}}; 
						\draw[black!80] (0.9,0) arc(0:{2*360/8}:0.9cm)node[]{
							\begin{tikzpicture} 
								\fill[red!50] (0,0) circle(0.25cm); 
						\end{tikzpicture}}; 
						\draw[black!80] (0.9,0)node[]{
							\begin{tikzpicture} 
								\fill[blue!70] (0,0) circle(0.25cm); 
						\end{tikzpicture}} arc(0:{360/8}:0.9cm)node[]{
							\begin{tikzpicture} 
								\fill[green!70] (0,0) circle(0.25cm); 
						\end{tikzpicture}}; 
				\end{tikzpicture}}; 
		\end{tikzpicture}}; 
\end{tikzpicture}}

\def\VitesseBulletx#1{
	\begin{tikzpicture} 
		\node at (0,0)[scale=#1]{
			\begin{tikzpicture} 
				\draw[black!75, line width=0.3mm] (0,0) circle(1.2cm and 0.35cm); 
				\fill[black!45] (0,0) circle(1.2cm and 0.35cm); \node at (0,0.22)[]{
					\begin{tikzpicture} 
						\fill[blue!45] (0,0) rectangle (1.8,0.9); 
				\end{tikzpicture}}; 
				\fill[black!60] (-1.2,0) arc(180:0:1.2cm and 0.35cm) -- (1.2,0.85) arc(0:180:1.2cm and 0.35cm) -- (-1.2,0); \draw[thick, line width=0.5mm] (-1,0.3) -- (-1,0.9); \draw[thick, line width=0.5mm] (1,0.3) -- (1,0.9); \node at (-0.85,0.6)[color=white]{\scriptsize $\bullet$}; \node at (-0.5,0.7)[color=red, scale=1.6]{\ding{170}}; 
				\node at (0.05,0.73)[color=red, scale=1.5, xscale=0.7]{$\sf 120$}; 
				\node at (0.68,0.7)[scale=0.3]{
					\begin{tikzpicture} 
						\draw[line width=1.5mm, rounded corners=2, red] (-0.06,0) -- (-0.35,-0.3) -- (-0.4,-0.75); \draw[line width=1.5mm, rounded corners=2, red] (0.06,0) -- (0.25,-0.4) -- (0.7,-0.4); 
						\draw[line width=1.5mm, rounded corners=2, red] (-0.06,0.68) -- (-0.35,0.5) -- (-0.7,0.8); 
						\draw[line width=1.5mm, rounded corners=2, red] (0.06,0.7) -- (0.45,0.65) -- (0.65,0.3); 
						\fill[red] (0,1.03) circle(5pt); 
						\fill[red, rounded corners=2] (-0.15,-0.06) rectangle (0.15,0.8); 
				\end{tikzpicture}}; 
				\draw[line width=0.3mm] (1.2,0) arc(0:180:1.2cm and 0.35cm); 
		\end{tikzpicture}}; 
\end{tikzpicture}}

\def\Tishort#1{
	\begin{tikzpicture} 
		\node at (0,0)[scale=#1]{
			\begin{tikzpicture} 
				\fill[black!10, rounded corners=3] (-1.45,1.4) -- (-1.5,0) to [bend right=20](1.5,0) -- (1.4,2.8) to [bend left=25](1,3.4) -- (0.95,4.2) -- (0.5,4.2) .. controls(0.2,3.7) and (-0.2,3.7) .. (-0.5,4.2) -- (-0.95,4.2) -- (-1,3.4) to [bend left](-1.4,2.8) -- (-1.45,1.4); 
				\draw[line width=0.6mm, rounded corners=3] (-1.45,1.4) -- (-1.5,0) to [bend right=20](1.5,0) -- (1.4,2.8) to [bend left=25](1,3.4) -- (0.95,4.2) -- (0.5,4.2) .. controls(0.2,3.7) and (-0.2,3.7) .. (-0.5,4.2) -- (-0.95,4.2) -- (-1,3.4) to [bend left](-1.4,2.8) -- (-1.45,1.4); 
				\fill[black!75] (0.5,4.2) .. controls(0.2,3.7) and (-0.2,3.7) .. (-0.5,4.2) -- (-0.75,4.2) .. controls(-0.6,3) and (0.6,3) .. (0.75,4.2);
				\draw[line width=0.5mm] (0.15,-0.3) -- (0.15,2);
				\node at (0.39,1.85)[scale=0.7]{
					\begin{tikzpicture} 
						\fill[black!75, rounded corners=2] (0,0.2) -- (0,0) to [bend right=22](0.7,0) -- (0.7,0.4) to [bend left=22](0,0.4) -- (0,0.2); 
				\end{tikzpicture}}; 
				\draw[line width=0.5mm] (-0.15,-0.3) -- (-0.15,2); \node at (-0.39,1.85)[scale=0.7]{
					\begin{tikzpicture} 
						\fill[black!75, rounded corners=2] (0,0.2) -- (0,0) to [bend right=22](0.7,0) -- (0.7,0.4) to [bend left=22](0,0.4) -- (0,0.2); 
				\end{tikzpicture}}; 
				\draw[line width=0.5mm, rounded corners=2] (0.35,-0.28) -- (0.35,1.21) -- (0.725,1.21) -- (0.8,2.75); 
				\node at (0.9,2.75)[scale=0.7]{
					\begin{tikzpicture} 
						\fill[black!75, rounded corners=2] (0,0.2) -- (0,0) to [bend right=22](0.7,0) -- (0.7,0.4) to [bend left=22](0,0.4) -- (0,0.2); 
				\end{tikzpicture}}; 
				\draw[line width=0.5mm, rounded corners=2] (-0.35,-0.28) -- (-0.35,1.21) -- (-0.725,1.21) -- (-0.8,2.75); 
				\node at (-0.9,2.75)[scale=0.7]{
					\begin{tikzpicture} 
						\fill[black!75, rounded corners=2] (0,0.2) -- (0,0) to [bend right=22](0.7,0) -- (0.7,0.4) to [bend left=22](0,0.4) -- (0,0.2); 
				\end{tikzpicture}}; 
				\draw[line width=0.5mm] (-0.5,-0.26) -- (-0.5,0.3) -- (-0.85,0.3) -- (-0.85,0.7); \node at (-1.08,0.7)[scale=0.7]{
					\begin{tikzpicture} 
						\fill[black!75, rounded corners=2] (0,0.2) -- (0,0) to [bend right=22](0.7,0) -- (0.7,0.4) to [bend left=22](0,0.4) -- (0,0.2); 
				\end{tikzpicture}}; 
				\node at (1.2,0.6)[scale=0.5]{
					\begin{tikzpicture} 
						\fill[black!10] (0,0) circle(1.15cm); 
						\draw[thick] (0,0) circle(1cm); 
						\draw[thick] (0,0) circle(0.9cm); 
						\draw[thick] (0,0) circle(0.8cm); 
						\fill[black!10] (-0.1,-1.1) rectangle (0.1,1.1); 
						\fill[black!10] (-1.1,-0.1) rectangle (1.1,0.1); 
						\draw[thick, rounded corners=2] (-0.1,-0.7) rectangle (0.1,0.7); 
						\draw[thick, rounded corners=2] (-0.5,0.3) rectangle (0.5,0.5); 
				\end{tikzpicture}}; 
		\end{tikzpicture}}; 
\end{tikzpicture}}

\def\RRoundAssociator#1#2#3#4{
	\begin{tikzpicture}
		\node at (0,0)[scale=#4]{	\begin{tikzpicture}
				\draw[line width=1mm, rounded corners=25, dashed, red] (0,3) rectangle (3.6,6.6);
				\node at (1.95,4.8)[]{#1};
				
				\draw[line width=1mm, rounded corners=25, dashed, red] (0,-1.8) rectangle (3.6,1.8);
				\node at (1.8,0)[]{#2};
				
				\draw[line width=1mm, rounded corners=25, dashed, red] (0,-3) rectangle (3.6,-6.6);
				\node at (1.8,-4.8)[]{#3};
				
				\draw[line width=1mm, red, dashed] (0,-4.8) -- (-5,0) -- (0,4.8) (-5,0) -- (0,0); 
				
				\node at (-0.25,4.5)[rotate=-45]{
					\begin{tikzpicture} 
						\fill[black!50] (0,0) circle(0.35cm and 0.4cm);
						\node at (0,0)[]{
							\begin{tikzpicture}[scale=0.7]
								\draw[line width=1mm, red] (0,0) circle(0.35cm and 0.4cm); 
						\end{tikzpicture}};
						\node at (0,0)[]{
							\begin{tikzpicture}[scale=0.4] 
								\fill[white] (0,0) circle(0.35cm and 0.4cm); 
						\end{tikzpicture}}; 
				\end{tikzpicture}};
				
				\node at (-0.25,0)[rotate=90]{
					\begin{tikzpicture}
						\fill[black!50] (0,0) circle(0.35cm and 0.4cm);
						\node at (0,0)[]{
							\begin{tikzpicture}[scale=0.7]
								\draw[line width=1mm, red] (0,0) circle(0.35cm and 0.4cm); 
						\end{tikzpicture}};
						\node at (0,0)[]{
							\begin{tikzpicture}[scale=0.4]
								\fill[white] (0,0) circle(0.35cm and 0.4cm); 
						\end{tikzpicture}}; 
				\end{tikzpicture}};
				
				
				\node at (-0.25,-4.5)[rotate=45]{
					\begin{tikzpicture} 
						\fill[black!50] (0,0) circle(0.35cm and 0.4cm);
						\node at (0,0)[]{
							\begin{tikzpicture}[scale=0.7]
								\draw[line width=1mm, red] (0,0) circle(0.35cm and 0.4cm); 
						\end{tikzpicture}};
						\node at (0,0)[]{
							\begin{tikzpicture}[scale=0.4] 
								\fill[white] (0,0) circle(0.35cm and 0.4cm); 
						\end{tikzpicture}}; 
				\end{tikzpicture}};
				
				\draw[line width=0.6mm, -Stealth] (3.6,4.8) -- (7,1.5);
				\draw[line width=0.6mm, -Stealth] (3.6,0) -- (7,0);
				\draw[line width=0.6mm, -Stealth] (3.6,-4.8) -- (7,-1.5);
		\end{tikzpicture}};
	\end{tikzpicture}
}

\def\Nuage#1#2{
	\begin{tikzpicture}
		\node at (0,0)[scale=#2]{
			\begin{tikzpicture}[scale=1.5]
				\draw[black!40,line width=0.8mm, fill=nuageColor] (-1,1)[rounded corners=21] -- (-1.5,1) -- (-1.5,0) -- (1.5,0) -- (1.5,1)[rounded corners=0] -- (1,1) .. controls(1,1.8) and (-0.25,2) .. (-0.5,1.4) to [bend right=70](-1,1);
				\draw[line width=0.8mm, black!40] (-0.5,1.4) to [bend left](-0.22,1.1); \draw[line width=0.8mm, black!40] (1,1) -- (0.7,1);
				\node at (0,0.15)[scale=1.1]{
					\begin{tikzpicture}
						\node at (0.75,1.1)[]{
							\begin{tikzpicture}
								\draw[line width=1.5mm, rounded corners=13, white] (0,0) -- (0,1.2) -- (0.9,1.2) -- (0.9,0);
						\end{tikzpicture}};
						\fill[blue!20] (0,0) rectangle (1.5,1.15);
						\node at (0.75,0.575)[]{
							\begin{tikzpicture}
								\fill[black!70, rounded corners=2] (0,0) rectangle (0.15,0.5); \fill[black!70] (0.075,0.4) circle(0.2cm);
						\end{tikzpicture}};
				\end{tikzpicture}};
				\node at (0.25,1.1)[scale=1.2]{\sf #1};
		\end{tikzpicture}};
\end{tikzpicture}}

\def\ContactIconeLoop#1{
	\begin{tikzpicture}
		\node at (0,0)[scale=#1]{
			\begin{tikzpicture}	
				\node at (0.75,-3.5)[scale=1.5]{\Huge \bf Researchers};	
				\node at (4.7,1.5)[scale=0.35]{
					\begin{tikzpicture} 
						\draw[line width=0.9mm] (0,0) circle(2cm); 
						\node at (0,0.35)[scale=1.1]{
							\begin{tikzpicture} 
								\node at (0,0)[scale=0.65, yscale=1.3, xscale=1]{
									\begin{tikzpicture}
										\fill[black!80, rounded corners=4] (0,0) -- (2,0) -- (2,1) -- (0.3,1.5) 	-- (0,0.8) -- (-0.3,1.5) -- (-2,1) -- (-2,0) -- (0,0); 
								\end{tikzpicture}}; 
								\node at (0,1.35)[scale=1.18]{
									\begin{tikzpicture} 
										\fill[black!80] (0.4,1) to [bend right=70](-0.4,1)[rounded corners=5] to 	[bend right=40](0,0) to [bend right=40](0.4,1); 
								\end{tikzpicture}}; 
						\end{tikzpicture}}; 
				\end{tikzpicture}}; 
				\node at (1.7,1.5)[scale=0.35]{
					\begin{tikzpicture} 
						\draw[line width=0.9mm] (0,0) circle(2cm); 
						\node at (0,0.35)[scale=1.1]{
							\begin{tikzpicture} 
								\node at (0,0)[scale=0.65, yscale=1.3, xscale=1]{
									\begin{tikzpicture}
										\fill[black!80, rounded corners=4] (0,0) -- (2,0) -- (2,1) -- 	(0.3,1.5) -- (0,0.8) -- (-0.3,1.5) -- (-2,1) -- (-2,0) -- (0,0);
								\end{tikzpicture}}; 
								\node at (0,1.35)[scale=1.18]{
									\begin{tikzpicture} 
										\fill[black!80] (0.4,1) to [bend right=70](-0.4,1)[rounded 	corners=5] to [bend right=40](0,0) to [bend right=40](0.4,1); 
								\end{tikzpicture}}; 
						\end{tikzpicture}}; 
				\end{tikzpicture}}; 
				\node at (3.7,-0.7)[scale=0.85]{
					\begin{tikzpicture}[scale=1.2] 
						\node at (-0.04,-0.04)[rotate=38]{
							\begin{tikzpicture} 
								\draw[line width=2.5mm] (0,0) circle(1.2cm); 
								\draw[line width=2.5mm] (0,-1.2) -- (0,-2); 
								\draw[line width=4.5mm] (0,-1.6) -- (0,-2); 
								\draw[line width=7.5mm] (0,-1.85) -- (0,-3); 
						\end{tikzpicture}}; 
						\draw[line width=0.8mm] (-2.03,0) -- (0,0); 
						\draw[line width=0.8mm] (-2.03,-0.15) -- (0,-0.15); 
						\draw[line width=3mm] (-0.1,0.1) -- (-0.1,1); 
						\draw[line width=3mm] (-0.55,0.1) -- (-0.55,1.4); 
						\draw[line width=3mm] (-1,0.1) -- (-1,0.65); 
						\draw[line width=3mm, white] (-1.45,0.1) -- (-1.45,0.9); 
						\draw[line width=1mm] (-1.35,0.1) -- (-1.35,0.9); \draw[line width=3mm] 	(-1.9,0.1) -- (-1.9,0.55);
				\end{tikzpicture}}; 
				\node at (1.7,-1.8)[scale=0.35]{
					\begin{tikzpicture} 
						\draw[line width=0.9mm] (0,0) circle(2cm); 
						\node at (0,0.35)[scale=1.1]{
							\begin{tikzpicture} 
								\node at (0,0)[scale=0.65, yscale=1.3, xscale=1]{
									\begin{tikzpicture}
										\fill[black!80, rounded corners=4] (0,0) -- (2,0) -- 	(2,1) -- (0.3,1.5) -- (0,0.8) -- (-0.3,1.5) -- (-2,1) -- (-2,0) -- (0,0);
								\end{tikzpicture}}; 
								\node at (0,1.35)[scale=1.18]{
									\begin{tikzpicture} 
										\fill[black!80] (0.4,1) to [bend 	right=70](-0.4,1)[rounded corners=5] to [bend right=40](0,0) to [bend right=40](0.4,1); 
								\end{tikzpicture}}; 
						\end{tikzpicture}}; 
				\end{tikzpicture}};	
				\node at (-0.8,1.5)[scale=0.95]{
					\begin{tikzpicture}
						\draw[line width=1.2mm, black!80] (0.12,0.2) -- (0.64,0.2) 	(0.76,0.2) -- (0.88,0.2);
						\draw[line width=1.2mm, black!80] (0.12,0.45) -- (0.88,0.45);
						\draw[line width=1.2mm, black!80] (0.12,0.7) -- (0.88,0.7); 
						\draw[line width=1.2mm, black!80] (0.12,0.95) -- (0.88,0.95); 
						\draw[line width=1.2mm, black!80] (-0.7,0.15) -- (0.16,0.96);
				\end{tikzpicture}};
				\node at (-3,1.5)[scale=0.7]{
					\begin{tikzpicture} 
						\draw[line width=1.2mm, black!80] (0,0) -- (2.25,0); 
						\draw[line width=3mm, black!80] (0.4,0) -- (0.4,1.2); 
						\draw[line width=3mm, black!80] (1.1,0) -- (1.1,0.8); 
						\draw[line width=3mm, black!80] (1.8,0) -- (1.8,0.5);	 
				\end{tikzpicture}};	
				\node at (-0.4,-0.3)[scale=0.8]{
					\begin{tikzpicture} 
						\node at (0,0)[]{
							\begin{tikzpicture} 
								\draw[line width=1.5mm] (0,0) -- (0,1); 
						\end{tikzpicture}}; 
						\node at (0,0)[rotate=45]{
							\begin{tikzpicture} 
								\draw[line width=1.5mm] (0,0) -- (0,1); 
						\end{tikzpicture}}; 
						\node at (0,0)[rotate=90]{
							\begin{tikzpicture} 
								\draw[line width=1.5mm] (0,0) -- (0,1); 
						\end{tikzpicture}}; 
						\node at (0,0)[rotate=135]{
							\begin{tikzpicture} 
								\draw[line width=1.5mm] (0,0) -- (0,1); 
						\end{tikzpicture}}; 
						\fill[white] (0,0) circle(0.27cm); 
						\draw[line width=1.5mm] (0,-0.7) -- (0,-1.3) (0,-1.45) -- 	(0,-1.65); 
				\end{tikzpicture}};	
				\node at (-1.8,-1)[scale=0.7]{
					\begin{tikzpicture}
						\fill[black!80] (0,0)[rounded corners=10] -- (1.3,0) -- 	(1.3,1.65)[rounded corners=0] -- (0.5,1.65) -- (0,0.7) -- (-0.5,1.65)[rounded corners=10] -- (-1.3,1.65) -- (-1.3,0) -- (0,0);
						\fill[black!80, rounded corners=3] (0.12,0.5) -- (0.05,1.25) -- 	(0.2,1.65) -- (-0.2,1.65) -- (-0.05,1.25) -- (-0.12,0.5); 
						\fill[black!80] (0,2.7) circle(0.88cm and 0.85cm); 
				\end{tikzpicture}};	
				\node at (-3.35,-0.6)[scale=0.9]{
					\begin{tikzpicture} 
						\fill[black!80] (0,0) rectangle (1,1.2); 
						\draw[line width=1.2mm, white] (0.12,0.2) -- (0.88,0.2); 
						\draw[line width=1.2mm, white] (0.12,0.45) -- (0.88,0.45); 
						\draw[line width=1.2mm, white] (0.12,0.7) -- (0.88,0.7); 
						\draw[line width=1.2mm, white] (0.12,0.95) -- (0.88,0.95); 
				\end{tikzpicture}}; 
		\end{tikzpicture}};
\end{tikzpicture}}

\def\IconsOne#1{
	\begin{tikzpicture}
		\node at (0,0)[scale=#1]{
			\begin{tikzpicture} 
				\draw[line width=0.9mm] (0,0) circle(2cm); 
				\node at (0,0.35)[scale=1.1]{
					\begin{tikzpicture} 
						\node at (0,0)[scale=0.65, yscale=1.3, xscale=1]{
							\begin{tikzpicture}
								\fill[black!80, rounded corners=4] (0,0) -- (2,0) -- (2,1) -- (0.3,1.5) 	-- (0,0.8) -- (-0.3,1.5) -- (-2,1) -- (-2,0) -- (0,0); 
						\end{tikzpicture}}; 
						\node at (0,1.35)[scale=1.18]{
							\begin{tikzpicture} 
								\fill[black!80] (0.4,1) to [bend right=70](-0.4,1)[rounded corners=5] to 	[bend right=40](0,0) to [bend right=40](0.4,1); 
						\end{tikzpicture}}; 
				\end{tikzpicture}}; 
		\end{tikzpicture}};
	\end{tikzpicture}
}

\def\IconsTwo#1{
	\begin{tikzpicture}
		\node at (0,0)[scale=#1]{\begin{tikzpicture}[scale=1.2] 
				\node at (-0.04,-0.04)[rotate=38]{
					\begin{tikzpicture} 
						\draw[line width=2.5mm] (0,0) circle(1.2cm); 
						\draw[line width=2.5mm] (0,-1.2) -- (0,-2); 
						\draw[line width=4.5mm] (0,-1.6) -- (0,-2); 
						\draw[line width=7.5mm] (0,-1.85) -- (0,-3); 
				\end{tikzpicture}}; 
				\draw[line width=0.8mm] (-2.03,0) -- (0,0); 
				\draw[line width=0.8mm] (-2.03,-0.15) -- (0,-0.15); 
				\draw[line width=3mm] (-0.1,0.1) -- (-0.1,1); 
				\draw[line width=3mm] (-0.55,0.1) -- (-0.55,1.4); 
				\draw[line width=3mm] (-1,0.1) -- (-1,0.65); 
				\draw[line width=3mm, white] (-1.45,0.1) -- (-1.45,0.9); 
				\draw[line width=1mm] (-1.35,0.1) -- (-1.35,0.9); \draw[line width=3mm] 	(-1.9,0.1) -- (-1.9,0.55);
		\end{tikzpicture}};
	\end{tikzpicture}
}

\def\IconsThree#1{
	\begin{tikzpicture}
		\node at (0,0)[scale=#1]{\begin{tikzpicture}
				\draw[line width=1.2mm, black!80] (0.12,0.2) -- (0.64,0.2) 	(0.76,0.2) -- (0.88,0.2);
				\draw[line width=1.2mm, black!80] (0.12,0.45) -- (0.88,0.45);
				\draw[line width=1.2mm, black!80] (0.12,0.7) -- (0.88,0.7); 
				\draw[line width=1.2mm, black!80] (0.12,0.95) -- (0.88,0.95); 
				\draw[line width=1.2mm, black!80] (-0.7,0.15) -- (0.16,0.96);
		\end{tikzpicture}};
	\end{tikzpicture}
}

\def\IconsFour#1{
	\begin{tikzpicture}
		\node at (0,0)[scale=#1]{\begin{tikzpicture} 
				\draw[line width=1.2mm, black!80] (0,0) -- (2.25,0); 
				\draw[line width=3mm, black!80] (0.4,0) -- (0.4,1.2); 
				\draw[line width=3mm, black!80] (1.1,0) -- (1.1,0.8); 
				\draw[line width=3mm, black!80] (1.8,0) -- (1.8,0.5);	 
		\end{tikzpicture}};
	\end{tikzpicture}
}

\def\IconsFive#1{
	\begin{tikzpicture}
		\node at (0,0)[scale=#1]{\begin{tikzpicture} 
				\node at (0,0)[]{
					\begin{tikzpicture} 
						\draw[line width=1.5mm] (0,0) -- (0,1); 
				\end{tikzpicture}}; 
				\node at (0,0)[rotate=45]{
					\begin{tikzpicture} 
						\draw[line width=1.5mm] (0,0) -- (0,1); 
				\end{tikzpicture}}; 
				\node at (0,0)[rotate=90]{
					\begin{tikzpicture} 
						\draw[line width=1.5mm] (0,0) -- (0,1); 
				\end{tikzpicture}}; 
				\node at (0,0)[rotate=135]{
					\begin{tikzpicture} 
						\draw[line width=1.5mm] (0,0) -- (0,1); 
				\end{tikzpicture}}; 
				\fill[white] (0,0) circle(0.27cm); 
				\draw[line width=1.5mm] (0,-0.7) -- (0,-1.3) (0,-1.45) -- 	(0,-1.65); 
		\end{tikzpicture}};
	\end{tikzpicture}
}

\def\IconsSix#1{
	\begin{tikzpicture}
		\node at (0,0)[scale=#1]{\begin{tikzpicture}
				\fill[black!80] (0,0)[rounded corners=10] -- (1.3,0) -- 	(1.3,1.65)[rounded corners=0] -- (0.5,1.65) -- (0,0.7) -- (-0.5,1.65)[rounded corners=10] -- (-1.3,1.65) -- (-1.3,0) -- (0,0);
				\fill[black!80, rounded corners=3] (0.12,0.5) -- (0.05,1.25) -- 	(0.2,1.65) -- (-0.2,1.65) -- (-0.05,1.25) -- (-0.12,0.5); 
				\fill[black!80] (0,2.7) circle(0.88cm and 0.85cm); 
		\end{tikzpicture}};
	\end{tikzpicture}
}

\def\IconsSeven#1{
	\begin{tikzpicture}
		\node at (0,0)[scale=#1]{\begin{tikzpicture} 
				\fill[black!80] (0,0) rectangle (1,1.2); 
				\draw[line width=1.2mm, white] (0.12,0.2) -- (0.88,0.2); 
				\draw[line width=1.2mm, white] (0.12,0.45) -- (0.88,0.45); 
				\draw[line width=1.2mm, white] (0.12,0.7) -- (0.88,0.7); 
				\draw[line width=1.2mm, white] (0.12,0.95) -- (0.88,0.95); 
		\end{tikzpicture}};
	\end{tikzpicture}
}

\def\IconForDROne#1{
	\begin{tikzpicture}
		\node at (0,0)[scale=#1]{
			\begin{tikzpicture} 
				\node at (0.75,-3.5)[]{\Huge \bf }; 
				\node at (-2,-2.1)[scale=0.7]{
					\begin{tikzpicture} 
						\draw[line width=0.5mm, fill=black] (0,0) to [bend right=15](1.5,0.5) to [bend right](0.5,2.25) -- (0,1.4) -- (-0.5,2.25) to [bend right](-1.5,0.5) to [bend right=15](0,0); 
						\node at (0.4,0.75)[]{
							\begin{tikzpicture}
								\draw[line width=1mm, white] (0,0.2) -- (0,-0.2); 
								\draw[line width=1mm, white] (-0.2,0) -- (0.2,0); 
						\end{tikzpicture}}; 
						\node at (-0.85,1.4)[scale=0.7]{
							\begin{tikzpicture} 
								\draw[line width=0.8mm, white, rounded corners=4] (-0.4,0)node[color=white]{\Large $\bullet$} -- (-0.6,0.15) to [bend left](0,1.25) to [bend left](0.6,0.15) -- (0.4,0)node[color=white]{\Large $\bullet$}; 
								\draw[line width=0.8mm, white] (0,1.2) to [bend left=15](0.3,2.25); 
						\end{tikzpicture}}; 
						\node at (0.8,1.65)[]{
							\begin{tikzpicture} 
								\draw[line width=1mm, white] (0,0) to [bend right=10](-0.2,1); 
								\draw[line width=1mm, white, fill=black] (0,0) circle(8pt); 
						\end{tikzpicture}}; 
						\draw[thick, fill=black] (0,3) circle(0.7cm); 
				\end{tikzpicture}}; 
				\node at (-2,-3.75)[scale=1.3]{\Huge \bf Doctor};

				\node at (-2,5.3)[]{\Familly{0.525}};
				\node at (2.8,5.3)[]{\Friends{0.55}};

				\node at (-2,1.65)[scale=0.6]{
					\begin{tikzpicture} 
						\draw[line width=0.2mm, fill=black] (0,0) -- (1.5,0) arc(0:180:1.5 and 1.4) -- (0,0); 
						\draw[thick, fill=black] (0,2.25) circle(0.7cm); 
						\draw[thick, fill=black] (0,3.13) to [bend left=15](0.6,2.9) -- (1,3.5) to [bend right=50](-1,3.4) -- (-0.6,2.9) to [bend left=15](0,3.13); 
						\node at (0,3.5)[]{
							\begin{tikzpicture} 
								\draw[line width=1.5mm, white] (-0.25,0) -- (0.25,0); 
								\draw[line width=1.5mm, white] (0,0.25) -- (0,-0.25); 
						\end{tikzpicture}};
				\end{tikzpicture}}; 
				\node at (-2,-0.15)[scale=1.3]{\Huge \bf Nurse};
				\node at (2.5,-2.1)[scale=0.7]{
					\begin{tikzpicture} 
						\draw[thick, fill=black] (0,0) -- (2,0) to [bend right=10](1.8,1.25) to [bend right=5](0.85,1.8) -- (0.6,1.65) -- (0,0.5) (0,0) -- (-2,0) to [bend left=10](-1.8,1.25) to [bend left=5](-0.85,1.8) -- (-0.6,1.65) -- (0,0.5); 
						\draw[thick, fill=black] (0,1.8) -- (-0.2,1.5) -- (-0.08,1.25) -- (-0.2,0.4) -- (0.2,0.4) -- (0.08,1.25) -- (0.2,1.5) -- (0,1.8); 
						\draw[fill=black] (0,2.9) circle(0.8cm and 0.85cm); 
						\node at (-1.5,1.2)[scale=0.8]{
							\begin{tikzpicture} 
								\node at (0.1,-0.1)[scale=1.5]{
									\begin{tikzpicture} 
										\draw[line width=1mm, white, fill=white] (0,0)[rounded corners=10] -- (1,0)[rounded corners=0] -- (0.2,1.6) -- (0.2,2.2)[rounded corners=2] -- (0.35,2.2) -- (0.35,2.4) -- (-0.35,2.4) -- (-0.35,2.2)[rounded corners=0] -- (-0.2,2.2) -- (-0.2,1.6)[rounded corners=10] -- (-1,0) -- (0,0); 
								\end{tikzpicture}};	
								\node at (0,0)[scale=1.2]{
									\begin{tikzpicture} 
										\draw[line width=1mm, fill=black] (0,0)[rounded corners=10] -- (1,0)[rounded corners=0] -- (0.2,1.6) -- (0.2,2.2)[rounded corners=2] -- (0.35,2.2) -- (0.35,2.4) -- (-0.35,2.4) -- (-0.35,2.2)[rounded corners=0] -- (-0.2,2.2) -- (-0.2,1.6)[rounded corners=10] -- (-1,0) -- (0,0); 
								\end{tikzpicture}}; 
								\node at (0,-0.75)[scale=1]{
									\begin{tikzpicture} 
										\draw[line width=1mm, fill=white] (0,0)[rounded corners=5] -- (1,0)[rounded corners=0] -- (0.4,1.1) -- (-0.4,1.1)[rounded corners=5] -- (-1,0) -- (0,0); 
								\end{tikzpicture}}; 
								\node at (-0.15,1.8)[scale=2]{$\bullet$}; 
								\node at (-0.3,2.2)[scale=2]{$\bullet$}; 
								\node at (-0.8,1.8)[scale=3]{$\bullet$}; 
						\end{tikzpicture}};  
				\end{tikzpicture}}; 
				\node at (2.9,-3.75)[scale=1.3]{\Huge \bf Lab Worker}; 
				\node at (2.5,1.6)[scale=0.7]{
					\begin{tikzpicture} 
						\draw[line width=0.2mm, fill=black] (0,0) -- (1.5,0) arc(0:180:1.5 and 1.4) -- (0,0); 
						\draw[thick, fill=black] (0,2.25) circle(0.7cm); 
						\node at (-0.8,0.5)[]{
							\begin{tikzpicture} 
								\draw[line width=2mm, white] (-0.35,0) -- (0.35,0); 
								\draw[line width=2mm, white] (0,0.35) -- (0,-0.35); 
						\end{tikzpicture}}; 
						\node at (0.8,0.6)[scale=0.7]{
							\begin{tikzpicture} 
								\fill[white] (0,0) circle(1.3cm); 
								\node at (0,0.2)[rotate=50, yscale=1.4, xscale=1.1]{
									\begin{tikzpicture} 
										\fill[black] (0.1,0)[rounded corners=7] -- (1,0) -- (1,0.5)[rounded corners=0] -- (0.1,0.5) -- (0.1,0); 
										\fill[black] (-0.1,0)[rounded corners=7] -- (-1,0) -- (-1,0.5)[rounded corners=0] -- (-0.1,0.5) -- (-0.1,0); 
								\end{tikzpicture}}; 
						\end{tikzpicture}}; 
				\end{tikzpicture}}; 
				\node at (2.9,0.1)[scale=1.3]{\Huge \bf Pharmacist};
		\end{tikzpicture}};
	\end{tikzpicture}
}

\def\IconForDRTwo#1{
	\begin{tikzpicture}
		\node at (0,0)[scale=#1]{
			\begin{tikzpicture}
				\node at (0,7)[]{\ContactIconeLoop{0.95}};
				\node at (0,1.5)[]{\Icon{0.9}};
			\end{tikzpicture}
		};
	\end{tikzpicture}
}

\def\Icon#1{
	\begin{tikzpicture}
		\node at (0,0)[scale=#1]{
			\begin{tikzpicture} 
				\node at (-2,{-2.1+1})[scale=0.7]{
					\begin{tikzpicture} 
						\draw[line width=0.5mm, fill=black] (0,0) to [bend right=15](1.5,0.5) to [bend right](0.5,2.25) -- (0,1.4) -- (-0.5,2.25) to [bend right](-1.5,0.5) to [bend right=15](0,0); 
						\node at (0.4,0.75)[]{
							\begin{tikzpicture}
								\draw[line width=1mm, white] (0,0.2) -- (0,-0.2); 
								\draw[line width=1mm, white] (-0.2,0) -- (0.2,0); 
						\end{tikzpicture}}; 
						\node at (-0.85,1.4)[scale=0.7]{
							\begin{tikzpicture} 
								\draw[line width=0.8mm, white, rounded corners=4] (-0.4,0)node[color=white]{\Large $\bullet$} -- (-0.6,0.15) to [bend left](0,1.25) to [bend left](0.6,0.15) -- (0.4,0)node[color=white]{\Large $\bullet$}; 
								\draw[line width=0.8mm, white] (0,1.2) to [bend left=15](0.3,2.25); 
						\end{tikzpicture}}; 
						\node at (0.8,1.65)[]{
							\begin{tikzpicture} 
								\draw[line width=1mm, white] (0,0) to [bend right=10](-0.2,1); 
								\draw[line width=1mm, white, fill=black] (0,0) circle(8pt); 
						\end{tikzpicture}}; 
						\draw[thick, fill=black] (0,3) circle(0.7cm); 
				\end{tikzpicture}}; 
				\node at (-2,{-3.75+1})[scale=1.3]{\Huge \bf Doctor};
				
				\node at (2.5,{-2.1+1})[scale=0.7]{
					\begin{tikzpicture} 
						\draw[thick, fill=black] (0,0) -- (2,0) to [bend right=10](1.8,1.25) to [bend right=5](0.85,1.8) -- (0.6,1.65) -- (0,0.5) (0,0) -- (-2,0) to [bend left=10](-1.8,1.25) to [bend left=5](-0.85,1.8) -- (-0.6,1.65) -- (0,0.5); 
						\draw[thick, fill=black] (0,1.8) -- (-0.2,1.5) -- (-0.08,1.25) -- (-0.2,0.4) -- (0.2,0.4) -- (0.08,1.25) -- (0.2,1.5) -- (0,1.8); 
						\draw[fill=black] (0,2.9) circle(0.8cm and 0.85cm); 
						\node at (-1.5,1.2)[scale=0.8]{
							\begin{tikzpicture} 
								\node at (0.1,-0.1)[scale=1.5]{
									\begin{tikzpicture} 
										\draw[line width=1mm, white, fill=white] (0,0)[rounded corners=10] -- (1,0)[rounded corners=0] -- (0.2,1.6) -- (0.2,2.2)[rounded corners=2] -- (0.35,2.2) -- (0.35,2.4) -- (-0.35,2.4) -- (-0.35,2.2)[rounded corners=0] -- (-0.2,2.2) -- (-0.2,1.6)[rounded corners=10] -- (-1,0) -- (0,0); 
								\end{tikzpicture}};	
								\node at (0,0)[scale=1.2]{
									\begin{tikzpicture} 
										\draw[line width=1mm, fill=black] (0,0)[rounded corners=10] -- (1,0)[rounded corners=0] -- (0.2,1.6) -- (0.2,2.2)[rounded corners=2] -- (0.35,2.2) -- (0.35,2.4) -- (-0.35,2.4) -- (-0.35,2.2)[rounded corners=0] -- (-0.2,2.2) -- (-0.2,1.6)[rounded corners=10] -- (-1,0) -- (0,0); 
								\end{tikzpicture}}; 
								\node at (0,-0.75)[scale=1]{
									\begin{tikzpicture} 
										\draw[line width=1mm, fill=white] (0,0)[rounded corners=5] -- (1,0)[rounded corners=0] -- (0.4,1.1) -- (-0.4,1.1)[rounded corners=5] -- (-1,0) -- (0,0); 
								\end{tikzpicture}}; 
								\node at (-0.15,1.8)[scale=2]{$\bullet$}; 
								\node at (-0.3,2.2)[scale=2]{$\bullet$}; 
								\node at (-0.8,1.8)[scale=3]{$\bullet$}; 
						\end{tikzpicture}};  
				\end{tikzpicture}}; 
				\node at (2.9,{-3.75+1})[scale=1.3]{\Huge \bf Lab Worker}; 
		\end{tikzpicture}};
\end{tikzpicture}}

\def\Pharmacist#1{
	\begin{tikzpicture}
		\node at (0,0)[scale=#1]{\begin{tikzpicture} 
				\draw[line width=0.2mm, fill=black] (0,0) -- (1.5,0) arc(0:180:1.5 and 1.4) -- (0,0); 
				\draw[thick, fill=black] (0,2.25) circle(0.7cm); 
				\node at (-0.8,0.5)[]{
					\begin{tikzpicture} 
						\draw[line width=2mm, white] (-0.35,0) -- (0.35,0); 
						\draw[line width=2mm, white] (0,0.35) -- (0,-0.35); 
				\end{tikzpicture}}; 
				\node at (0.8,0.6)[scale=0.7]{
					\begin{tikzpicture} 
						\fill[white] (0,0) circle(1.3cm); 
						\node at (0,0.2)[rotate=50, yscale=1.4, xscale=1.1]{
							\begin{tikzpicture} 
								\fill[black] (0.1,0)[rounded corners=7] -- (1,0) -- (1,0.5)[rounded corners=0] -- (0.1,0.5) -- (0.1,0); 
								\fill[black] (-0.1,0)[rounded corners=7] -- (-1,0) -- (-1,0.5)[rounded corners=0] -- (-0.1,0.5) -- (-0.1,0); 
						\end{tikzpicture}}; 
				\end{tikzpicture}}; 
		\end{tikzpicture}};
	\end{tikzpicture}
}

\def\Doctor#1{
	\begin{tikzpicture}
		\node at (0,0)[scale=#1]{\begin{tikzpicture} 
				\draw[line width=0.5mm, fill=black] (0,0) to [bend right=15](1.5,0.5) to [bend right](0.5,2.25) -- (0,1.4) -- (-0.5,2.25) to [bend right](-1.5,0.5) to [bend right=15](0,0); 
				\node at (0.4,0.75)[]{
					\begin{tikzpicture}
						\draw[line width=1mm, white] (0,0.2) -- (0,-0.2); 
						\draw[line width=1mm, white] (-0.2,0) -- (0.2,0); 
				\end{tikzpicture}}; 
				\node at (-0.85,1.4)[scale=0.7]{
					\begin{tikzpicture} 
						\draw[line width=0.8mm, white, rounded corners=4] (-0.4,0)node[color=white]{\Large $\bullet$} -- (-0.6,0.15) to [bend left](0,1.25) to [bend left](0.6,0.15) -- (0.4,0)node[color=white]{\Large $\bullet$}; 
						\draw[line width=0.8mm, white] (0,1.2) to [bend left=15](0.3,2.25); 
				\end{tikzpicture}}; 
				\node at (0.8,1.65)[]{
					\begin{tikzpicture} 
						\draw[line width=1mm, white] (0,0) to [bend right=10](-0.2,1); 
						\draw[line width=1mm, white, fill=black] (0,0) circle(8pt); 
				\end{tikzpicture}}; 
				\draw[thick, fill=black] (0,3) circle(0.7cm); 
		\end{tikzpicture}};
	\end{tikzpicture}
}

\def\Nurse#1{
	\begin{tikzpicture}
		\node at (0,0)[scale=#1]{\begin{tikzpicture} 
				\draw[line width=0.2mm, fill=black] (0,0) -- (1.5,0) arc(0:180:1.5 and 1.4) -- (0,0); 
				\draw[thick, fill=black] (0,2.25) circle(0.7cm); 
				\draw[thick, fill=black] (0,3.13) to [bend left=15](0.6,2.9) -- (1,3.5) to [bend right=50](-1,3.4) -- (-0.6,2.9) to [bend left=15](0,3.13); 
				\node at (0,3.5)[]{
					\begin{tikzpicture} 
						\draw[line width=1.5mm, white] (-0.25,0) -- (0.25,0); 
						\draw[line width=1.5mm, white] (0,0.25) -- (0,-0.25); 
				\end{tikzpicture}};
		\end{tikzpicture}};
	\end{tikzpicture}
}

\def\LabWorker#1{
	\begin{tikzpicture}
		\node at (0,0)[]{\begin{tikzpicture} 
				\draw[thick, fill=black] (0,0) -- (2,0) to [bend right=10](1.8,1.25) to [bend right=5](0.85,1.8) -- (0.6,1.65) -- (0,0.5) (0,0) -- (-2,0) to [bend left=10](-1.8,1.25) to [bend left=5](-0.85,1.8) -- (-0.6,1.65) -- (0,0.5); 
				\draw[thick, fill=red] (0,1.8) -- (-0.2,1.5) -- (-0.08,1.25) -- (-0.2,0.4) -- (0.2,0.4) -- (0.08,1.25) -- (0.2,1.5) -- (0,1.8); 
				\draw[fill=black] (0,2.9) circle(0.8cm and 0.85cm); 
				\node at (-1.5,1.2)[scale=0.8]{
					\begin{tikzpicture} 
						\node at (0.1,-0.1)[scale=1.5]{
							\begin{tikzpicture} 
								\draw[line width=1mm, white, fill=white] (0,0)[rounded corners=10] -- (1,0)[rounded corners=0] -- (0.2,1.6) -- (0.2,2.2)[rounded corners=2] -- (0.35,2.2) -- (0.35,2.4) -- (-0.35,2.4) -- (-0.35,2.2)[rounded corners=0] -- (-0.2,2.2) -- (-0.2,1.6)[rounded corners=10] -- (-1,0) -- (0,0); 
						\end{tikzpicture}};	
						\node at (0,0)[scale=1.2]{
							\begin{tikzpicture} 
								\draw[line width=1mm, fill=black] (0,0)[rounded corners=10] -- (1,0)[rounded corners=0] -- (0.2,1.6) -- (0.2,2.2)[rounded corners=2] -- (0.35,2.2) -- (0.35,2.4) -- (-0.35,2.4) -- (-0.35,2.2)[rounded corners=0] -- (-0.2,2.2) -- (-0.2,1.6)[rounded corners=10] -- (-1,0) -- (0,0); 
						\end{tikzpicture}}; 
						\node at (0,-0.75)[scale=1]{
							\begin{tikzpicture} 
								\draw[line width=1mm, fill=white] (0,0)[rounded corners=5] -- (1,0)[rounded corners=0] -- (0.4,1.1) -- (-0.4,1.1)[rounded corners=5] -- (-1,0) -- (0,0); 
						\end{tikzpicture}}; 
						\node at (-0.15,1.8)[scale=2]{$\bullet$}; 
						\node at (-0.3,2.2)[scale=2]{$\bullet$}; 
						\node at (-0.8,1.8)[scale=3]{$\bullet$}; 
				\end{tikzpicture}};  
		\end{tikzpicture}};
	\end{tikzpicture}
}

\def\Familly#1{
	\begin{tikzpicture}
		\node at (0,0)[scale=#1]{
			\begin{tikzpicture}
				\node at (-1.6,0)[scale=0.5]{\ManAvatarx{1}};
				\node at (0,0)[scale=0.5]{\ManAvatarx{1}};
				\node at (1.6,0)[scale=0.5]{\ManAvatarx{1}};
				\node at (0,-4)[scale=2.4]{\Huge \textbf{Familly}};
			\end{tikzpicture}
		};
	\end{tikzpicture}
}

\def\Friends#1{
	\begin{tikzpicture}
		\node at (0,0)[scale=#1]{
			\begin{tikzpicture}
				\node at (0,0)[scale=0.5]{\ManAvatarx{1}};
				\node at (-2.2,0)[]{\ManAvatarHandUp{0.5}};
				\node at (-0.8,-3.7)[scale=2.4]{\Huge \textbf{Friends}};
			\end{tikzpicture}
		};
	\end{tikzpicture}
}


\def\DoFunction#1{
	\begin{tikzpicture}
		\node at (0,0)[scale=#1]{\begin{tikzpicture}
				\node at (6,0)[]{\RRoundAssociator{\Tishort{0.7}}{\VitesseBulletx{1.1}}{\MagnificClockx{0.8}}{1}};
				\node at (0,0)[scale=0.9]{\ManAvatarx{0.8}};
				\node at (13.1,0)[scale=0.9]{\phonex{0.4}};
		\end{tikzpicture}};
	\end{tikzpicture}
}

\def\SPFunction#1#2{
	\begin{tikzpicture}
		\node at (0,0)[scale=#1]{
			\begin{tikzpicture}
				\node at (0,3.4)[]{\Nuage{\Huge \textbf{#2}}{1}};
			\end{tikzpicture}
		};
	\end{tikzpicture}
}

\def\CPFunction#1#2{
	\begin{tikzpicture}
		\node at (0,0)[scale=#1]{
			\begin{tikzpicture}
				\node at (0,-3.2)[]{\Nuage{\Huge \textbf{#2}}{1}};
			\end{tikzpicture}
		};
	\end{tikzpicture}
}

\def\SPCPFunction#1#2#3{
	\begin{tikzpicture}
		\node at (0,0)[scale=#1]{
			\begin{tikzpicture}
				\node at (0,3.4)[]{\Nuage{\Huge \textbf{#2}}{1}};
				\draw[line width=2mm, Triangle-Triangle] (0,1.5) -- (0,-1.5);
				\node at (0,-3.2)[]{\Nuage{\Huge \textbf{#3}}{1}};
			\end{tikzpicture}
		};
	\end{tikzpicture}
}

\def\SPCPServeurFunction#1#2#3{
	\begin{tikzpicture}
		\node at (0,0)[scale=#1]{
			\begin{tikzpicture}
				\node at (0,3.3)[]{\Nuage{\Huge \textbf{#2}}{0.8}};
				\node at (0.4,-3.4)[]{};
				\node at (2,-3.1)[]{};
				\node at (2,-5)[]{};
			\end{tikzpicture}
		};
	\end{tikzpicture}
}

\def\SRFunction#1#2{
	\begin{tikzpicture}
		\draw[line width=1mm, Latex-Latex, #2] (-#1,0) -- (#1,0);
	\end{tikzpicture}
}

\def\SDFunction#1#2{
	
	\begin{tikzpicture}
		\draw[line width=0.7mm, -Latex, #2] (-#1,0) -- (#1,0);
	\end{tikzpicture}
}

\def\RedLeftRightArrow#1{
	\begin{tikzpicture}[xscale=#1]
		\draw[line width=1.2mm, Triangle-Triangle, red] (0,0) -- (1,0);
	\end{tikzpicture}
}

\def\BlackLeftRightArrow#1{
	\begin{tikzpicture}[xscale=#1]
		\draw[line width=1.2mm, Triangle-Triangle, black] (0,0) -- (1,0);
	\end{tikzpicture}
}

\def\BlackRightArrow#1{
	\begin{tikzpicture}[xscale=#1]
		\draw[line width=1mm, -Triangle] (0,0) -- (1,0);
	\end{tikzpicture}
}

\def\Titledefx#1#2{
	\begin{tikzpicture}[xscale=#2, yscale=2] \fill[titleColor, rounded corners=5] (0,0) rectangle (3.5,0.5); \node at (1.75,0.25)[color=black]{\sf \textbf{\Huge #1}}; \end{tikzpicture}}

\def\FirstSeparateFigure#1{
	\begin{tikzpicture}
		\node at (0,0)[scale=#1]{\begin{tikzpicture}
				\node at (-2,2)[]{\Titledefx{\textbf{DO}}{0.85}};
				\node at (11.75,4)[]{\Titledefx{\textbf{DRs}}{0.85}};
				\node at (0,0)[]{\DoFunction{0.3}};
				\node at (6.5,-2.05)[]{\SPCPServeurFunction{0.65}{SP}{CP}};
				\node at (12.5,-0.5)[xscale=0.95]{\IconForDROne{0.65}};
				\node at (3.7,0)[]{\BlackRightArrow{2.5}};
				\node at (8.7,0)[]{\RedLeftRightArrow{2.1}};
				\node at (7.5,-7.7)[]{};
		\end{tikzpicture}};
	\end{tikzpicture}
}

\def\SecondSeparateFigure#1{
	\begin{tikzpicture}
		\node at (0,0)[scale=#1]{\begin{tikzpicture}
				\node at (7.5,-5.3)[]{};
				\node at (4,7.5)[]{\FourthSeparateFigure{0.7}};
				\node at (-2,3.7)[]{\Titledefx{\textbf{DO}}{0.85}};
				\node at (0,1.5)[]{\DoFunction{0.3}};
				\node at (7.8,-0.5)
				[]{\SPCPServeurFunction{0.65}{SP}{CP}};
				\node at (4.5,2)[]{\BlackRightArrow{3.5}};
				\node at (4.5,1.2)[]{\RedLeftRightArrow{3.5}};
		\end{tikzpicture}};
	\end{tikzpicture}
}

\def\ThirdSeparateFigure#1{
	\begin{tikzpicture}
		\node at (0,0)[scale=#1]{\begin{tikzpicture}
				\node at (-1,4.35)[]{\Titledefx{\textbf{DOs}}{0.85}};
				\node at (12.75,2.8)[]{\Titledefx{\textbf{DRs}}{0.85}};
				
				\node at (-1,1.5)[]{\DoFunction{0.3}};
				\node at (-1,-3)[]{\DoFunction{0.3}};
				\node at (-1,-7.5)[]{\DoFunction{0.3}};
				
				\node at (6.6,-3.5)[]{\SPCPServeurFunction{0.65}{SP}{CP}};
				
				\node at (13.5,-1.5)[xscale=0.95]{\IconForDRTwo{0.65}};
				
				\node at (3.3,0)[rotate=-38]{\BlackRightArrow{4}};
				\node at (3.3,-2.25)[rotate=23]{\BlackRightArrow{3.5}};
				\node at (3.6,-4.7)[rotate=55]{\BlackRightArrow{6.5}};
				\node at (9.25,-1.4)[]{\RedLeftRightArrow{2.8}};
		\end{tikzpicture}};
	\end{tikzpicture}
}

\def\FourthSeparateFigure#1{
	\begin{tikzpicture}
		\node at (0,0)[scale=#1]{\begin{tikzpicture}
				\node at (-60,0)[right]{\Huge \sf\textbf{Send request /access  result}};
				\node at (-60,-1.5)[right]{\Huge \sf\textbf{Upload raw data}};
				\node at (-44,0)[]{\RedLeftRightArrow{4}};
				\node at (-44,-1.5)[]{\BlackRightArrow{4}};
		\end{tikzpicture}};
	\end{tikzpicture}
}

\def\RectangleBlanc#1{
	\begin{tikzpicture}
		\node at (0,0)[scale=#1]{
			\begin{tikzpicture}
				\draw[thick, color=white, fill=white] (-3,-7) rectangle (15,5);
			\end{tikzpicture}
		};
	\end{tikzpicture}
}


\maketitle

\begin{abstract}
\added{Wearable devices have been widely adopted not only to offer personalised services but also for community health monitoring. Modern healthcare systems demand more support for privacy-preserving data processing and sharing requirements under three main use cases: (i) Data owners (DO) requesting processing of their own data for personal health monitoring. (ii) Multiple data requesters (DRs) such as doctors, family, and friends require data processing of a single data owner for health checkups. (iii) Multiple DRs (ex., researchers, scientists) require data processing of multiple data owners to monitor community or social health indications of a particular disease or pandemic. 
Most existing solutions either provide secure data processing or fine-grained access control separately; few existing works consider secure fine-grained access control over the encrypted computational results. However, they lack data owner access control and do not efficiently support the above three use cases, which is unsuitable for resource-constrained IoT devices.} \substituted{In this paper}{Therefore}, we propose
a novel, efficient, privacy-preserving, and flexible fine-grain for \textbf{S}ingle (DO) \textbf{A}nd \textbf{M}ultiple (DOs) data \textbf{A}ggregation and sharing scheme in a user-centric access control named SAMA. Specifically, we deploy two key ideas.
First, it uses a multi-key partial homomorphic cryptosystem
to allow flexibility in accommodating the aggregation of data
originating from both a single data owner or multiple data
owners while preserving privacy during the processing. It also
uses ciphertext-policy attribute-based encryption (CP-ABE) to support fine-grain sharing with multiple data requesters based on user-centric access control. Formal security analysis shows
that SAMA supports data confidentiality and authorisation.
The proposed scheme has also been analysed in terms of computational and communication overheads. Our experimental results demonstrate that SAMA supports privacy-preserving
flexible data aggregation more efficiently than the relevant
state-of-the-art solutions that meet the processing and sharing
needs of a modern IoT wearable healthcare system.

\end{abstract}

\begin{IEEEkeywords}
IoT, Security, Privacy, Multi-key homomorphic encryption, Attribute-based encryption, Access control.
\end{IEEEkeywords}


%
\IEEEpeerreviewmaketitle

\section{Introduction} \label{sec:intro}
%
%
%
%

\IEEEPARstart
{N}{owdays} wearable devices are equipped with sensors and communication capabilities to collect various data from users such as data related to their health (e.g., heart rate, oxygen saturation), activities (e.g., steps count, sleep quality) and environment (e.g., location, humidity levels)~\cite{motti2020wearable}. 
Modern healthcare systems can utilize this data to run analytic models. Then, these models could be used to improve services to (i) individuals, e.g., personalised treatments, remote patient monitoring, and early disease diagnosis by detecting anomalies, and (ii) the wider public, e.g., predicting the spread of disease by analysing data from multiple individuals~\cite{islam2020wearable}. 

\begin{figure*}[h!]
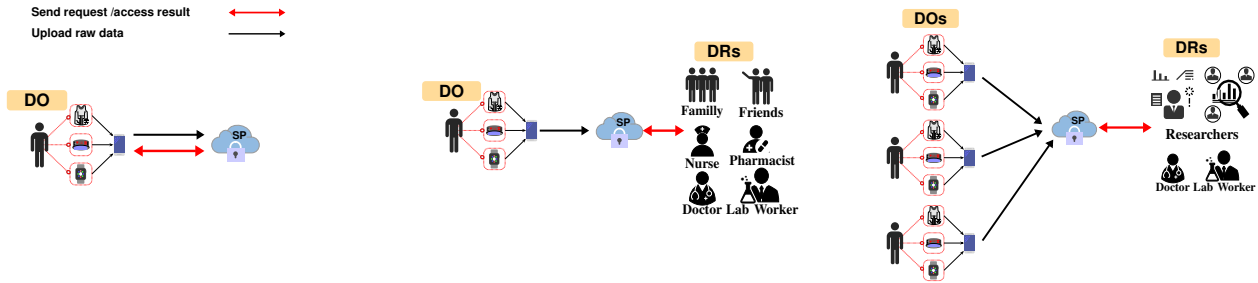

		\centering
		
		\begin{subfigure}[b]{0.3\textwidth}
			\centering
			\SecondSeparateFigure{0.27}
			\caption{A data owner requests aggregation of own data (DO-DO).}
			\label{fig:DRs-DOs}
		\end{subfigure}
		\hfill
		\begin{subfigure}[b]{0.3\textwidth}
			\begin{adjustbox}{width=\linewidth} 
				\FirstSeparateFigure{0.3}
			\end{adjustbox} 
			\caption{Data requesters request aggregation data of a single data owner (DRs-DO).}
			\label{fig:DRs-DOs}
		\end{subfigure}
		\hfill
		\begin{subfigure}[b]{0.3\textwidth}
			\begin{adjustbox}{width=\linewidth} 
				\ThirdSeparateFigure{0.3}
			\end{adjustbox} 
			\caption{Data requesters request aggregation data of multiple data owners (DRs-DOs).}
			\label{fig:DRs-DOs}
		\end{subfigure}
		\caption{Data aggregation requests use cases. 
  }
	\label{fig:three_use_cases}
\end{figure*}

Data requesters such as organisations (e.g., healthcare
providers and research institutions) and individuals (e.g., family members, friends, and data owners themselves) could also benefit from having access
to the results of these analytic models. 
Therefore, health systems should be able to support data processing and sharing in the following three cases: (i) data owners (DO) accessing the results of analytics run on their own data (DO-DO) and (ii) data requesters (DRs) 
accessing the results of analytics run on data of a single data owner (DRs-DO) or (iii) 
on data of multiple data owners (DRs-DOs). These cases are shown in Fig.~\ref{fig:three_use_cases}. Moreover, due to the resource-constrained nature of IoT wearable devices on the DO side, there is a need to address the above three cases in an efficient manner.

Healthcare provision via wearable devices has led a large number of applications to deploy cloud service providers (CSP) to run these analytic models. 
This comes with concerns over user privacy. 
First, although collected through secure channels, service providers typically process data in plaintext. This comes with risks of data leaks to unauthorised parties~\cite{zhou2015security}. Second, users usually have no control over who has access to their data~\cite{shafagh2020droplet, wang2016sieve, li2012scalable}. Note that unauthorised exposure of personal health data also violates GDPR~\cite{REGULATI80:online} and HIPPA~\cite{act1996health} regulations, which advocate for users’ privacy protection and access control. 
Hence, it is essential to achieve flexible data processing and sharing in a privacy-preserving manner and efficiently support the three use cases while adopting a user-centric approach. Such an approach protects users’ data from unauthorised access and gives control over the data in the hands of data owners rather than service providers~\cite{safavi2014conceptual}.


There are already attempts in the literature to achieve privacy-preserving processing and sharing of data mainly based on homomorphic encryption (HE) or attribute-based encryption (ABE) schemes. However, none of the existing solutions can support all three cases: DO-DO, DRs-DO, and DRs-DOs.

Most existing solutions for secure flexible data processing are mainly based on single-key HE schemes~\cite{pang2020privacy, ding2017encrypted, ara2017secure, wang2019achieve, erkin2012generating, li2022lightweight, zhang2015foresee}, which come with drawbacks. First, user data is encrypted with the same public key that belongs either to one of the data requesters or to a third-party cloud provider. 
If the corresponding private key is leaked, this can lead to serious privacy issues~\cite{aloufi2021computing, zhang2021privacy}. Second, these solutions restrict DOs from accessing and retrieving their data as they cannot access the corresponding private key~\cite{wang2014tale}, hence cannot fulfill the DO-DO case. 
Third, HE-based solutions by default do not generally support flexible data sharing with multiple data requesters~\cite{ding2017privacy, mustafa2015dep2sa}.
A trivial solution to overcome these drawbacks would be for each data requester and data owner to encrypt their data with the homomorphic public key of the requester. This is an inefficient solution as it comes with additional costs for the data owner. 

A few multi-key HE solutions~\cite{lopez2012fly, chen2021non, aloufi2019blindfolded} have been proposed to address the drawbacks of single-key HE schemes. Some are based on fully HE (FHE) schemes, which are considered unsuitable for resource-constrained devices, while others are based on partial HE (PHE) schemes. 
Although they address the data owners access case (DO-DO), none support fine-grain data sharing capabilities with multiple requesters~\cite{pang2020privacy, zhang2021privacy}, thereby failing to accommodate the remaining two cases -- DRs-DO and DRs-DOs. Solutions based on ABE~\cite{li2021efficient, wang2016sieve, li2012scalable, alshehri2012secure, narayan2010privacy}, on the other hand, provide secure data sharing with multiple requesters. However, ABE is intended for data sharing, so they do not support secure data processing. Hence, none of the approaches on its own can support both secure flexible data processing and sharing.

The scheme proposed in~\cite{ding2017privacy} combines PHE and ABE schemes to address the DRs-DOs case. However, the solution uses a single-key HE scheme, bearing all the drawbacks of such schemes mentioned earlier. Moreover, despite supporting flexible data sharing through ABE, it fails to support personalised processing and access control settings by data owners. It leaves this control to third-party cloud servers. Therefore, supporting privacy-preserving data processing 
along with fine-grain sharing 
that encompass the three cases shown in~Fig.~\ref{fig:three_use_cases} in an efficient way (i.e., suitable for resource-constrained devices) is still an open issue.

To fill this research gap, we address the drawbacks of~\cite{ding2017privacy} by proposing a secure and privacy-preserving flexible data aggregation and sharing scheme named SAMA. 
To the best of our knowledge, SAMA is the first attempt to combine multi-key PHE with ABE that supports flexible privacy-preserving data processing along with fine-grain sharing with a focus on user-centric access control, and yet it is suitable for resource-constrained devices. To this end, the novel contributions of this work are two-fold:

\begin{itemize}

\item We propose a scheme called SAMA, which combines multi-key PHE and CP-ABE to offer flexible data processing with fine-grain sharing capabilities. To the best of our knowledge, SAMA is the first attempt to achieve data processing and fine-grain sharing with data owner access. It allows data owners to upload their data only once and yet supports all three use cases: DO-DO, DRs-DO, and DRs-DOs. This addresses the needs of modern healthcare applications in an efficient manner suitable for resource-constrained devices. SAMA also offers a user-centric access policy setting that allows data owners to set access control to their outsourced data with two different access policies.

\item We investigate SAMA both theoretically in terms of security and experimentally in terms of computational and communication costs through simulations. Our results show that SAMA is more efficient than scheme~\cite{ding2017privacy} in terms of computational and communication cost. 

\end{itemize}

The rest of the paper is organised as follows. Section~\ref{sec:related_work} discusses related work. Section~\ref{sec:preliminaries} introduces design preliminaries and the main building blocks used. Section~\ref{sec:SAMA_scheme} details the design of SAMA. Sections~\ref{sec:security_analysi} and~\ref{sec:performance_evaluation} detail SAMA's security analysis and performance evaluation, respectively. Section~\ref{sec:conclusion} concludes the paper. 

\begin{table*}
\caption{Comparison of SAMA with related work in use-cases support. 
} 
\label{table:usecase_comp}
\begin{center}
\scriptsize
\begin{tabular}{l c c c c c c c c c c}
\toprule
\textbf{} & \textbf{\cite{ruj2013decentralized}} & \textbf{\cite{mustafa2015dep2sa}} & \textbf{\cite{bhowmik2023eeppda}}&
\textbf{\cite{zhang2022distributed}} & \textbf{\cite{tang2019efficient}} & 
\textbf{\cite{pang2020privacy}} &
\textbf{\cite{zhang2021privacy}} &
\textbf{\cite{liu2019secure}} & 
\textbf{\cite{ding2017privacy}} &
\textbf{SAMA} \\

\midrule




DO-DO support & \xmark & \xmark & \xmark &  \xmark & \xmark & \cmark  & \cmark  & \xmark  & \xmark  & \cmark\\
DRs-DO support & \xmark & \xmark & \cmark & \xmark &\cmark & \xmark & \xmark & \xmark & \xmark & \cmark\\
DRs-DOs support &  \cmark &  \cmark &  \xmark &  \cmark  & \xmark & \xmark &  \xmark & \cmark & \cmark  & \cmark\\











\bottomrule
\end{tabular}

\end{center}
\end{table*}

\section{Related Work}
\label{sec:related_work}

There are efforts to protect users' privacy while their data is being processed using single- or multi-key HE schemes.  
Most solutions with single-key PHE schemes~\cite{ara2017secure, wang2019achieve, erkin2012generating, zhang2015foresee} limit data owners' access to data, and might cause privacy issues.
They have also considered secure data processing provided only by a single user or multiple users.  
Other schemes~\cite{aloufi2019blindfolded, lopez2012fly, chen2021non} are based on FHE schemes which are not suitable for resource-constrained devices~\cite{pang2020privacy}. A few multi-key PHE schemes~\cite{pang2020privacy, zhang2021privacy} may be suitable for resource-constrained devices. However, they lack flexible data sharing capabilities. Therefore, multi-key PHE schemes alone are not suitable for addressing privacy-preserving data processing and sharing.




To support secure access control, there are proposals~\cite{li2021efficient, wang2016sieve, li2012scalable, wang2023ciphertext, alshehri2012secure, narayan2010privacy} adopting ABE schemes~\cite{sahai2005fuzzy}. These proposals allow users to choose who can access their data, hence supporting fine-grain access control and multiple data requesters' access. ABE schemes can be classified into two types: ciphertext-policy ABE (CP‐ABE)~\cite{bethencourt2007ciphertext} and key‐policy ABE (KP‐ABE)~\cite{goyal2006attribute} schemes. 
In the CP‐ABE scheme, the access structure is embedded with ciphertexts, and users' attributes are embedded with the users' private keys. In contrast, with the KP‐ABE scheme, the access structure is associated with users' private keys, and the ciphertext is associated with attributes.
Therefore, with the KP‐ABE schemes, users do not have control over who can access the data; they can only control attribute assignments~\cite{bethencourt2007ciphertext}. In summary, ABE schemes on their own do not support computations over encrypted data.

There are some existing proposals that combine secure data processing with access control.
Ding et al.~\cite{ding2017privacy} and~\cite{ding2020extended} proposed a flexible access control scheme over encrypted user data's computation results by combining ABE with single-key PHE techniques. The scheme is suitable for resource-constrained devices. However, it cannot support data owners' access over their data (DO-DO) as data is encrypted with a cloud key, thereby 
it does not provide flexible processing for all three use cases despite providing a fine-grain sharing option.

Ruj and Nayak~\cite{ruj2013decentralized} combined Paillier HE with ABE to support privacy-preserving data aggregation and access control in a smart grid.
However, the aggregated data needs to be decrypted and then re-encrypted with an access policy by a trusted authority. 
Mustafa et al.~\cite{mustafa2015dep2sa} designed a multi-recipient system called DEP2SA which combines HE and selective data aggregation in a smart grid. Zhang et al.~\cite{zhang2022distributed} proposed a privacy-preserving data aggregation scheme with fine-grain access control for the smart grid, which combined HE with proxy re-encryption. Liu et al.~\cite{liu2019secure} designed a scheme to support secure computation of aggregate statistics using Multi-Party Computation (MPC) with fine-grain access control of outsourcing IoT data using CP-ABE. However, using MPC technique requires high user active interactions for processing and more communication overhead as the number of the generated shares depends on the number of fog nodes. 
These schemes support only the DRs-DOs case and fail to support the other two use cases. Moreover, they do not support a user-centric access policy.

Bhowmik and Banerjee~\cite{bhowmik2023eeppda} developed an efficient privacy-preserving aggregation method in which data are aggregated at edge servers en route to a cloud server. Then, the aggregation results are accessed by authorised medical professionals. Tang et al.~\cite{tang2019efficient} proposed privacy-preserving fog-assisted health data sharing that supports a flexible user-centric approach using ABE. 
However, this scheme requires heavy processing at the user side which might not be suitable for resource-constrained devices. In addition, both these schemes support only DRs-DO use case.

Pang and Wang~\cite{pang2020privacy} and Zhang et al.~\cite{zhang2021privacy} proposed privacy-preserving operations on outsourced data from multiple parties under multi-key environments. The proposals support sharing of processed data only with a data requester, thereby supporting only DO-DO use case. Moreover, they do not support fine-grain data sharing with multiple DRs. 

In summary, most existing solutions that address privacy-preserving data processing do not offer fine-grain data sharing. A few solutions combining both cannot efficiently accommodate all three use cases, as shown in Table~\ref{table:usecase_comp}. Therefore, we aim to address this knowledge gap by 
designing a SAMA scheme that efficiently supports flexible, secure data processing with fine-grain sharing to accommodate all three use cases under user-centric access control that offers two access control policies (APs) and (APm) to DOs.


\section{Preliminaries}
\label{sec:preliminaries}


\begin{figure}[t]
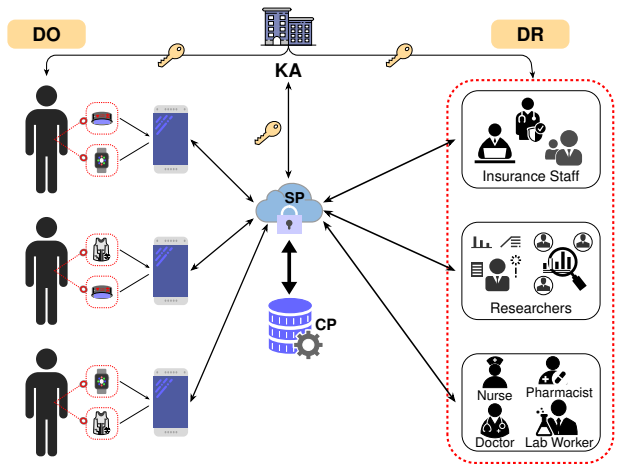

    \centering
    \MyFinalDiagramxx{0.35}{1.7}
    \caption{System model of the SAMA scheme. 
    }
    \label{fig:system_model}
\end{figure}

\subsection{System Model}
The system model used by SAMA consists of the following entities (see Fig.~\ref{fig:system_model}). 
\emph{Data owners ($DO$)} are individuals who possess \emph{wearables} and are willing to process their data for their own personal benefits or the collective benefit of society and share their data and processing result with various \emph{data requesters}. Data owners' wearable data is usually collected and shared via their \emph{smartphone (gateway)}.
\emph{Data requesters ($DRs$)} are data consumers 
who wish to utilise data owners' wearable data in order to provide (personalised) services to $DOs$ or society. Example $DRs$ could be individuals (e.g., data owners themselves, family members, friends) and organisations (e.g., hospitals, research centres).
\emph{Service provider ($SP$)} 
provides $DO$ with data storage, manages access requests, and process $DOs$' data, while
\emph{Computational party ($CP$)} cooperates with $SP$ in data computations and access control. 
A \emph{Key Authority} (KA) plays the role of a key management organisation.


The list of acronyms used throughout the paper is given in Table~\ref{table:Acronyms}.


\subsection{Threat Model}
The threat model of the proposed SAMA scheme is as follows.
    {DOs}, DRs, SP and CP 
    are semi-honest (honest-but-curious) entities. They follow the protocol as per the specifications, yet they are curious about the sensitive information of DOs or their data aggregation.
    The {KA} is considered a trustworthy entity. It performs all its duties honestly and never colludes with any other entities.
        The external entities 
        are considered to be untrustworthy, hence malicious. They may utilize different network eavesdropping attacks, modify data in transit, or try to gain unauthorised access to disrupt the system.

   Herein, we define the capabilities of adversary $ \mathbcal{A}$ as follows.
 $\mathbcal{A}$ may compromise the SP and try to guess the raw data from the encrypted data sent by DOs or CP or to be sent for DRs. 
Adversary $\mathbcal{A}$ should be restricted from compromising CP, DOs and {DRs}. If $\mathbcal{A}$ compromises the DOs, it can gain access to plaintext. Similarly, if $\mathbcal{A}$ compromises the DRs, the decrypted processing result can be obtained. Moreover, if $\mathbcal{A}$ compromises the CP, it can access the strong secret key and thereby access the raw data from DOs. This threat model is representative and common in adversaries used in other state-of-the-art schemes~\cite{pang2020privacy} and~\cite{ding2017privacy}.

\begin{table}[t]
\caption{Acronyms}
\label{table:Acronyms}
\centering
\footnotesize
\begin{tabular}{@{}llcc@{}}
\toprule
\textbf{Acronym} & \textbf{Meaning}  \\
\midrule
\acrshort{do}& \acrlong{do}\\
\acrshort{dr}& \acrlong{dr}\\
\acrshort{sp}& \acrlong{sp}\\
\acrshort{cp}& \acrlong{cp} 
\\
\acrshort{ka}& \acrlong{ka}\\
\acrshort{he}& \acrlong{he}
\\
\acrshort{fhe}& \acrlong{fhe}\\
\acrshort{phe}& \acrlong{phe}\\
\acrshort{pe}& \acrlong{pe}\\
\acrshort{vp}& \acrlong{vp}\\
\acrshort{abe}& \acrlong{abe}\\
\acrshort{cpabe}& \acrlong{cpabe}\\
\acrshort{do-do}& \acrlong{do-do}\\
\acrshort{drs-do}& \acrlong{drs-do}\\
\acrshort{drs-dos}& \acrlong{drs-dos}\\

\bottomrule
\end{tabular}
\end{table}

\subsection{Assumptions}
We consider the following assumptions in our design.
\begin{itemize}
\item The communication channels among all entities are encrypted and authenticated. 
\item The SP and CP do not collide with each other and any other entities or external adversaries as they have a legal responsibility to prevent leakage of the DOs' sensitive data.
\item  All entities' identities are verified by {KA} before obtaining their cryptographic public/private keys. 
\end{itemize}

\subsection{Design Requirements}
The proposed system should satisfy the following functional, security and privacy, and performance requirements.

\subsubsection{Functional Requirements} 
\begin{itemize}


\item Flexible data processing requests:
     SAMA should support all of the following three use cases:  
     (i) data owners accessing the aggregation results of their own data (DO-DO), (ii) data requesters accessing the aggregation results of a single data owner (DRs-DO), and (iii) of multiple data owners (DRs-DOs).


    \item Fine-grain access control: SAMA should support a flexible access policy for data owners and facilitate granting different access rights to a set of data requesters based on the access policy set by data owners.

     \item User-centric: each data owner should control who is authorised to access the raw data collected from their wearables as well as the aggregated data that contains their raw data in (DRs-DO) and (DRs-DOs) use cases.  
       


    
      
    \end{itemize}

\subsubsection {Security and Privacy Requirements} 
    \begin{itemize}

    \item Data confidentiality: DOs' raw and aggregated data 
    should be protected from unauthorised disclosure in transit, at rest, and while processing. 

    \item Authorisation: Only authorised DRs should be able to access DOs' aggregated data.

        \end{itemize}
    
\subsubsection{Performance Requirements}


\begin{itemize}
  
     \item Efficiency: SAMA should be viable for devices with limited computational capabilities. 
\end{itemize}

\subsection{Building Blocks}
\label{sec:building_blocks}

This section reviews briefly the Paillier cryptosystem~\cite{paillier1999public}, the Variant-Paillier in Multi-key cryptosystem~\cite{pang2020privacy}, and 
CP-ABE~\cite{bethencourt2007ciphertext}, which are used in the design of SAMA. The notations used in the paper are given in Table~\ref{table:notations}.

\subsubsection{Paillier Cryptosystem} 
It is a practical and semantically secure additive homomorphic encryption scheme~\cite{paillier1999public}. 

\textit{Paillier in Single-Key Environment}
\label{sec:build-block-Paillier}
consists of three algorithms: key generation algorithm $(\kgen_{PE})$, the encryption algorithm$ (\enc_{PE})$, and decryption algorithm$ (\dec_{PE})$.

\begin{itemize} 
\item $\kgen_{PE}$($k$) $\xrightarrow{}ppk, psk$: 
Given a security parameter $k$, select two large prime numbers $\pp$ and $\qq$.
Compute $\nn =  \pp\cdot\qq$, and $\secpar = lcm(\pp-1,\qq-1)$. 
Define $L(x)= (x-1)/ \nn$. 
Select a generator g $\in  \ZZ^*_{\nn^2}.$  
Compute $\mu =(L$(g$^\secpar\mod\nn^2))^{-1}\mod\nn$.
The public key is $ppk=(\nn,$g) and the private key is $psk = (\secpar,\mu)$.

\item $\enc_{PE}(ppk,\mm)\xrightarrow{}\cc$: 
Given a message $\mm\in\ZZ$ and a public key $ppk=(\nn,$g), choose a random number $\rr \in {\ZZ_\nn^*}$, and compute $\cc= \enc_{PE}(ppk, \mm)$ = g$^\mm\cdot{\rr^\nn}\mod \nn^2$. 

\item $\dec_{PE}(psk,\cc)\xrightarrow{}\mm$: 
Given a ciphertext $\cc$ and a private key $psk = (\secpar,\mu)$, 
recover the message $\mm$ = $\dec_{PE}(psk, \cc) = L(\cc^\secpar\mod\nn^2)\cdot\mu\mod\nn$. 
\newline
\end{itemize}

\textit{Variant-Paillier in Multi-Key Environment}\label{sec:build-block-Var-Paillier}~\cite{pang2020privacy} is a variation of the Paillier cryptosystem~\cite{paillier1999public}, which makes it compatible to work in multiple users (DOs) environment by generating a different public-private key pair for each DO with two trapdoor decryption algorithms. The scheme comprises: key generation $(\kgen_{VP})$, encryption $(\enc_{VP})$, decryption with a weak secret key $(\dec_{wsk})$, and decryption with a strong secret key $(\dec_{ssk})$.

\begin{itemize}
\item $\kgen_{VP}(k) \xrightarrow{}vpk,wsk,ssk$:
Given a security parameter $k$, 
choose $k+1$ small odd prime factors $u, v_1, \ldots, v_i, \ldots, v_k$ and choose two large prime factors $v_p$ and $v_q$ in which $p$ and $q$ are large primes with the same bit length.
Compute $p$ and $q$ as
$p = 2uv_1v_2 \cdots v_i \cdots v_kv_p + 1$ and
$q = 2uv_1v_2 \cdots v_i \cdots v_kv_q + 1$.
Calculate $n =p\cdot q$ and $\secpar = lcm(p-1,q-1)$. 
Choose $t$ as a number or a product of multiple numbers from the set $( v_1, v_2, \ldots, v_i, \ldots, v_k)$, and $t\vert\secpar$ naturally exists. 
Choose a random integer $g\in \ZZ^\ast_{n^2}$ that satisfies $g^{utn} =1 \mod n^2$, and $\gcd(L(g^\secpar\mod n^2),n)=1$.
Define $L(x)= (x-1)/n$. Compute $h = g^{n\times\secpar/ t}\mod n^2$.
The public key is $vpk$ = $(n, g, h)$, the weak secret key is $wsk = t$ and the strong secret key is $ssk = \secpar$

\item $\enc_{VP}(vpk,m)\xrightarrow{}c$: 
Given a message $m\in\ZZ_n$ and a public key $vpk$ = $(n, g, h)$, choose a random number $r\in\ZZ_n$, and compute the ciphertext $c$ as $c = \enc_{VP}(vpk,m) = g^mh^r \mod n^2$.

\item $W\dec_{VP}(wsk,c)\xrightarrow{}m$: 
The decryption algorithm with a weak secret key decrypts only the ciphertext encrypted with the associated public key.
Given $wsk$ and $c$, the ciphertext can be decrypted as $m = W\dec_{VP}(wsk,c) = \frac{L(c^t\mod n^2)}{L(g^t\mod n^2)}\mod n$. 

\item $S\dec_{VP}(ssk,c)\xrightarrow{}m$:
The decryption algorithm with a strong key decrypts the ciphertexts encrypted with any public key of the scheme.
Given $ssk$ and $c$, 
the ciphertext can be decrypted as $m= S\dec_{VP}(ssk,c) = L(c^\secpar\mod n^2)\cdot\mu\mod n$.
\[\frac{L(c^\secpar\mod n^2)}{L(g^\secpar\mod n^2)}\mod n = \frac{L(g^{\secpar m}\mod n^2)}{L(g^\secpar \mod n^2)}\mod n\]
\end{itemize}

\begin{table}[t]
\centering
\caption {Notations} 
\label{table:notations}
\scriptsize
\begin{tabular}{@{}llcc@{}}
\toprule
\textbf{Symbol} & \textbf{Meaning}  \\
\midrule
$DO_i$ &  $ith$ $DO$, $i=\{1,\dots,N\}$\\
$m_i$ & raw data provided by $DO_i$\\
$r_i$ & random number generated by $SP$ for each $DO_i$\\
$N_{DO}$, $N_{DR}$ & number of $DOs$, $DRs$ \\
$N_{req}$ & number of data points requested for aggregation \\
$N_m$ & number of messages received by $DO_i$ \\
$N$ & number of $DOs$ in DRs-DOs case\\
\midrule
$ssk$ & strong secret key in VP-HE\\
$vpk_i,wsk_i$ & VP-HE key pair (public key, weak secret key) of $DO_i$\\
$\enc_{VP}$, $\dec_{VP}$ & encryption, decryption using VP-HE\\
$[m_i]$ & $m_i$ encrypted by the $vpk_i$ of $DO_i$\\
$[r_i]$ & random number encrypted by the $vpk_i$ of $DO_i$\\
\midrule
$\enc_{PE}$, $\dec_{PE}$ & encryption, decryption using PE\\
$ppk_j,psk_j$ & PE public, private key pair used by $DR$ \\
\midrule
$\pk$, $MK$, $\sk$ & public parameters, master key, secret key in CP-ABE\\
$\enc_{ABE}$, $\dec_{ABE}$ & encryption, decryption using CP-ABE\\
$AP_S/AP_M$ & single- and multiple-$DO(s)$ data access policy\\
$BiPair$ & cost of bilinear pairing in ABE\\ $|\gamma|+1$ & number of attributes in the access policy tree\\
$\vartheta$ & number of attributes needed to satisfy the access policy \\
\bottomrule
\end{tabular}
\end{table}

\subsubsection{Ciphertext-Policy Attribute Based Encryption}  \label{sec:build-block-CP-ABE}
The CP-ABE is a type of public-key encryption in which the ciphertext is associated with an access policy and $DO$'s private key is dependent upon attributes to support fine-grain access control~\cite{pang2020privacy}. It consists of four main algorithms: a setup algorithm $(Setup)$, encryption algorithm $(\enc_{ABE})$, key generation algorithm $(\kgen_{ABE})$, and decryption algorithm $(\dec_{ABE})$.

\begin{itemize}
\item $Setup(k,U) \xrightarrow{}\pk, mk$: Given a security parameter $k$ 
 and a universe of attributes $U$, the setup algorithm outputs the public parameters $\pk$ and a master key $mk$.

\item $\enc_{ABE}(\pk, M, A)\xrightarrow{}C$: Given public parameters $\pk$, a message $M$, and an access structure $A$ over the universe of attributes, the encryption algorithm outputs a ciphertext $C$ which implicitly contains $A$.  

\item $\kgen_{ABE}(mk, s)\xrightarrow{}\sk$: Given a master key $mk$ and a set of attributes $s$ which describe the key, the key generation algorithm outputs a private key $\sk$. 

\item $\dec_{ABE}(\pk, C, \sk)\xrightarrow{}M$: Given public parameters  $\pk$, a ciphertext $C$, which includes an access policy $A$, and a private key $\sk$, using a decryption algorithm, $DO$ can decrypt the ciphertext and get a message $M$ only if the attributes associated with the private key satisfy $A$. 
\end{itemize}

\section{The SAMA Scheme}
\label{sec:SAMA_scheme}

\subsection{Overview of the SAMA Scheme}

\begin{figure}
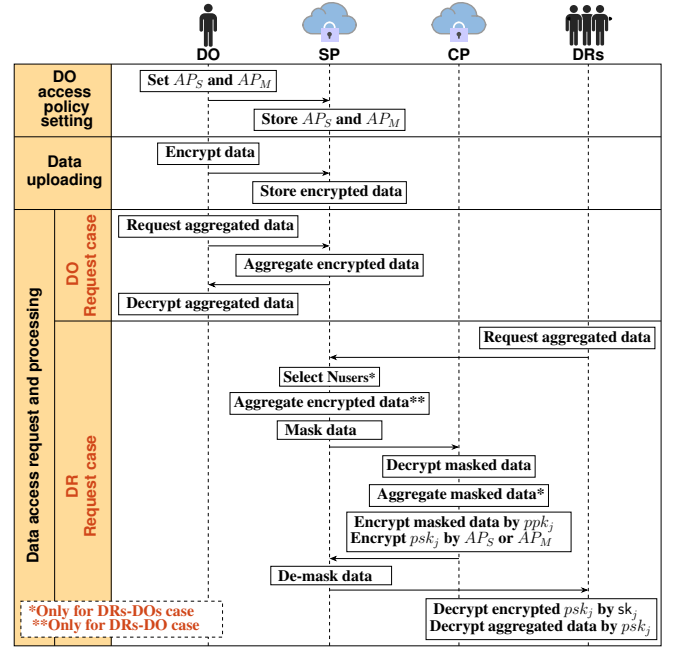

    \centering
    \Extrafigure{0.43}{\Large}
    \caption{An overview of the SAMA scheme.}
    \label{fig:SAMA_overview}
\end{figure}

SAMA combines the VP-HE and CP-ABE schemes and consists of three phases: (i) DO access policy setting, (ii) data uploading, and (iii) data access request and processing. The overview of these phases is shown Fig.~\ref{fig:SAMA_overview}.

During the first phase, to achieve a user-centric fine-grain access policy functionality, DOs define two types of policies, single ($AP_S$) and multiple ($AP_M$) access policies, and send them to SP. This allows SP to process and share DOs' data with multiple DRs according to DOs' preferences. 
In the next phase, DOs encrypt data only once with their \acrshort{vp} public key and send the resulting ciphertext to SP. 
During the last phase, SP receives requests to grant access to the (aggregated) data of DOs. Both SP and CP process these requests, and the results are shared with the requester. 
There can be three types of requests: from DOs for accessing aggregation of their own data (DO-DO) or from the DRs requesting (aggregated) data of either a single DO (DRs-DO) or multiple DOs (DRs-DOs).

Upon receiving a request from a DO, SP aggregates the DO's encrypted data and returns the result. The DO uses its own \acrshort{vp} weak secret key to obtain their aggregated result.
If the request comes from a DR for aggregation of a single DO's data, SP aggregates the DO's encrypted data, masks it, and sends the result to CP. CP performs strong decryption to obtain the masked data, encrypts this result (masked aggregated data) with a HE public key and encrypts the corresponding private key using CP-ABE with $AP_S$, and sends both ciphertexts to SP. If the request is for data of multiple DOs, the process is slightly different. SP gets the encrypted data of DOs, masks them, and sends the masked encrypted data to CP. CP performs strong decryption on the received ciphertexts, aggregates the results (masked data), encrypts the result with a HE public key, and encrypts the corresponding private key using CP-ABE with $AP_M$. In both cases, CP sends both ciphertexts to SP. SP then performs de-masking on the received ciphertext and sends the encrypted result (aggregated data) and CP-ABE ciphertext to DR. 
Finally, DRs who satisfy the access policy decrypt CP-ABE ciphertext and obtain the HE private key to access DOs aggregate data.

\subsection{System Initialisation}

\subsubsection{System Parameters Setup}

In this phase, system parameters of the encryption schemes (which determine the key lengths) are set.

\begin{itemize}
\item VP-HE setup: The KA sets a security parameter $k$ and chooses two large prime numbers $p$ and $q$ such that $L(p)=L(q)= k$. $L$ is the bit length of the input data. 

\item Paillier setup: Given the security parameter $k$, 
the KA then chooses two large prime numbers $\pp$ and $\qq$. 
Then, the key generation algorithm is initiated as in Section.~\ref{sec:build-block-Paillier}

\item ABE setup: Given the security parameter $k$, the KA generates $U$ attributes, which are used to generate $\pk$ and $mk$ using the $Setup$ algorithm described in Section~\ref{sec:build-block-CP-ABE}. 

\end{itemize}

\subsubsection{System Key Generation and Distribution}
This phase is divided into three steps outlined below. 
\begin{itemize}
\item \textit VP-HE Key Generation:
The KA generates a unique $ssk$ and distinct variant Paillier homomorphic public/private key pair $(vpk_i, wsk_i)$ for every DO, $DO_i$, $i=1,\ldots,N_{DO}$, using the $KGen_{VP}$ algorithm in Section.~\ref{sec:build-block-Var-Paillier}. 

\item Paillier Key Generation:
The KA generates a distinct Paillier homomorphic public/private key pair $(ppk_j, psk_j)$, for each request that comes from the same or any DR, using the $KGen_{PE}$ algorithm described in Section.~\ref{sec:build-block-Paillier}.


\item ABE Key Generation:
The KA generates a distinct private key $\sk_j$ for every $DR_j$, using $KGen_{ABE}$ (Section.~\ref{sec:build-block-CP-ABE}). 
$DR_j$ obtains $\sk_j$, which embeds her/his attributes/roles. 
\end{itemize}

\subsection{SAMA in Detail}
SAMA consists of three phases: DO access policy setting, 
data uploading, and data access request and processing.

\subsubsection{DO Access Policy Setting} 
It allows data owners to set their access policy for data aggregation and sharing requirements and share it with SP. It has three steps: a) define the access policy, b) activate notifications, and c) update the access policy.

\textit{a) Define access policy:} 
Each data owner defines two types of access policy: 
$AP_{S}$ -- single-data owner and $AP_{M}$ -- multiple-data owners data aggregation and sharing access policy. 
\textit{
$AP_S$} allows data owners to control who can access the aggregated results of their own data. 
Only authorised DRs with specific attributes satisfying the access policy can have access to the final aggregated result.
\textit{
$AP_M$} allows data owners to set whether they agree their data to be aggregated with other data owners' data and the result to be shared. 
Each data owner defines his/her sharing preferences and gives consent to allow the use of their individual data in aggregation along with other data owners' 
 data. 
$AP_{M}$ does not authorise SP to share any specific individual raw data with anyone. It only allows SP to use the encrypted data of DOs whose sharing preferences match with the attributes of DRs who requested data access. 



\textit{b) Activate notification:} 
DOs can select to receive regular notifications: a summary of all 
requests for their data received by SP. 
They can check who has requested access to their data and whose requests were granted/rejected. 
Regular notifications can be switched on/off by DOs and be received as daily/weekly/monthly summaries. 

\textit{c) Update access policy:} SP provides DOs with the ability to update their access policy periodically or on demand. DOs can also update their pre-defined policies ($AP_{S}$ or $AP_{M}$). 

\subsubsection{Data Uploading}
During this phase, DOs upload their data to SP regularly. They encrypt their wearable data $m_i$ with their variant-Paillier public key, $vpk_i$, to obtain $C_{vpk_i}=\enc_{VP}(vpk_i,m_i)$ 
and send the encrypted data to SP. This phase is the same for single and multi-DOs data sharing. 

\subsubsection{Data Access Request and Processing}
There can be three different types of data access requests for data owners' aggregated data: 
a) Data owner requests access to their own (aggregated) data (DO-DO), and DRs request access to aggregated data of b) a single DO (DRs-DO) and c) multiple DOs (DRs-DOs). DO requests are directly handled by SP, while DR requests are handled by both SP and CP. 

\emph {a) DO-DO use case:}
A DO requests SP to aggregate his/her own encrypted wearable data and provide the result. Upon receiving the request to aggregate $N_{req}$ data points, SP aggregates the DO's data (i.e., it performs additive homomorphic operations by multiplying the encrypted data owner data) to get $[\sum_{i=1}^{N_{req}} m_i]_{vpk_i} = \prod_{i=1}^{N_{req}} C_{vpk_i}$, where $[m]$ denotes encrypted data. The result then is sent to the DO who can decrypt $[\sum_{i=1}^{N_{req}} m_i]_{vpk_i}$ with his weak secret key to get the aggregated data, $\sum_{i=1}^{N_{req}} m_i ~mod~n =W\dec_{VP}(wsk_i,[\sum_{i=1}^{N_{req}} m_i]_{vpk_i})$. For simplicity and improved readability, in the rest of the paper, we will omit $mod~n$ in the aggregation results.

\emph {b) DRs-DO use case:}
A DR requests access to the aggregated data of a single DO, e.g., a doctor requires access to aggregated data of a patient. The aggregated data can be accessed only by DRs whose attributes satisfy the fine-grain access policy $AP_S$ set by the DO 
(see Algorithm~\ref{alg:the_alg1}).

\textit{(i) Handling DR request}: After DR has issued a request to access the aggregated data, 
SP performs the same additive homomorphic operations as in Step a) explained earlier. The result is a ciphertext of the aggregated data: $[\sum_{i=1}^{N_{req}} m_i]_{vpk_i}$.

\begin{algorithm}[t]
	\scriptsize
			\caption{SAMA: DRs-DO use case. }
			\DontPrintSemicolon
			
			\KwInput{Two ciphertexts $[m_1]_{vpk_i}$ and $[m_2]_{vpk_i}$}
			\KwOutput{The ciphertexts $[m_1+m_2]_{ppk_j}$}
			
			\SetKwProg{Fn}{Steps~by~\textbf{SP}:}{}{}
			\Fn{}{
				$[m_1+m_2]_{vpk_i} \leftarrow [m_1]_{vpk_i} \times [m_2]_{vpk_i}$ \\
			$r_{i}\overset{R}{\leftarrow}\mathbb{Z}_n$  
            \\ 
			$[r_{i}]_{vpk_i} \leftarrow \enc_{vp}(vpk_i,r_{i})$\\
				 $ [m_1+m_2+r_{i}]_{vpk_i} = [m_1+m_2]_{vpk_i} \times[r_{i}]_{vpk_i}$ \\
				Send $[m_1+m_2+r_{i}]_{vpk_i}$ and $AP_S$ to $CP$\\
			}

			\SetKwProg{Fn}{Steps~by~\textbf{CP}:}{}{}
			\Fn{}{
				$(m_1+m_2+r_{i}) \leftarrow S\dec_{VP}(ssk,[m_1+m_2+r_{i}]_{vpk_i})$\\
   $[m_1+m_2+r_{i}]_{ppk_j}\leftarrow \enc_{PE}(ppk_j,(m_1+m_2+r_{i}))$\\
   	$[psk_j]_{AP_S}\leftarrow\enc_{ABE}(\pk,psk_j,AP_S)$ \\ 
    Send $[m_1+m_2+r_{i}]_{ppk_j}$ and $[psk_j]_{AP_S}$ to SP
			}
			
			\SetKwProg{Fn}{Steps~by~\textbf{SP}:}{}{}
			\Fn{}{ 
			$[r_{i}]_{ppk_j}=\enc_{PE}(ppk_j,r_{i})$\\
			$[-r_{i}]_{ppk_j} \leftarrow ([r_{i}]_{ppk_j})^{n-1}$ \\
			
			  $[m_1+m_2]_{ppk_j} = [m_1+m_2+r_{i}]_{ppk_j}\times [-r_{i}]_{ppk_j}$\\
			  Send $[m_1+m_2]_{ppk_i}$ and $[psk_j]_{AP_S}$ to DR\\
			}

    	  \label{alg:the_alg1}

		\end{algorithm}

\textit{(ii) Masking}: SP masks the aggregated data. Specifically, it generates a random number and encrypts it with the DO's VP-HE public key, $vpk_i$, $[r_{i}]_{vpk_i} = \enc_{vp}(vpk_i,r_{i})$. The ciphertext is then multiplied with the ciphertext of the aggregated data to get a ciphertext of the masked aggregated data, $[\sum_{i=1}^{N_{req}} m_i + r_{i}]_{vpk_i}=[\sum_{i=1}^{N_{req}} m_i]_{vpk_i}*[r_{i}]_{vpk_i}$. The result is sent to CP along with the $AP_{S}$ set by the DO.  

\textit{(iii) Preparing the processing result}: 
CP decrypts the result using its strong decryption key to get the masked aggregate data, $\sum_{i=1}^{N_{req}} m_i + r_{i}=S\dec_{VP}(ssk,[\sum_{i=1}^{N_{req}} m_i + r_{i}]_{vpk_i})$.
A new Paillier key pair ($ppk_j,psk_j$) is generated by KA and sent to CP. $ppk_j$ is used to encrypt the masked aggregated data to get $[\sum_{i=1}^{N_{req}} m_i + r_{i}]_{ppk_j} = \enc_{PE}(ppk_j,\sum_{i=1}^{N_{req}} m_i + r_{i})$, while the $psk_j$ is encrypted by the DO defined access policy ($AP_S$) to get $[psk_j]_{AP_S} = \enc_{ABE}(\pk,psk_j,AP_S)$. 
Both ciphertexts are sent to SP.

\textit{(iv) De-masking}: SP initiates a de-masking process. It encrypts the random number $r_{i}$ (used in the masking process) with $ppk_j$ to obtain $[r_{i}]_{ppk_j}=\enc_{PE}(ppk_j,r_{i})$. Then, it calculates the additive inverse of $[r_{i}]_{ppk_j}$, generating $[-r_{i}]_{ppk_j} = [r_{i}]_{ppk_j}^{n-1}$. Finally, it de-masks the aggregated data: $[\sum_{i=1}^{N_{req}} m_i]_{ppk_j} = [\sum_{i=1}^{N_{req}} m_i + r_{i}]_{ppk_j}*[-r_{i}]_{ppk_j}$. Finally, SP sends $[\sum_{i=1}^{N_{req}} m_i]_{ppk_j}$ and $[psk_j]_{AP_S}$ to DR. 

\textit{(v) DR access the processing result}: DR can access the processing result only if the DR's key attributes satisfy the DO' $AP_S$. Hence, DR can decrypt and obtain $psk_j$ by using its ABE secret key $psk_j = \dec_{ABE}(\pk,[psk_j]_{AP_S},\sk)$.  Finally, it uses $psk_j$ to obtain the initially requested aggregated data of the DO:  $\sum_{i=1}^{N_{req}} m_i = \dec_{PE}(psk_j,[\sum_{i=1}^{N_{req}} m_i]_{ppk_j})$. 

\emph {c) DRs-DOs use case:}
A DR requests access to aggregated data of multiple DOs. 
The aggregated data can be accessed only by DRs whose attributes satisfy the $AP_M$ (see Alg.~\ref{alg:the_alg2}).

\textit{ (i) Handling DR request:} Upon receiving a data access request, SP initiates the process by comparing DO' $AP_{M}$ with DR attributes. It then selects DOs whose $AP_{M}$ matches with DR request. Let us assume SP selects $N$ data owners. 

\textit{ (ii) Masking:} SP generates random number for every DO's data and encrypts them with the corresponding DO's variant Paillier public key, $[r_{i}]_{vpk_i}=\enc_{vp}(vpk_i,r_{i})$. Next, each encrypted random number is multiplied with the respective DO's encrypted data to obtain $[m_i+r_{i}]_{vpk_i}=[m_i]_{vpk_i}*[r_i]_{vpk_i}$. Finally, the $N$ masked ciphertexts are sent to CP with the $AP_M$ set by the DO for further processing.

		\begin{algorithm}[t]
		\scriptsize
              \caption{SAMA: DRs-DOs use case.}
			\DontPrintSemicolon
			
			\KwInput{Two ciphertexts $[m_1]_{vpk_1}$ and $[m_2]_{vpk_2}$}
			\KwOutput{The ciphertexts $[m_1+m_2]_{ppk_j}$}
			  \label{alg:the_alg2}
			\SetKwProg{Fn}{Steps by~\textbf{SP}:}{}{}
			\Fn{}{
				$r_{1},r_{2} \overset{R}{\leftarrow}\mathbb{Z}_n$\\
				$[r_{1}]_{vpk_1}\leftarrow\enc_{vp}(vpk_1,r_{1})$\\
				$[r_{2}]_{vpk_2}\leftarrow\enc_{vp}(vpk_2,r_{2})$\\
			 $[m_1+r_{1}]_{vpk_1}=[m_1]_{vpk_1}\times[r_{1}]_{vpk_1}$\\
			 $[m_2+r_{2}]_{vpk_2}=[m_2]_{vpk_2}\times [r_{2}]_{vpk_2}$\\
				Send $[m_1+r_{1}]_{vpk_1}$, $[m_2+r_{2}]_{vpk_2}$, and $AP_M$ to CP\\
		
			}
			
			\SetKwProg{Fn}{Steps~by~\textbf{CP}:}{}{}
			\Fn{}{
				$(m_1+r_{1})\leftarrow S\dec_{VP}(ssk,[m_1+r_{1}]_{vpk_1})$\\
				$(m_2+r_{2})\leftarrow S\dec_{VP}(ssk,[m_2+r_{2}]_{vpk_2})$\\
				$(m_1+r_{1}+m_2+r_{2}) = (m_1+r_{1})+(m_2+r_{2})$
				
	           $[m_1+r_{1}+m_2+r_{2}]_{ppk_j}\leftarrow \enc_{PE}(ppk_j,(m_1+r_{1}+m_2+r_{2}))$\\
	      $[psk_j]_{AP_M}\leftarrow\enc_{ABE}(\pk,psk_j,AP_M)$\\
           Send $[m_1+r_{1}+m_2+r_{2}]_{ppk_j}$ and $[psk_j]_{AP_M}$ to SP
			}
			\SetKwProg{Fn}{Steps~by~\textbf{SP}:}{}{} 
			\Fn{}{
			$(r_{1}+r_{2})=r_{1}+r_{2}$ \\
        	$[r_{1}+r_{2}]_{ppk_j}\leftarrow\enc_{vp}(ppk_j,(r_{1}+r_{2}))$\\
        	$[-(r_{1}+r_{2})]_{ppk_j}\leftarrow([r_{1}+r_{2}]_{ppk_j})^{n-1}$\\
			
			  $[m_1+m_2]_{ppk_j} =[m_1+m_2+r_{1}+r_{2}]_{ppk_j}\times[-(r_{1}+r_{2})]_{ppk_j}$ \\
			  
			  Send $[m_1+m_2]_{ppk_i}$ and $[psk_j]_{AP_M}$ to DR\\
        	
				}
			
			
		\end{algorithm}

\textit{ (iii) Preparing the processing result:}
    First, CP decrypts all the received masked ciphertexts with the variant Paillier strong secret key to obtain the individual DOs' masked data: $ m_i+r_{i} = \dec_{vp}(ssk,[m_i+r_{i}]_{vpk_i})$. Then, it performs an addition operation to get the masked aggregation: $ \sum_{i=1}^{N} m_i + \sum_{i=1}^{N} r_{i}  = \sum_{i=1}^{N}(m_i + r_{i})$.
    Second, KA generates a new Paillier public-private key (${ppk_j,psk_j}$) for every authorised DR request received. 
    Third, CP encrypts the masked result using the Paillier public key, $[\sum_{i=1}^{N} m_i + \sum_{i=1}^{N} r_{i}]_{ppk_j}$= $\enc_{PE}(ppk_j,(\sum_{i=1}^{N} m_i + \sum_{i=1}^{N} r_{i}))$, while the corresponding private key $psk_j$ is encrypted with the common $AP_M$:  $ [psk_j]_{AP_M} = \enc_{ABE}(pk,psk_j,AP_M)$.
    CP sends both ciphertexts $[\sum_{i=1}^{N} m_i + \sum_{i=1}^{N} r_{i}]_{ppk_j}$, $[psk_j]_{AP_M}$ to SP.

\textit{ (iv) De-masking:} SP aggregates all the random numbers (used for masking) and encrypts the result, $[\sum_{i=1}^{N}r_{i}]_{ppk_j}=\enc_{PE}(ppk_j,\sum_{i=1}^{N}r_{i})$, computes additive inverse of $[\sum_{i=1}^{N}r_{i}]_{ppk_j}$, $[-\sum_{i=1}^{N}r_{i}]_{ppk_j} = [\sum_{i=1}^{N}r_{i}]_{ppk_j}^{n-1}$, and de-masks the result: 
$[\sum_{i=1}^{N}m_i]_{ppk_j}=([\sum_{i=1}^{N}m_i + \sum_{i=1}^{N}r_{i}]_{ppk_j})*([-\sum_{i=1}^{N}r_{i}]_{ppk_j})$.

\textit{ (v) DR access the processing result:} DR decrypts $[psk_j]_{AP_M}$ using $\sk$ if the DR's key satisfies the access policy: $ psk_j = \dec_{ABE}(\pk,[psk_j]_{AP_M},\sk)$. Finally, it uses $psk_j $ to access the data:  $ \sum_{i=1}^{N}m_i = \dec_{PE}(psk,[\sum_{i=1}^{N}m_i]_{ppk_j})$.

\section{Security Analysis}
\label{sec:security_analysi}

\subsection{Security of the SAMA Scheme} 
\label{sec:Cryptsystem_security}


The security analysis of SAMA is based on a simulation paradigm with the presence of semi-honest (honest-but-curious and non-colluding) adversaries. To prove the execution view of the $IDEAL$ world is computationally indistinguishable from the execution view of the $REAL$ world, we construct four simulators ($Sim_{DO}$, $Sim_{SP}$, $Sim_{CP}$, and $Sim_{DR}$) which represent DO, SP, CP, and DR. They simulate the execution of the following adversaries $Adv_{DO}$, $Adv_{SP}$, $Adv_{CP}$, and $Adv_{DR}$ that compromise DO, SP, CP, and DR, respectively. KA is excluded as a trustworthy entity.

$Theorem~1$. SAMA can securely retrieve in plaintext the result of the addition operation over encrypted data in presence of semi-honest adversary models and threat attacks.

$Proof$: We prove the security of the SAMA scheme by considering the case with two data inputs. 

\subsubsection{$Sim_{DO}$}

$Sim_{DO}$ encrypts the provided inputs $m_1$ and $m_2$ using VP-HE and returns both ciphertexts to $Adv_{DO}$. 
The simulation view of the $IDEAL$ world of $Adv_{DO}$ is computationally indistinguishable from the $REAL$ world view owing to the semantic security of VP-HE.

\subsubsection{$Sim_{SP}$}

$Sim_{SP}$ simulates $Adv_{SP}$ in DRs-DO and DRs-DOs cases. In the former case, $Sim_{SP}$ multiplies the provided ciphertexts and then encrypts a random number $r$ with VP-HE. It then multiplies the encrypted random number with the result of the multiplication of the ciphertexts. Later, this random number is encrypted with the public key of the Paillier scheme and its ciphertext is raised to $n-1$ and multiplied with the given ciphertext. 
In the later case, $Sim_{SP}$ generates two random numbers $r_1$ and $r_2$, encrypts them with the VP-HE public key and multiplies the encrypted random numbers with the ciphertexts (encrypted $m_1$ and $m_2$), respectively. Later, the same random numbers are encrypted with the Paillier public key, and the results are raised to $n-1$ and multiplied with the given ciphertext. 
In both cases, the $Adv_{SP}$ receives the output ciphertexts from $Sim_{SP}$. Therefore, the $REAL$ and $IDEAL$ views of $Adv_{SP}$ are computationally indistinguishable owing to the semantic security of VP-HE and Paillier encryption.

\subsubsection{$Sim_{CP}$}

The execution view of CP in the $REAL$ world is given by both ciphertext of $(m_1+r_1)$ and $(m_2+r_2)$, which are used to obtain $m_1+r_1$ and $m_2+r_2$ by executing decryption with the strong secret key on these ciphertexts ($r_1$ and $r_2$ are random integers in $\ZZ_n$). The execution view of CP in the $IDEAL$ world has two ciphertexts randomly selected in the $\ZZ_{n^2}$. $Sim_{CP}$ simulates $Adv_{CP}$ in both DRs-DO and DRs-DOs cases. 
In the former case, $Sim_{CP}$ simulates $Adv_{CP}$ as follows. 
It runs the strong decryption algorithm and obtains $m'_1+m'_2+r'$ and then the decryption result undergoes further encryption by the public key of Paillier scheme to obtain a new ciphertext. 
In the later case, $Sim_{CP}$ runs the strong decryption algorithm and obtains $m'_1+r'_1$ and $m'_2+r'_2$. It aggregates the decryption results, and then the aggregated result is further encrypted by the Paillier public key to obtain a ciphertext.
In both cases, a randomly generated number is encrypted with CP-ABE. Then, the two ciphertexts (generated by the Paillier and CP-ABE schemes) are provided by $Sim_{CP}$ to $Adv_{CP}$.
These ciphertexts are computationally indistinguishable between the $REAL$ and $IDEAL$ world of $Adv_{CP}$ since the CP is honest and the semantic security of VP-HE and Paillier cryptosystem, and the security of CP-ABE.

\subsubsection{$Sim_{DR}$}

$Sim_{DR}$ randomly selects chosen ciphertexts (besides not having access to challenged data), decrypts, and sends them to $Adv_{DR}$ to gain data information. The view of $Adv_{DR}$ is the decrypted result without any other information irrespective of how many times the adversary accesses $Sim_{DR}$. Due to the security of CP-ABE and the semantic security of the Paillier scheme, both $REAL$ and $IDEAL$ world views are indistinguishable. Since the DO data encryption process and DR decryption process are common for both DRs-DO and DRs-DOs in SAMA, the security proof of $Adv_{DO}$ and $Adv_{DR}$ is common for both cases. 

\subsection{Analysis against Security and Privacy Requirements}
\subsubsection{Data Confidentiality}

Every DO encrypts his/her data using his/her VP-HE public key $vpk_i$. SP then performs homomorphic addition operation over encrypted data, and delivers the ciphertext with the encrypted private Paillier key using CP-ABE to DR. Only authorised DRs can obtain the key and hence have access to the DO data. Furthermore, SAMA conceals DOs' raw data by adding random numbers at SP, i.e., masking the data, hence preserving the privacy of the DO(s) data at CP. Moreover, the Paillier cryptosystem is semantically secure and the CP-ABE is secure under the generic elliptic curve bi-linear group model as discussed in~\ref{sec:Cryptsystem_security}. In addition, the communication channels among all the entities (DO, SP, CP, and DR) are secure (e.g., encrypted using SSL). 
Therefore, only the authorised entities (i.e., DO/DR) can access the result and all the unauthorised internal/external entities who might eavesdrop messages sent and/or collect information can only access ciphertexts.

\subsubsection{Authorisation}

SAMA uses CP-ABE to implement secure access control, where the processing result is encrypted by data owner's defined access policies and the decryption key is associated with the attributes of the requesters. 
The user-centric access policy has been applied, which allows DO to define their access policies to securely and selectively grant DRs access to the processing result. Thus, the processing result is encrypted using $AP_S$ and $AP_M$, which are access policies set by DO to determine their preferences for sharing data processing results. Hence, the private key of DR (\sk) is required to decrypt the encrypted processing result using CP-ABE and only the authorised DR who satisfies the access policy can access the key and thereby decrypt the processing result.
Thus, using CP-ABE, SAMA provides user-centric access control and only authorised DR can access the processing result. 

\begin{table*}
\caption{Comparison of SAMA with related work.} 
\label{table:req_comp}
\begin{center}
\scriptsize
\begin{tabular}{l c c c c c c c c c c}
\toprule
\textbf{Requirement / Feature} & \textbf{\cite{ruj2013decentralized}} & \textbf{\cite{mustafa2015dep2sa}} &\textbf{\cite{bhowmik2023eeppda}} &
\textbf{\cite{zhang2022distributed}} & 
\textbf{\cite{tang2019efficient}} & \textbf{\cite{pang2020privacy}} &
{\textbf{\cite{zhang2021privacy}}} &
{\textbf{\cite{liu2019secure}}} & 
{\textbf{\cite{ding2017privacy}}} &
\textbf{SAMA} \\

\midrule




Flexible data processing request  & \xmark  & \xmark & \xmark & \xmark & \xmark & \xmark & \xmark &  \xmark & \xmark & \cmark \\

Fine grain data sharing  & \cmark & \xmark & \xmark & \xmark & \cmark  & \xmark & \xmark & \cmark& \cmark& \cmark \\

User-centric & \xmark & \xmark & \xmark  & \xmark &\cmark & \xmark &\xmark & \cmark & \xmark &\cmark\\

\midrule
Data confidentiality &\cmark$^*$& \cmark & \cmark & \cmark &\cmark  & \cmark  & \cmark  & \cmark& \cmark &  \cmark \\

Authorisation & \cmark & \cmark & \cmark & \cmark & \cmark  & \cmark  & \cmark & \cmark& \cmark & \cmark \\

\midrule
Multi-key setting   & \xmark & \xmark & \xmark  & \xmark & \xmark & \cmark &\cmark & MPC  & \xmark &\cmark \\


Trust level on service providers & FT$^*$ & ST & ST & ST & ST & ST & ST & ST & ST & ST\\

Supported operations & ADD & ADD & ADD & ADD & CLASS & MULT, COMP & All & ADD, MULT& All & ADD \\ 






\bottomrule
\end{tabular}
\begin{flushleft}
*Agg. results are decrypted by a trusted party. { 
\newline ST - Semi-trusted; FT - Fully-trusted; ADD - Addition; CLASS - Classification; MULT - Multiplication; DIV - Division; COMP - Comparison; All - All operation without classification.} 
\end{flushleft}

\end{center}
\end{table*}

\subsection{Comparison with Related Work}

Table~\ref{table:req_comp} provides a comparison of SAMA with the closely related existing scheme that combine both data processing and sharing with respect to our design requirements.

In terms of flexible data processing requests, SAMA is the only scheme that supports all three use cases (DO-DO, DRs-DO, and DRs-DOs), whereas, other schemes support only either one or two of the use cases. For a complete breakdown of which scheme supports which cases, we refer the reader to Table~\ref{table:usecase_comp}. 

In SAMA, to achieve fine grain data sharing, the resource-intensive CP-ABE is outsourced to the cloud, reducing the burden on data owners 
in a way suitable for resource-constrained devices similar to \cite{ding2017privacy, tang2019efficient, liu2015secure}. 
Furthermore, like SAMA, the schemes~\cite{liu2019secure, tang2019efficient} offer a user-centric access policy approach. However, they all provide only one access policy setting option for data owners to control their own data processing and sharing (DRs-DO). Unlike other schemes, SAMA allows data owners to control whether their data is included with other multiple data owners' data (DRs-DOs).
All schemes achieve data confidentiality using one of these techniques (homomorphic encryption, MPC, and symmetric encryption), however, the scheme \cite{ruj2013decentralized} does not achieve end-to-end encryption.
For authorisation, all schemes allow only authorised entities to access data.

As discussed earlier, the schemes~\cite{ruj2013decentralized,mustafa2015dep2sa,ding2017privacy,bhowmik2023eeppda} use public key-based PHE schemes in a single-key setting, which requires data to be encrypted with a third party public key, thereby having multiple DOs encrypt their own data with a key that does not belong to them. Hence, these schemes may not be suitable for highly sensitive data applications. Further, using a single-key HE setting does not allow data owners to access their own data directly, as the data are not encrypted with their public keys. 


In summary, there are limited efforts to explore the integration of privacy-preserving flexible data processing with fine-grain sharing under user-centric access control. The state-of-the-art research either addresses the issue of privacy preserving data processing and sharing separately or cannot satisfy the diverse demand of modern health systems to accommodate all three use cases efficiently, as shown in Table 
\ref{table:req_comp}. Our experimental evaluation and comparison with~\cite{ding2017privacy} (see next section) demonstrates that SAMA is more efficient in terms of computation and communication cost. To make SAMA suitable for resource-constrained devices, we ensure that data owners encrypt their data only once with their public key to satisfy all three use cases without further interaction with the cloud.

\section{Performance Evaluation}
\label{sec:performance_evaluation}

In this section, we evaluate SAMA in terms of computational complexity and communication overheads. 

\subsection{Computational Complexity}
The computationally expensive operations considered in the SAMA scheme are modular exponentiation and multiplication operations, denoted as $ModExp$ and $ModMul$, respectively. We ignore the fixed numbers of modular additions in our analysis as their computational cost is negligible compared to $ModExp$ and $ModMul$. 
In our analyses we also use the following parameters: 
$BiPair$ -- cost of bilinear pairing in ABE; $|\gamma|+1$ -- number of attributes in the access policy tree; $\vartheta$ -- number of attributes needed to satisfy the access policy. 

\subsubsection{Computational Complexity of HE Data Aggregation}
\label{subsec:Computational_Complexity}

In our analysis, we split the computational complexity into four parts: the complexity of each entity.


\textit{Computations at DOs:}
At each reporting time slot, each DO encrypt their data by their VP-HE public key $vpk_i$ to generate a ciphertext for data processing/analysing. This encryption requires two modular exponentiation operations and one modular multiplication. Hence the computational complexity at the DO side is $2ModExp + ModMul$. 

\textit{Computations at SP and CP:}
\label{subsubsec:Computational_Complexity_of_CSP}
This includes operations performed by SP and CP. As these operations are slightly different for the DRs-DO and DRs-DOs cases, we analyse them separately.  

For the DRs-DO case, SP performs additive homomorphic encryption on the received data owner ciphertexts $((N_m - 1)ModMul)$
, generates a random number $r$, encrypts it with the $DO$'s VP public key $vpk_i$ $(2ModExp + ModMul)$, multiplies the results of the homomorphic addition with the encrypted random number $(ModMul)$
and sends it to $CP$. Next, $SP$ re-encrypts the generated random number $r$ by $ppk_j$ $(2ModExp + ModMul)$, calculates the additive inverse of $r$ and then multiplies it with the encrypted processing result to remove the masking from the original data $(ModMul)$. Thus, $SP$ performs in total: $4ModExp + (N_{m}+3)ModMul$.
$CP$ performs strong decryption using $ssk$ on the received ciphertexts $(ModExp + ModMul)$.
It then encrypts the aggregated masked result with $ppk_j$ $(2ModExp + ModMul)$, and encrypts $psk_j$ with CP-ABE using $AP_S$ $((|\gamma|+1)Exp)$. Hence, $CP$ performs in total: $3ModExp + 2ModMul + (|\gamma|+1)Exp$.
In total, the cost at $SP$ and $CP$ in a DRs-DO case is: $7ModExp + (N_{m}+5)ModMul + (|\gamma|+1)Exp$.

For the DRs-DOs case, $SP$ generates a random number for every $DO$'s data, encrypts them using the VP public key of the corresponding $DO$, $vpk_i$, $(N_{m}(2ModExp + ModMul))$, and then multiplies the resulting ciphertexts with the ciphertexts received from $DO$s $(N_{m}ModMul)$.
Later, it aggregates all the generated random numbers,  
encrypts it using $ppk_j$ $(2ModExp + ModMul)$, calculates the additive inverse of the aggregation result, and then multiplies the aggregation result ciphertext with the received ciphertext from ${CP}$ to remove the masking from the original data $(ModMul)$. 
Thus, the computational cost of $SP$ in DRs-DOs case is: 
$(2N_{m}+2)ModExp + (2N_{m}+2)ModMul$.
$CP$ performs strong decryption using $ssk$ for all $N$ received ciphertexts $(N_{m}(ModExp + ModMul))$
, and then aggregates the decryption result. 
Next, it encrypts the addition result with a Paillier public key $ppk_j$ $(2ModExp + ModMul)$
, and then encrypts $psk$ with CP-ABE using $AP_M$ $(|\gamma|+1)Exp)$.
Hence, the total computation cost of $SP$ and $CP$ in DRs-DOs case is: 
$(N_{m}+2)ModExp + (N_{m}+1)ModMul + (|\gamma|+1)Exp$.
Therefore, in total, computational complexity of both  in DRs-DO case is: 
$(3N_{m} + 4)ModExp + (3N_{m} + 3)ModMul + (|\gamma|+1)Exp$.

\textit{Computations at $DRs$:}
In DRs-DO and DRs-DOs use cases, a $DR$ decrypts the ABE ciphertext using his/her $\sk$ to obtain the Paillier decryption key $psk_j$ (at most $\vartheta BiPair$), and then uses it to decrypt the encrypted processing result ($ModExp + ModMul$). In total, this gives a computational cost at $DR$: $(ModExp + ModMul + \vartheta BiPair)$. 

We compare the total computational costs of each entity in SAMA with the addition scheme of~\cite{ding2017privacy} in Table~\ref{table:Computation_Cost}.

\begin{table}[t]
\centering
\caption {Computation Cost.} 
\label{table:Computation_Cost}
\scriptsize
\begin{tabular}{p{1.1cm} p{6.5cm}}
\toprule
 & \textbf{SAMA: DRs-DO case} \\
\midrule
$DO$  &  $2ModExp+ModMul$   \\

$SP+CP$  & $7ModExp + (N_{m}+5)ModMul + (|\gamma|+1)Exp$ \\

$DR$ & $ModExp$ +  $ModMul$ +  $\vartheta BiPair$  \\

\midrule

& \textbf{SAMA: DRs-DOs case} \\

\midrule

$DO$  &  $2ModExp+ModMul$  \\

$SP+CP$  & $(3N_{m}+4)ModExp + (3N_{m}+3)ModMul + (|\gamma|+1)Exp$ \\

$DR$ & $ModExp$ +  $ModMul$ +  $\vartheta BiPair$ \\

\midrule

& \textbf{Scheme in~\cite{ding2017privacy}: DRs-DOs case (ADD operation)} \\

\midrule

$DO$  &  $2ModExp+ModMul$   \\

$SP+CP$  & $ 9ModExp + (N_{m}+4)ModMul  + 2(|\gamma|+1)Exp$\\

$DR$ &  $ModExp$ + $ModMul + ModInverse + \vartheta BiPair$ \\

\bottomrule
\end{tabular}
\medskip
\newline 
Note: all modular exponentiation and modular multiplications modulo are under $n^2$
\end{table}


\subsubsection{Computational Complexity of Access Control} 
We assume there are $|U|$ universal attributes, in which $|\gamma|$ attributes are in the access policy tree $\tau$, and at most $\vartheta$ attributes should be satisfied in $\tau$ to decrypt the ciphertext.
The $Setup()$ will generate the public parameters using the given system parameters and attributes $U$. This requires $|U|+1$ exponentiations and one bi-linear pairing.
The $\enc_{ABE}()$ requires two exponential operations for each leaf in the ciphertext’s access tree $\tau$, which needs $(|\gamma|+1)Exp$, whereas the $\kgen_{ABE}()$ algorithm requires two exponential operations for every attribute given to the $DO$. Also, the private key consists of two group elements for every attribute. 
Finally, $\dec_{ABE}()$ requires two pairings for every leaf of the access tree $\tau$ matched by a private key attribute and at most one exponentiation for each node along a path from that leaf to the root node.


\subsection{Communication Overhead}

\begin{table}[t]
\caption{Communication Overhead.} 
\label{table:Communication_Overhead}
\scriptsize
\centering
\begin{tabular}{p{2cm} p{5cm}}
\toprule
\textbf{} & \textbf{SAMA: DRs-DO case}  \\
\midrule

$DOs$-to-$SP$  & $2NL(n)$    \\

{$SP$}-to-{$CP$} &  $4L(n)+(|\gamma|+1)\mathcal{L}$ \\

$SP$-to-$DR$ & $2L(n)+(|\gamma|+1)\mathcal{L}$  \\

\midrule
\textbf{} & \textbf{SAMA: DRs-DOs case} \\
\midrule

$DOs$-to-$SP$  & $ 2NL(n)$   \\

{$SP$}-to-{$CP$} & $2(N+1)L(n) +(|\gamma|+1)\mathcal{L}$ \\

$SP$-to-$DR$ & $2L(n)+(|\gamma|+1)\mathcal{L}$ \\

\midrule
\textbf{} & \textbf{Scheme in~\cite{ding2017privacy}: DRs-DOs case (ADD operation)} \\
\midrule

$DOs$-to-$SP$ & $ 4NL(n)$   \\

{$SP$}-to-{$CP$} & $8L(n)+(|\gamma|+1)\mathcal{L}$\\

$SP$-to-$DR$ & $4L(n)+(|\gamma|+1)\mathcal{L}$ \\

\bottomrule
\end{tabular}
\end{table}

There are two types of communication overhead incurred in the SAMA scheme: overhead due to occasional data communication and overhead due to regular data communication. The former overhead captures the data sent occasionally, e.g., AP $(AP_S, AP_M)$ uploads/updates and notifications. The latter overhead includes the regular data communication patterns within SAMA, such as data upload, data requests, and data exchanged between cloud providers when data is being processed. Since the former overhead is negligible compared to the latter overhead, here we focus only on the communication overhead due to regular data communication patterns.

 
To ease the analyses, we divide the communication overhead introduced by the SAMA scheme into three parts: overhead incurred (1) between DOs and SP denoted as (DOs-to-SP), (2) between SP and CP (SP-to-CP), and (3) between SP and DRs (SP-to-DRs).


\subsubsection{DOs-to-SP} This is a common step for DRs-DO and DRs-DOs use cases. At each data reporting time slot, each data owner $DO_i$ sends one ciphertext to SP. As each ciphertext has a length of $2L(n)$ (operations are performed under $mod$ $n^2$), the total communication overhead for this part in the DRs-DO and DRs-DOs use cases is $N2L(n)$. 

\subsubsection{SP-to-CP} 

The communication between $SP$-to-$CP$ in the DRs-DO use case is as follows.
SP sends one ciphertext of length $2L(n)$, which is the masked aggregated DO's data, to CP. Then, CP sends one ciphertext of $2L(n)$ to SP, which is the masked encrypted processing result, and one CP-ABE ciphertext of $(|\gamma|+1)\mathcal{L}$, where $\mathcal{L}$ is the bit length of elements in ABE. Therefore, the total communication among SP-to-CP in the DRs-DO case is: $4L(n) + (|\gamma|+1)\mathcal{L}$.

The communication between SP-to-CP in DRs-DOs is as follows.
SP sends $N$ ciphertext (masked of encrypted DO's data) of length $2L(n)$ to CP, which is $2NL(n)$.
Then, similar to the DRs-DO scenario, $CP$ sends one ciphertext of $2L(n)$ and $(|\gamma|+1)\mathcal{L}$ of the CP-ABE ciphertext to SP.
The total communication cost among SP-to-CP in DRs-DOs case is:
$(N+1)2L(n) + (|\gamma|+1)\mathcal{L}$.  


\subsubsection{SP-to-DRs} 
In the DRs-DO and DRs-DOs, SP sends to DRs one ciphertext of length $2L(n)$ (the encrypted processing result) and one CP-ABE ciphertext of length $(|\gamma|+1)\mathcal{L}$. Thus, The communication between SP and the DRs is: $2L(n) + (|\gamma|+1)\mathcal{L}$.

A comparison between the communication overhead of the SAMA scheme and the addition scheme proposed in~\cite{ding2017privacy} is shown in Table~\ref{table:Communication_Overhead}. Overall, SAMA has a lower communication overhead than the Addition scheme in~\cite{ding2017privacy} at the DO and DR side, while, the overhead between SP-to-CP in DRs-DOs of the SAMA scheme is higher than~\cite{ding2017privacy}. 

\subsection{Experimental Results}

We present the experimental results of SAMA in three different settings: (1) computational cost of the data processing operations, (2) computational cost of the data access operations, and (3) communication overheads within SAMA.

For the computational cost, we have implemented the SAMA scheme to test its computational performances by conducting experiments with Java Pairing-Based Cryptography (jPBC) 
\cite{ISCC:DecIov11} and Java Realization for Ciphertext-Policy Attribute-Based Encryption (cpabe)\cite{wang2012java} libraries on a laptop with Intel Core i7-7660U CPU 2.50GHz and 8GB RAM. We ran each experiment 500 times and took the average values. We set the length of $n$ to 1024 bits, $m$ to 250 bits, and $r$ to 500 bits according to \cite{ding2017privacy} to make the comparison compatible. 
We show the computation evaluation for the DRs-DO and DRs-DOs use cases for all entities separately and specifically SP and CP as they perform different sets of computations in each case as described in Section~\ref{subsubsub:data_processing_efficency}. In addition, the efficiency of user-centric access control and communication overhead among the entities are shown in Section~\ref{subsubsub:access_control_efficency} and Section~\ref{subsubsub:Communication_cost_efficency} respectively.

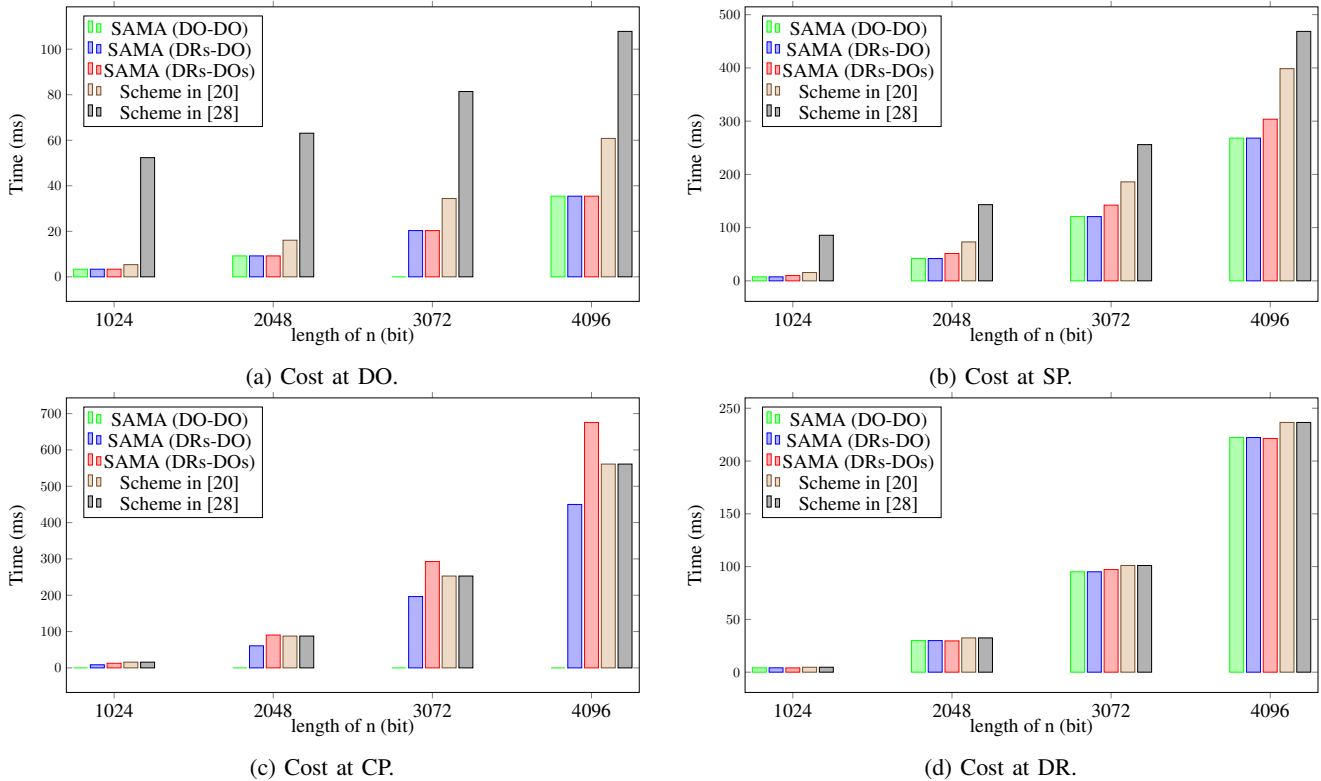
\begin{figure}[t!]
\centering
\begin{subfigure}[b]{0.24\textwidth}
\begin{adjustbox}{width=\linewidth} 
\begin{tikzpicture}
\begin{axis} [ybar,height=7cm,width=9cm,
      xticklabels={\Large{1024},\Large{2048},\Large{3072},\Large{4096}},
      xtick={1,2,3,4},
            ylabel= \Large{Time (ms)},
            xlabel=\Large{length of n (bit)},
            legend pos=north west,
          ]

\addplot[blue,fill=blue!30!white] coordinates {
     (1,3.354) 
    (2,9.198) 
    (3,20.326)
    (4,35.43)
    };

\addplot coordinates {
     (1,3.354) 
    (2,9.198) 
    (3,20.326)
    (4,35.43)
    };

\addplot[brown!60!black,fill=brown!30!white] coordinates {
     (1,5.354) 
    (2,16.088) 
    (3,34.396)
     (4,60.802)}; 
\legend{\Large{SAMA (DRs-DO)}, \Large{SAMA (DRs-DOs)}, \Large{Scheme in \cite{ding2017privacy}}} 
\end{axis}
\end{tikzpicture}
\end{adjustbox}
\caption{Cost at DO.}
\label{fig_different_n_lengths_user}
\end{subfigure}
\hfill
\begin{subfigure}[b]{0.24\textwidth}
\begin{adjustbox}{width=\linewidth} 
\begin{tikzpicture}
\begin{axis}  [ybar,height=7cm,width=9cm,
      xtick={1,2,3,4},
      xticklabels={\Large{1024},\Large{2048},\Large{3072},\Large{4096}},
      legend pos=north west,
            ylabel= \Large{Time (ms)},
            xlabel=\Large{length of n (bit)}]

\addplot[blue,fill=blue!30!white] coordinates { 
    (1,7.532) 
    (2,42.016) 
    (3,120.708)
     (4,268.14)
};
\addplot coordinates { 
    (1,9.99) 
    (2,51.378) 
    (3,142.222)
    (4,303.646)};
\addplot[brown!60!black,fill=brown!30!white] coordinates { 
    (1,15.646) 
    (2,73.124) 
    (3,185.93) 
     (4,398.608)
};

\legend{\Large{SAMA (DRs-DO)}, \Large{SAMA (DRs-DOs)}, \Large{Scheme in \cite{ding2017privacy}}} 
\end{axis}
\end{tikzpicture}
\end{adjustbox} 
\caption{Cost at SP.}
\label{fig_different_n_lengths_SP}
\end{subfigure}
   \hfill
\begin{subfigure}[b]{0.24\textwidth}
\begin{adjustbox}{width=\linewidth} 
\begin{tikzpicture}
\begin{axis}  [ybar,height=7cm,width=9cm,
      xtick={1,2,3,4},
      xticklabels={\Large{1024},\Large{2048},\Large{3072},\Large{4096}},
      legend pos=north west,
       ylabel= \Large{Time (ms)},
       xlabel=\Large{length of n (bit)}]
       
\addplot[blue,fill=blue!30!white] coordinates {
    (1,8.572) 
    (2,60.77) 
    (3,196.392) 
     (4,449.94)};
           
\addplot coordinates {
    (1,12.67) 
    (2,90.44)
    (3,292.794)
     (4,675.714)};

\addplot[brown!60!black,fill=brown!30!white] coordinates {
    (1,15.8) 
    (2,87.622) 
    (3,252.864)
     (4,561.094)}; 

\legend{\Large{SAMA (DRs-DO)}, \Large{SAMA (DRs-DOs)}, \Large{Scheme in \cite{ding2017privacy}} } \end{axis}
\end{tikzpicture}
\end{adjustbox}
\caption{Cost at CP.}
\label{fig_different_n_lengths_CP}
\end{subfigure}
  \hfill
\begin{subfigure}[b]{0.24\textwidth}
\begin{adjustbox}{width=\linewidth} 
\begin{tikzpicture}
\begin{axis} [ybar,height=7cm,width=9cm,
      xticklabels={\Large{1024},\Large{2048},\Large{3072},\Large{4096}},
      xtick={1,2,3,4},
            ylabel= \Large{Time (ms)},
            xlabel=\Large{length of n (bit)},
            legend pos=north west]

\addplot[blue,fill=blue!30!white] coordinates {
     (1,4.132) 
    (2,29.914) 
    (3,95.116) 
    (4,222.276)};
    
\addplot coordinates {
     (1,4.068) 
    (2,29.612) 
    (3,97.216)
    (4,221.4)};
    
\addplot[brown!60!black,fill=brown!30!white] coordinates {
     (1,4.68) 
    (2,32.46) 
    (3,101.05) 
    (4,236.56)};

\legend{\Large{SAMA (DRs-DO)}, \Large{SAMA (DRs-DOs)}, \Large{Scheme in \cite{ding2017privacy}}}
\end{axis}
\end{tikzpicture}
\end{adjustbox}
\caption{Cost at DR.}
\label{fig_different_n_lengths_DR}
\end{subfigure}
\hfill
\caption{SAMA scheme cost with the different lengths of $n$.}
\label{fig_different lengths of $n$}
\end{figure}

\subsubsection{Computational Cost of Data Processing}
\label{subsubsub:data_processing_efficency}

We evaluate the computational cost for DO, SP, CP, and DR in both DRs-DO and DRs-DOs scenarios and compare with the related work~\cite{ding2017privacy} (DRs-DOs) in terms of different lengths of $n$. In addition, we show the computational cost of DRs-DO and DRs-DOs cases with a variable number of messages and DOs, respectively. 

\textit{(i) Influence of different lengths of $n$ on data processing:}
Figure~\ref{fig_different lengths of $n$} shows the influence of the different lengths of $n$ on data processing of two messages, where $n$=1024, 2048, 3072, and 4096 bits. We can observe that in Fig.~\ref{fig_different_n_lengths_user}, the computational cost of SAMA is low on the data owner side and the lowest among all the other entities because the data owner only needs to encrypt data once, which is suitable for resource-constrained devices. Note, although DOs encrypt their data once to support both DRs-DO and DRs-DOs cases, we show two separate bars for consistency. Moreover, since in the encryption, there is an extra addition and multiplication that depends on $n$ (key size) in~\cite{ding2017privacy} compared to SAMA, the experimental results show that SAMA's data owner side encryption is better than~\cite{ding2017privacy}. In our DRs-DO and DRs-DOs cases, $SP$ achieves better computational efficiency compared to the DSP scheme in~\cite{ding2017privacy}, as shown in Fig.~\ref{fig_different_n_lengths_SP}. The computational efficiency of CP in our DRs-DO is better than the CP of the scheme in~\cite{ding2017privacy} as shown in Fig.~\ref{fig_different_n_lengths_CP} Whereas the computational efficiency of $CP$ is slightly lower in our DRs-DOs compared to the CP of the scheme in~\cite{ding2017privacy}. The operation time of DR, as shown in Fig.~\ref{fig_different_n_lengths_DR} is marginally better than the scheme in~\cite{ding2017privacy}, since the decryption of~\cite{ding2017privacy} can not be optimised by pre-computation as the decryption is dependent on ciphertexts. Therefore, there is an extra $modInverse$ operation compared to SAMA. Whereas in SAMA, the denominator of decryption needs to be computed only once and it is not dependent on the ciphertext, hence it can be pre-computed.

We can observe that the computation cost is linearly increasing with the increase of the bit length of n among all of the entities; DO, SP, CP, and DR. However, as expected SP and CP computation costs increase much more rapidly with the increase of bit length of $n$ compared to the DO and DR in case of the DRs-DOs. The computational performance evaluation shown in Fig.~\ref{fig_different lengths of $n$} is consistent with our analysis in Section~\ref{subsec:Computational_Complexity}. In general, the above tests prove that the most computation costs are undertaken at SP and CP and DO/DR do not have much computation overhead. This result shows the practical advantage of SAMA with DOs and DRs. 
Also, overall our scheme performs better in terms of efficiency compared to the scheme in~\cite{ding2017privacy}, which supports only DRs-DOs case.



\begin{figure}[t!]
\centering
\begin{subfigure}[b]{0.24\textwidth}
\begin{adjustbox}{width=\linewidth} 
\begin{tikzpicture}
\begin{axis} [ybar,height=7cm,width=9cm,
      xticklabels={\Large{10},\Large{100},\Large{1000},\Large{10000}}, ymode=log,
      xtick={1,2,3,4},
            ylabel= \Large{Time (ms)},
            xlabel=\Large{No. of data messages / DOs},
            legend pos=north west]
\addplot coordinates {
 (1,32) 
 (2,252) 
 (3,2438) 
 (4,24782) 
};

\addplot coordinates {
    (1,2.5) 
    (2,2.5) 
    (3,2.5) 
    (4,2.5) 
};
\addplot[brown!60!black,fill=brown!30!white] coordinates {
     (1,2.6) 
    (2,2.6) 
    (3,2.6)
     (4,2.6)}; 

\legend{\Large{SAMA (DRs-DO)},\Large{SAMA (DRs-DOs)}, \Large{Scheme in \cite{ding2017privacy}}} 
\end{axis}
\end{tikzpicture}
\end{adjustbox}
\caption{Operation time at DO}
\label{userfig}
\end{subfigure}
\hfill
\begin{subfigure}[b]{0.24\textwidth}
\begin{adjustbox}{width=\linewidth} 
\begin{tikzpicture}
\begin{axis}  [ybar,height=7cm,width=9cm, xtick={1,2,3,4}, xticklabels={\Large{10},\Large{100},\Large{1000},\Large{10000}},legend pos=north west, ymode=log, 
ylabel= \Large{Time (ms)},xlabel=\Large{No. of data messages / DOs}]
\addplot coordinates {
    (1,15) 
    (2,16) 
    (3,31) 
    (4,108) 
};
\addplot coordinates {
    (1,26) 
    (2,251) 
    (3,2470) 
    (4,24923) 
};
\addplot[brown!60!black,fill=brown!30!white] coordinates {
     (1,7.905+13.16) 
    (2,10.015+9.955) 
    (3,35.57+10.07)
     (4,448.035+23.07)}; 

\legend{\Large{SAMA (DRs-DO)},\Large{SAMA (DRs-DOs)},\Large{Scheme in \cite{ding2017privacy}}} 
\end{axis}
\end{tikzpicture}
\end{adjustbox} 
\caption{Operation time at SP}
\end{subfigure}
 \hfill
\begin{subfigure}[b]{0.24\textwidth}
\begin{adjustbox}{width=\linewidth} 
\begin{tikzpicture}
\begin{axis}  [ybar,height=7cm,width=9cm,
      xtick={1,2,3,4},
      xticklabels={\Large{10},\Large{100},\Large{1000},\Large{10000}}, ymode=log, legend pos=north west,ylabel= \Large{Time (ms)}, xlabel=\Large{No. of data messages / DOs}]
    
\addplot coordinates {
    (1,8.5) 
    (2,8.5) 
    (3,8.5) 
    (4,8.5) 
};

\addplot coordinates {
    (1,46) 
    (2,419) 
    (3,4018) 
    (4,39668) 
};

\addplot[brown!60!black,fill=brown!30!white] coordinates {
     (1,17.075) 
    (2,14.385) 
    (3,14.48)
     (4,34.655)};  
     
\legend{\Large{SAMA (DRs-DO)},\Large{SAMA (DRs-DOs)},\Large{Scheme in \cite{ding2017privacy}} }
\end{axis}
\end{tikzpicture}
\end{adjustbox}
\caption{Operation time at CP}
\end{subfigure}
  \hfill
\begin{subfigure}[b]{0.24\textwidth}
\begin{adjustbox}{width=\linewidth} 
\begin{tikzpicture}
\begin{axis} [ybar,height=7cm,width=9cm, ymode=log, 
      xticklabels={\Large{10},\Large{100},\Large{1000},\Large{10000}},
       ymin=1,
       ymax=10,
      xtick={1,2,3,4}, 
   ylabel= \Large{Time (ms)},
    xlabel=\Large{No. of data messages / $DO$s},
            legend pos=north west]
\addplot coordinates {
     (1,4) 
    (2,4) 
    (3,4) 
    (4,4) 
};    
\addplot[red,fill=red!30!white]  coordinates {
     (1,4) 
    (2,4) 
    (3,4) 
    (4,4) 
};
\addplot[brown!60!black,fill=brown!30!white] coordinates {
     (1,4.1) 
    (2,4.1) 
    (3,4.1)
     (4,4.1)}; 

\legend{\Large{SAMA (DRs-DO)}, \Large{SAMA (DRs-DOs)},\Large{Scheme in \cite{ding2017privacy}}} 
\end{axis}
\end{tikzpicture}
\end{adjustbox}
\caption{Operation time at $DR$}
\end{subfigure}
\hfill
\caption{Cost with different numbers of messages/DOs.}
\label{single computation}
\end{figure}
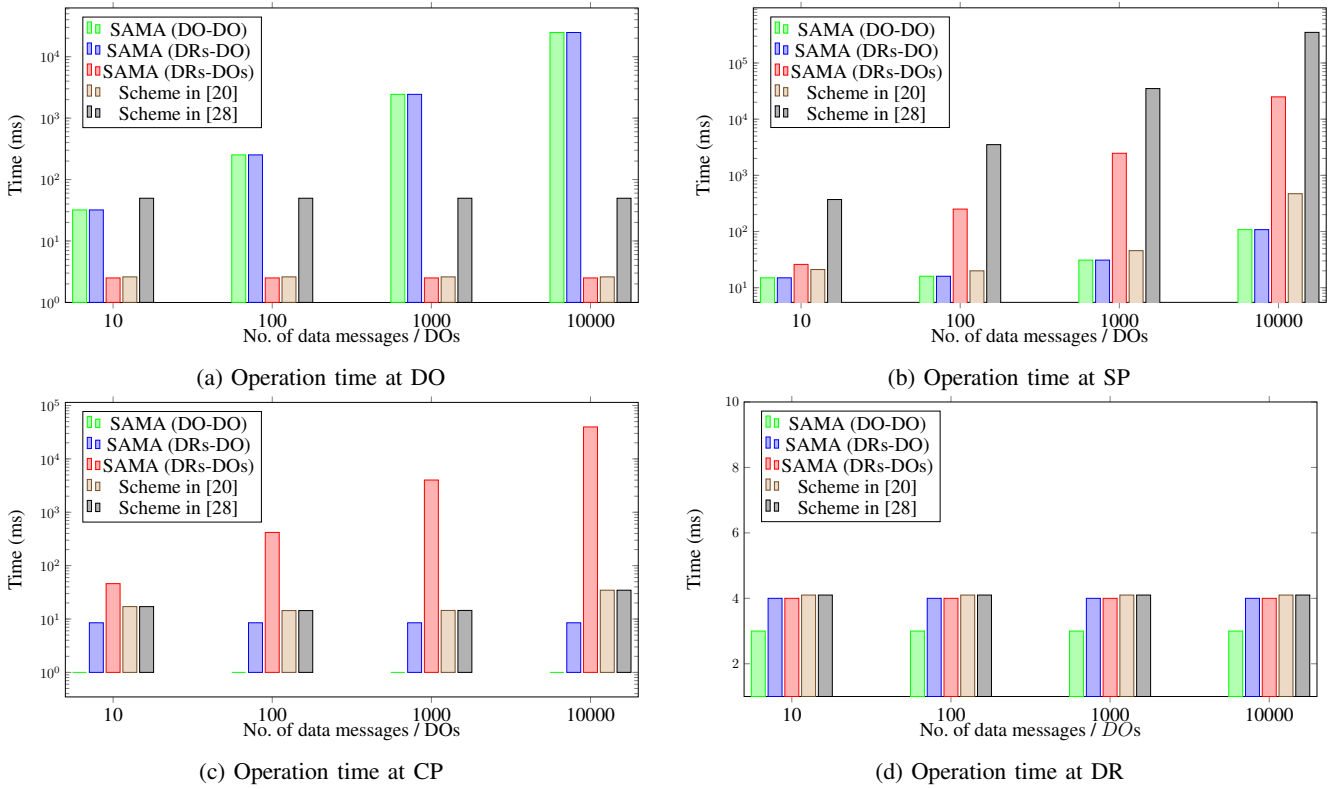

\begin{figure}[t!]
\centering
\begin{adjustbox}{width=0.7\linewidth} 
\begin{tikzpicture}
\begin{axis}  [ybar,height=7
cm,width=12cm,
           xtick={1,2,3,4,5},
          legend pos=north west, xticklabels={\Large{2},\Large{4},\Large{6},\Large{8},\Large{10}},
            ylabel= \Large{Time (ms)},
            xlabel=\Large{No. of attributes}]
\addplot[teal,fill=teal!60!white]  plot coordinates  {(1,61)(2,69)(3,70)(4,71)(5,82)}; 
\addplot plot[pink!70!black,fill=magenta!60!white]  coordinates {(1,201)(2,330)(3,488)(4,628)(5,729)};
\addplot[titleColor!70!black,fill=titleColor] coordinates {(1,214)(2,358)(3,507)(4,653)(5,757)};
\addplot plot coordinates {(1,46)(2,48)(3,49)(4,59)(5,44)};
\legend{\Large{Setup},\Large{$\kgen_{ABE}$}, \Large{$\enc_{ABE}$},\Large{$\dec_{ABE}$}}  
\end{axis}
\end{tikzpicture}
\end{adjustbox} 
\caption{CP-ABE time with the different numbers of attributes.}
\label{cpabe}
\end{figure}
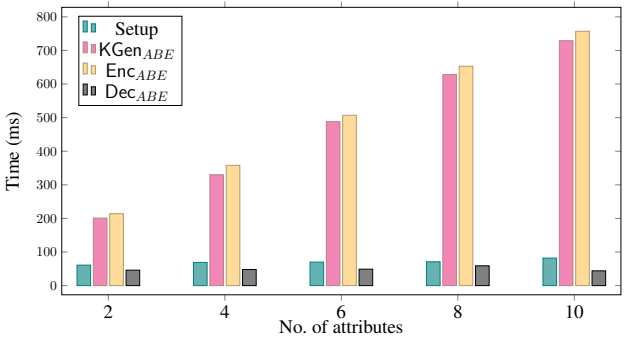

\textit{(ii) Performance of SAMA (DRs-DO case) with a variable number of provided messages:}
We tested the computation of SAMA's DRs-DO case by varying the number of data messages provided by a single DO, as shown in Fig.~\ref{single computation}. As can be seen from Fig.~\ref{userfig}, the operational time increases with the increase of the number of messages. However, only DR's and CP's operation times are independent of the number of messages because, in the DRs-DO case, the processed result is decrypted only once, regardless of the number of messages that are processed at the SP. 

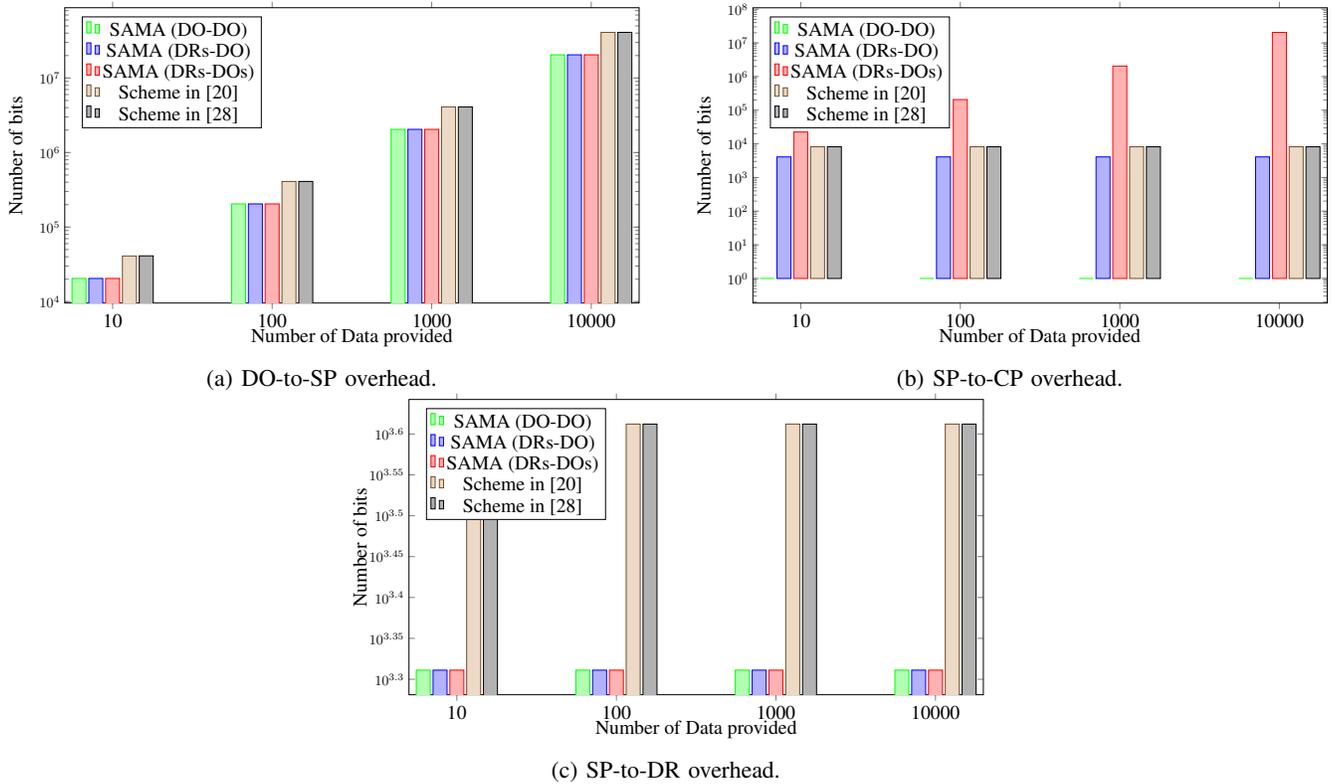
\begin{figure}[t]
\centering
\begin{subfigure}[b]{0.24\textwidth}
\begin{adjustbox}{width=\linewidth} 
\begin{tikzpicture}
\begin{axis} [ybar,height=7cm,width=9cm,
 xtick={1,2,3,4}, ymode=log, legend pos=north west,
 xticklabels={\Large{10},\Large{100},\Large{1000},\Large{10000}},
ylabel= \Large{Number of bits}, xlabel=\Large{Number of Data provided}]

\addplot coordinates {
   (1,20480) 
    (2,204800) 
    (3,2048000) 
    (4,20480000) 
};

\addplot coordinates {
   (1,20480) 
    (2,204800) 
    (3,2048000) 
    (4,20480000) 
};

\addplot[brown!60!black,fill=brown!30!white]  coordinates {
   (1,40960) 
    (2,409600) 
    (3,4096000) 
    (4,40960000)
};

\legend{\Large{SAMA (DRs-DO)}, \Large{SAMA (DRs-DOs)}, \Large{Scheme in~\cite{ding2017privacy}}} 
\end{axis}
\end{tikzpicture}
\end{adjustbox} 
\caption{{DO-to-SP} overhead.}
\label{commuser}
\end{subfigure}
\hfill
\begin{subfigure}[b]{0.24\textwidth}
\begin{adjustbox}{width=\linewidth} 
\begin{tikzpicture}
\begin{axis}  [ybar,height=7cm,width=9cm,
 xtick={1,2,3,4}, ymode=log, legend pos=north west,
 xticklabels={\Large{10},\Large{100},\Large{1000},\Large{10000}},
ylabel= \Large{Number of bits}, xlabel=\Large{Number of Data provided} ]

\addplot  coordinates {
   (1,4096) 
    (2,4096) 
    (3,4096) 
    (4,4096) 
};

\addplot[red,fill=red!30!white] coordinates {
   (1,22528) 
    (2,206848) 
    (3,2050048) 
    (4,20482048) 
};

\addplot[brown!60!black,fill=brown!30!white]  coordinates {
   (1,8192) 
    (2,8192) 
    (3,8192) 
    (4,8192)
};
\legend{\Large{SAMA (DRs-DO)}, \Large{SAMA (DRs-DOs)}, \Large{Scheme in~\cite{ding2017privacy}}}

\legend{\Large{SAMA (DRs-DO)}, \Large{SAMA (DRs-DOs)}, \Large{Scheme in~\cite{ding2017privacy}}} 
\end{axis}
\end{tikzpicture}
\end{adjustbox} 
\caption{{SP-to-CP} overhead.}
\label{commcsp}
\end{subfigure}
   \hfill
\begin{subfigure}[b]{0.24\textwidth}
\begin{adjustbox}{width=\linewidth} 
\begin{tikzpicture}
\begin{axis}  [ybar,height=7cm,width=9cm,
 xtick={1,2,3,4}, ymode=log, legend pos=north west,
 xticklabels={\Large{10},\Large{100},\Large{1000},\Large{10000}},
ylabel=\Large{Number of bits},xlabel=\Large{Number of Data provided}]

\addplot coordinates {
   (1,2048) 
    (2,2048) 
    (3,2048) 
    (4,2048) 
};

\addplot[red,fill=red!30!white]  coordinates {
   (1,2048) 
    (2,2048) 
    (3,2048) 
    (4,2048) 
};

\addplot[brown!60!black,fill=brown!30!white]  coordinates {
   (1,4096) 
    (2,4096) 
    (3,4096) 
    (4,4096)
};
\legend{\Large{SAMA (DRs-DO)}, \Large{SAMA (DRs-DOs)}, \Large{Scheme in~\cite{ding2017privacy}}} 
\end{axis}
\end{tikzpicture}
\end{adjustbox} 
\caption{{SP-to-DR} overhead.}
\label{commdr}
\end{subfigure}
\caption{Communication overhead of SAMA.}
\label{communiction}
\end{figure}

\textit{(iii) Performance of SAMA (DRs-DOs case) with a variable number of data owners:}
We tested the performance of SAMA's DRs-DOs case by varying the number of DOs ($N_{
DO}$= 10, 100, 1000, 10000) and fixing each DO to generate only one message for data processing. As expected, the SP and CP have more operation time compared to the DO and DR. Moreover, as shown in Fig.~\ref{single computation}, 
the $SP$ and $CP$ operation time are higher in the DRs-DOs case compared to the~\cite{ding2017privacy} and DRs-DO case. Since VP-HE is similar to PE used in~\cite{ding2017privacy} that supports only single-key homomorphic addition, our DRs-DO case processing is comparable to the ~\cite{ding2017privacy} and our DRs-DOs computation time is higher only at the cloud side (SP and CP) due to separate masking and de-masking for each of the messages. 


\subsubsection{Efficiency of User-Centric Access Control}
\label{subsubsub:access_control_efficency}

We tested the computational efficiency of CP-ABE by varying the number of attributes from two to ten that are involved in the access policy as shown in Fig.~\ref{cpabe}. The $Setup$ algorithm is relatively constant as it does not depend on the number of attributes. In addition, the $\dec_{ABE}$ in the test was set to require only one attribute needed to satisfy the access policy tree, therefore, the operation time of $\dec_{ABE}$ is constant.
The computational costs of $\enc_{ABE}$ and $\kgen_{ABE}$ are linearly increasing with the increase in number of the attributes. Although employing CP-ABE achieves user-centric fine-grain access control, there is an additional computation overhead incurred. 

\subsubsection{Communication Efficiency} 
\label{subsubsub:Communication_cost_efficency}

The communication overhead among the entities is shown in Fig.~\ref{communiction} and it is evaluated by fixing the key size length $n=1024$ bits and varying the number of messages to be computed.
It is evident from Fig.~\ref{commuser}, the DO-to-SP communications at the SAMA scheme reduce the communication overhead by $50\%$ 
compared to the scheme in~\cite{ding2017privacy}. Furthermore, it is essential to note that the scheme in \cite{ding2017privacy} supports only DRs-DOs case by encrypting data with CSP's public key. To support DO-DO case, they need to re-encrypt the same data again with the DO's public key as mentioned in~\cite{ding2017privacy}. 
Therefore, if we also compare the DRs-DO communication overhead of the scheme in~\cite{ding2017privacy} at the $DO$ side, our scheme reduces the communication overhead by $75\%$. 
At SAMA, a DO has to encrypt wearable data only once for DRs-DO and DRs-DOs cases compared to the scheme in~\cite{ding2017privacy}, which requires encrypting the data owner's data twice to support both DRs-DO and DRs-DOs.
In addition, the scheme in~\cite{ding2017privacy} generates two ciphertexts for every data encryption, which increases communication overhead on the DO side. While in the SAMA scheme, only one ciphertext is generated. Clearly, we reduced the communication overhead significantly on the DO side, which suits the resource-constrained devices. These results are consistent with the results obtained in~\cite{pang2020privacy}, which compares the communication overhead of the two HE algorithms: BCP and VP-HE. They found that the communication cost of BCP is about twice that of VP-HE, which was used in~\cite{ding2017privacy}.

Fig.~\ref{commcsp} depicts the communication overhead among the cloud servers: (SP-to-CP and CP-to-SP). Although our DRs-DO case achieves better communication efficiency compared to~\cite{ding2017privacy}, our DRs-DOs case communication performance is significantly higher than the DRs-DOs case of~\cite{ding2017privacy}. However, since SP and CP are not limited in resources, they can afford to support this higher communication overhead for DRs-DOs case. Moreover, the frequency of DRs-DOs case is relatively less than DRs-DO case in most wearable and healthcare use cases that are more personalized. We achieve better communication efficiency with the most frequent DRs-DO case. Therefore, our scheme is suitable mainly for applications that require more frequent DRs-DO cases than DRs-DOs cases such as wearables and outsourced personalized healthcare data processing.

The communication overhead of the SP-to-DR part is shown in Fig.~\ref{commdr}. As DRs access only the processed result, there is less communication overhead between the SP and DR. It is clear that our DRs-DO and DRs-DOs cases perform better than the scheme in~\cite{ding2017privacy} which supports only DRs-DOs case. Therefore, overall our scheme has significantly less total communication overhead compared to~\cite{ding2017privacy}.

 
\section{Conclusion}
\label{sec:conclusion}

In this paper, we have designed and evaluated a novel flexible data processing with a fine-grain sharing scheme called SAMA that also supports user-centric access control. It addresses the diverse demand for modern wearable healthcare applications by accommodating all three main use-case scenarios: DO-DO, DRs-DOs, and DRs-DOs. To achieve this, SAMA combines the use of multi-key VP-HE and CP-ABE to support efficient and privacy-preserving aggregation with fine-grain sharing. In addition, SAMA is suitable for resource-constrained devices like IoTs as we ensure that data owners encrypt their data only once with their own public key to satisfy all three use case processing requirements without any further interaction with the cloud. Our experimental evaluation and comparison with the closest previous work~\cite{ding2017privacy} demonstrate that SAMA is more efficient in terms of computation and communication cost. Further, the resource-intensive CP-ABE is outsourced to the cloud, reducing the burden on data owners and yet allowing them to set two different access policies ($AP_S$ and $AP_M$) to control their data and thereby achieve user-centric access policy. Our security analysis through the simulation paradigm shows that SAMA is secure and fulfills the specified set of security and privacy requirements.


\ifCLASSOPTIONcaptionsoff
  \newpage
\fi

\bibliographystyle{IEEEtran}
\bibliography{thebibliography.bib}

\end{document}